\documentclass[a4paper,11pt]{article}
\pdfoutput=1 

\usepackage{jcappub} 

\usepackage[T1]{fontenc} 

\usepackage{graphicx}
\usepackage{float}
\usepackage{hyperref}
\usepackage{xstring}
\usepackage[dvipsnames]{xcolor}
\usepackage{amssymb, amsmath, mathtools}
\usepackage{listings}
\usepackage{lscape}
\usepackage{lipsum}
\usepackage{multicol}
\usepackage{physics}
\usepackage{natbib}
\usepackage{txfonts}
\usepackage[small]{caption}
\usepackage{xspace}
\usepackage{comment}
\usepackage[normalem]{ulem} 

\newcommand{\pot}[2]{#1 \times 10^{#2}}

\newcommand{\boost}{x\partial_x}

\newcommand{\boostO}{\hat{\mathcal{O}}_x}

\newcommand{\KompO}{\hat{\mathcal{K}}_x}

\newcommand{\DiffO}{\hat{\mathcal{D}}_x}
\newcommand{\DiffstarO}{\hat{\mathcal{D}}^*_x}

\newcommand{\KompOp}{\hat{\mathcal{K}}_{x'}}

\newcommand{\DiffstarOp}{\hat{\mathcal{D}}^*_{x'}}

\newcommand{\nbb}{n_{\rm bb}}
\newcommand{\Gspec}{{G}}
\newcommand{\Yspec}{{{Y}}}
\newcommand{\Ynspec}[1]{{Y}_{#1}}
\newcommand{\Mspec}{{M}}

\newcommand{\COBEF}{{\it COBE/FIRAS}\xspace}
\newcommand{\Planck}{{\it Planck}\xspace}
\newcommand{\LiteBird}{{\it LiteBird}\xspace}

\newcommand{\Mpc}{{\rm Mpc}}
\newcommand{\GHz}{{\rm GHz}}

\newcommand{\expf}[1]{{{\rm e}^{#1}}}

\newcommand{\Tz}{{T_{z}}}

\newcommand{\TCMB}{T_{\rm CMB}}

\newcommand{\vbetah}{{\hat{{\boldsymbol\beta}}}}

\newcommand{\vgh}{{\hat{\boldsymbol\gamma}}}

\newcommand{\xc}{x_{\rm c}}

\newcommand{\id}{{\,\rm d}}

\newcommand{\beq}{\begin{equation}}   %

\newcommand{\eeq}{\end{equation}}   %

\newcommand{\beqa}{\begin{eqnarray}}   %

\newcommand{\eeqa}{\end{eqnarray}}   %

\newcommand{\bealf}[1]{\begin{align} #1 \end{align}}

\newcommand{\beal}{\begin{align}}
\newcommand{\enal}{\end{align}}

\newcommand{\bspl}{\begin{split}}

\newcommand{\espl}{\end{split}}

\newcommand{\bsub}{\begin{subequations}}

\newcommand{\esub}{\end{subequations}}

\newcommand{\bmulti}{\begin{multline}}   %

\newcommand{\beqm}{\begin{mathletters}}   %

\newcommand{\eeqm}{\end{mathletters}}   %

\newcommand{\kB}{k_{\rm B}}

\newcommand{\me}{m_{\rm e}}

\newcommand{\Te}{T_{\rm e}}

\newcommand{\The}{\theta_{\rm e}}

\newcommand{\Thg}{\theta_{\gamma}}

\newcommand{\vek} [1]{\mbox{\boldmath${#1}$\unboldmath}}

\newcommand{\Thz}{\theta_{z}}

\newcommand{\Tin}{T_{\rm in}}

\newcommand{\Thetaref}{{\Theta_{z}}}
\newcommand{\Thetabar}{\bar{\Theta}}

\newcommand{\oOx}[1]{{x^{#1}\partial_x^{#1}}}

\usepackage{bbold}
\usepackage{bbm}
\def\i{\mathbbm{i}}

\title{Improved frequency hierarchy treatment for anisotropic spectral distortions}

\author[a]{Jens Chluba}
\author[a]{Sara Evangelista}
\author[b]{Tom Daman}
\author[c]{and Geoff Vasil}

\affiliation[a]{Jodrell Bank Centre for Astrophysics, School of Physics and Astronomy, The University of Manchester, Oxford Road, Manchester, M13 9PL, U.K.}

\affiliation[b]{Instituut-Lorentz for Theoretical Physics,
Leiden University, 2333 CA Leiden, The Netherlands}

\affiliation[c]{School of Mathematics and the Maxwell Institute for Mathematical Sciences,
University of Edinburgh, EH9 3FD, U.K.}

\date{Feb 2026}

\begin{document}

\abstract{Spectral distortion anisotropies of the cosmic microwave background (CMB) provide a new probe of the early Universe that can be accessed using traditional CMB imaging techniques. It is possible to compute the creation and evolution of anisotropic signals for various scenarios using the frequency hierarchy method recently developed for {\tt CosmoTherm}. However, the current treatment is not perfect and some approximations had to be made. Here, we carefully construct a modified form for the evolution equations that has the full equilibrium solutions built into the formulation. We improve the formalism to account for i) additional stimulated scattering effects, ii) kinematic corrections to the thermalization terms, iii) corrections to the standard perturbation variables and iv) direct photon sources. These effects could not be captured with the original formulation of the frequency hierarchy method, but are indeed important for cleanly separating real distortions from temperature signals. However, we show that previous results are not altered significantly when compared to the improved formulation presented here. As a new worked example, which could not be treated before, we also illustrate how possible changes in the temperature-redshift relation would create spectral distortion anisotropies in the pre-recombination era. The theoretical methods presented here are also an important step towards being able to consistently predict the CMB spectral distortion anisotropies in photon-dark photon and photon-axion conversion scenarios.}

\maketitle
\flushbottom

\newpage

\section{Introduction}
\vspace{-2mm}
Spectral distortions of the cosmic microwave background (CMB) have now been recognized as a new probe of the early Universe \citep[see][for distortion physics]{Sunyaev1970mu, Burigana1991, Hu1993, Sunyaev2009, Chluba2011therm, Sunyaev2013, Tashiro2014, deZotti2015, Lucca2020}. The science case is rich \citep[e.g.,][]{Chluba2019BAAS, Chluba2021Voyage} and new sub-orbital CMB spectrometers such as TMS \citep{Jose2020TMS}, COSMO \citep{Masi2021}, BISOU \citep{BISOU} are expected to advance the field in the next $\simeq 5-10$ years. Together these promise to improve the long-standing limits obtained with \COBEF \citep{Mather1994, Fixsen1996} by more than one order of magnitude and thereby detect the largest expected $\Lambda$CDM distortion caused by the hot gas in the Universe \citep{Hill2015, Chluba2016}.
Simultaneously, future CMB spectral distortion space mission concepts like 
{\it PIXIE} \citep{Kogut2011PIXIE, Kogut2025PIXIE} and {\it FOSSIL} are being actively discussed by the community, aiming to open an entirely new window to the early universe by measuring the average energy spectrum of the CMB. 

However, independent information about the early Universe can also be extracted by considering CMB spectral anisotropies: variations of the CMB frequency spectrum across the sky, encoding full {\it spectro-spatial} information. Recently, a generalized Boltzmann treatment that allows us to predict anisotropic distortion signals for a range of scenarios was developed \citep{chluba_spectro-spatial_2023-II, kite_spectro-spatial_2023-III}. This defines new science targets \citep[e.g.,][]{Pajer2012, Ganc2012, Biagetti2013, Remazeilles2018muT, Zegeye2023S4, kite_spectro-spatial_2023-III} that are accessible even with traditional CMB mappers such as \Planck, ACT, \LiteBird and the Simons Observatory, as no absolute measurements of the CMB are required. Indeed, some first constraints using existing data have already been considered \citep[e.g.,][]{Rotti2022muT, Bianchini2022} and in the future even a combination with non-CMB data from SKA could prove powerful \citep{Zegeye2024}.

The challenge of computing anisotropic spectral distortion signals is that a brute force approach that considers the full spectro-spatial evolution of the CMB is computationally prohibitive. However, using an efficient spectral decomposition of the distortion signals \citep{chluba_spectro-spatial_2023-I}, the problem can be significantly reduced \citep{chluba_spectro-spatial_2023-II}, allowing us to perform detailed computations of various distortion scenarios using the cosmological thermalization code {\tt CosmoTherm} \citep{Chluba2011therm, chluba_spectro-spatial_2023-I, kite_spectro-spatial_2023-III} by applying the so-called {\it frequency-hierarchy method} \citep{chluba_spectro-spatial_2023-II}. 

The current version of the frequency hierarchy framework developed in \citep{chluba_spectro-spatial_2023-II, kite_spectro-spatial_2023-III} is not perfect, and several approximations were made. One approximation is related to stimulated electron scattering terms, another to kinematic corrections to the thermalization effects \citep[see Appendices of][for details]{chluba_spectro-spatial_2023-II}. These scattering effects are indeed not as important for pure distortion scenarios, where changes to the CMB temperature field are insignificant. However, when wishing to calculate scenarios with significant sources of temperature fluctuations, one could expect these corrections to become more noticeable. Some scenarios of interest in this context are due to photon to dark photon or axion conversion \citep{Mirizzi2005SN, Mirizzi:2009iz}, for which CMB distortions deliver competitive constraints \citep[e.g.,][]{Chluba2024DP, Cyr2024Axions}.
As we will explain here, stimulated scattering effects and kinematic corrections indeed have to be included for a consistent evolution of the field towards full thermodynamic equilibrium. This has now allowed a direct computation of the signals in photon to dark photon conversion scenarios  \citep{Evangelista2026DP}.

To include these effects, we carefully separate the evolution of distortion variables and corrections to the standard perturbation variables, independently solving for the latter. We use operator identities to recast the related scattering terms using the basis of \citep{chluba_spectro-spatial_2023-II}. We also consider possible sources of photons as a natural extension to the framework. With these generalizations, we demonstrate that the previous predictions are only modified slightly, while new scenarios can now be calculated.
Additional effects relating to i) polarized spectral distortion transport, ii) second order perturbations, and iii) anisotropic scattering/thermalization effects are expected to be subdominant. The methods developed in \citep{Pitrou2009, Bartolo2007JCAP, Chluba2012, Ota2017} could open a way to also include these contributions, but a lot more work is needed and left to future investigations.

In Sect.~\ref{sec:prelim}, we provide more background information about the frequency hierarchy method. 
This heavily draws on references \citep{chluba_spectro-spatial_2023-I, chluba_spectro-spatial_2023-II, kite_spectro-spatial_2023-III}, where the main frequency hierarchy approach was introduced. We refer the reader to these works should they not be familiar with the general approach.
In Sect.~\ref{sec:eq_solutions_general}, we discuss general aspects of the expected equilibrium solutions to the thermalization problem in the presence of anisotropies. In Sect.~\ref{sec:improvements}, we improve the current description of thermalization terms by adding previously neglected stimulated scattering effects. In Sect.~\ref{sec:k_corrections}, we then add kinematic corrections and in Sect.~\ref{sec:Photon sources} we consider sources of photons. Section~\ref{sec:consistency} explains how to incorporate the corrections to the standard perturbation variables and also demonstrates the consistency of the new formulation with respect to changes of the CMB temperature. 
In Sect.~\ref{sec:FH_eq} we provide a compressed summary of the new system, also directly relating it to the line-of-sight integral approach that is required for computing the signal cross power spectra.
In Sect.~\ref{sec:illustrations} we illustrate the photon transfer functions and signal power spectra for a few examples, but leave a more detailed exploration of constraints to another paper. We then conclude in Sect.~\ref{sec:conclusions}.

\vspace{-3mm}
\section{Preliminaries and notation}
\label{sec:prelim}
\vspace{-2.5mm}
Before going into the details of the derivations, we shall clarify some of the basic framework and notation. 
Throughout, we closely follow the formalism of \citep{chluba_spectro-spatial_2023-I} and \citep{chluba_spectro-spatial_2023-II}. The major step taken in those papers was to convert the full frequency-dependent photon Boltzmann equations into a reduced system of equations that describes the evolution of the distortion amplitudes for a chosen spectral basis. The goal is to evolve $\Delta n=n-\nbb$, which is the photon occupation number correction with respect to the reference blackbody, $\nbb=1/(\expf{x}-1)$, for dimensionless frequency $x=h\nu/\kB\Tz$ relative to the reference temperature, $\Tz=T_0(1+z)$. This reference temperature can be conveniently chosen and merely allows us to eliminate the redshifting effect from the problem as long as $\Tz\propto (1+z)$ \cite{Chluba2011therm}.
We note that generally $\Tz\neq \TCMB$, where $\TCMB=T_{\rm CMB, 0}(1+z)$ is the CMB blackbody temperature scaled backwards from today's value, $T_{\rm CMB, 0}=2.7255$~K \citep{Fixsen1996, Fixsen2009}. The results of the computation should not depend on the chosen value for $\Tz$ (as long as non-linear effects do not become important), which is one of the consistency checks for the equations that we will carry out in Sect.~\ref{sec:consistency}.

To discretize the problem, we use the  basis functions \citep{chluba_spectro-spatial_2023-I}
\bealf{
\label{eq:basis}
&\Gspec(x)
=\frac{x\,\expf{x}}{(\expf{x}-1)^2},
\quad 
\Yspec(x)
=\Gspec(x)\left[x\frac{\expf{x}+1}{\expf{x}-1}-4\right],
\quad 
Y_k=(1/4)^k \,\boostO^k Y,
\quad 
\Mspec(x)=\Gspec(x)\left[\frac{1}{\beta_M}-\frac{1}{x}\right],
\nonumber \\[2mm]
&\Delta n= \vek{B} \cdot \vek{y}, 
\quad
\vek{B}=(\Gspec, Y_0, Y_1, \cdots, Y_N, M)^T, 
\quad
\vek{y}=(\Theta, y_0, y_1, \cdots, y_N, \mu)^T, 
}
where $\boostO=-x\partial_x$ is the energy shift generator, $G=\boostO \nbb$ is the spectrum of a temperature shift, $Y$ and $M$ describe the classic $y$ and $\mu$-distortions, and $Y_n$ are boosts of $Y$ with $Y_0\equiv Y$. The spectral amplitudes, $\vek{y}$, can have spatial dependence and depend on the observation direction. We will ultimately go to Fourier space and perform a Legendre Transform to solve the problem, where we will use $\vek{y}_\ell=(\Theta_\ell, y_{0,\ell}, y_{1,\ell}, \cdots, y_{N,\ell}, \mu_{\ell})^T$ to describe the multipoles of $\vek{y}$.

The basis functions have several important properties built in. Firstly, we shall introduce the normalized number and energy density integrals as
\bealf{
\label{eq:basis_int_f}
\epsilon_{N, f}&=\frac{\int x^2 f(x) \id x}{N_{\nbb}}, \qquad 
\epsilon_{f}=\frac{\int x^3 f(x) \id x}{ E_{\nbb}}
}
where $N_{\nbb}=\int x^2 \nbb \id x= G_2\approx 2.4041$ and $E_{\nbb}=\int x^3 \nbb \id x= G_3\approx 6.4939$ are the corresponding blackbody integrals. By construction, we then find
\bealf{
\label{eq:basis_int}
\vek{b}_N&=\frac{\int x^2 \vek{B} \id x}{N_{\nbb}}=(3, 0, 0, \cdots, 0, 0)^T, \qquad 
\vek{b}_\epsilon=\frac{\int x^3 \vek{B} \id x}{E_{\nbb}}
=(4, 4, 4, \cdots, 4, \epsilon_{M})^T
}
with $\epsilon_{M}=1/1.4007$. Here, we note that only $G$ carries photon number and energy density, while the other basis functions only carry energy density, ensuring photon number conservation in scattering problems.
The number and energy density projectors, $\vek{b}_N$ and $\vek{b}_\epsilon$, can be used to compute the relative number and energy density perturbations of a given spectral state vector, $\vek{y}$, by using \footnote{Although the notation might seem contradictory, with symbols $\epsilon_f \neq \epsilon_\ell$, there is never a case were we need $f(x)=\ell$ in any evaluation, such that these should be clear from the context.}
\bealf{
\label{eq:projections_N_rho}
\epsilon_{N}=\vek{b}_{N}\cdot \vek{y}\equiv 3\Theta, 
\qquad 
\epsilon=\sum_f \epsilon_f = \vek{b}_{\epsilon}\cdot \vek{y},
\qquad
\epsilon_{N, \ell}=\vek{b}_{N}\cdot \vek{y}_\ell\equiv 3\Theta_\ell, 
\qquad 
\epsilon_{\ell}=\vek{b}_{\epsilon}\cdot \vek{y}_\ell.
}
We will also denote perturbation orders with superscripts like $\vek{y}^{(i)}$, where $\vek{y}^{(0)}$ denotes the background level and $\vek{y}^{(1)}$ the first order spatial fluctuations. We shall also use $\vek{e}_G=(1, 0, \ldots, 0)^T$, $\vek{e}_Y=(0, 1, 0, \ldots, 0)^T$, $\vek{e}_{Y_1}=(0, 0, 1, 0, \ldots, 0)^T$, $\vek{e}_{M}=(0, \ldots, 0, 1)^T$ etc to denote the unit vector of a given spectral basis.

In connection to the Compton scattering process, we will encounter the Compton energy exchange integrals
\bealf{
\label{eq:Compton_exchange_E}
\eta_{f}=\frac{\int x^3 w_y(x) \, f(x) \id x}{4 E_{\nbb}}, 
\qquad 
\vek{b}_{\rm \Theta}=(\eta_G, \eta_Y, \eta_{Y_1}, \ldots, \eta_{Y_{15}}, \eta_{M})^T,
\qquad 
\Theta_{\rm C}=\vek{b}_{\rm \Theta}\cdot \vek{y}_0
}
where $w_y=x(1+2\nbb)-4$ and we also defined the equilibrium temperature projector, $\vek{b}_{\rm \Theta}$, which directly yields the related Compton equilibrium temperature, $\Theta_{\rm C}$, given the decomposition of $f$ in terms of $\vek{y}$ (see Sect.~\ref{sec:T_eq}). The integrals $\eta_{f}$ determine what temperature the electrons take in a given radiation field. Some values of the integrals are $\eta_{G}=1$, $\eta_{\Yspec}\approx 5.3996$ and $\eta_\Mspec\approx0.4561$. If only $\Theta_0$ is present, then $\Theta_{\rm C}=\Theta_0$, while if one also has a non-zero chemical potential, then $\Theta_{\rm C}=\Theta_0+\eta_M \mu_0$. We note that the Compton equilibrium temperature is solely determined by the {\it monopole} of the photon distribution, which is why we used $\vek{y}_0$ above.

\section{Equilibrium solutions for the anisotropic transport problem}
\label{sec:eq_solutions_general}
In the expanding universe, the anisotropic transfer problem of CMB photons in the presence of distortions is defined by the evolution equations \citep{chluba_spectro-spatial_2023-II}
\bsub
\label{eq:evol_transport}
\bealf{
&\frac{\partial \Delta n^{(0)}}{\partial \eta}
={\rm C}^{(0)}_{\rm th}[\Delta n]
\\
&\frac{\partial \Delta n^{(1)}}{\partial \eta}+\vgh\cdot \nabla \Delta n^{(1)}+
\boostO n^{(0)} \left(\frac{\partial \Phi^{(1)}}{\partial \eta}+ \vgh \cdot \nabla\Psi^{(1)} \right)= 
{\rm C}^{(1)}_{\rm T}[\Delta n]+{\rm C}^{(1)}_{\rm th}[\Delta n]
+\Psi^{(1)} {\rm C}^{(0)}_{\rm th}[\Delta n]
\\
&{\rm C}^{(1)}_{\rm T}[\Delta n]=\tau'\left[\Delta n^{(1)}_0+\frac{1}{10}\,\Delta n^{(1)}_2 - \Delta n^{(1)} + \beta^{(1)}\chi\,\boostO n^{(0)}\right]
}
\esub
where $\vgh$ defines the direction of the photon; $\chi=\vgh\cdot \vbetah$ is the direction cosine with the baryon velocity vector; $\boostO=-x\partial_x$ is the energy shift generator; $n=\nbb+\Delta n^{(0)}+\Delta n^{(1)}$ is the photon occupation number with background, $\Delta n^{(0)}$, and first order perturbations, $\Delta n^{(1)}$; $\nbb=1/(\expf{x}-1)$ is the average blackbody spectrum, and $\eta=\int c\id t/a$ is the conformal time variable. We furthermore use
${\rm C}_{\rm th}[\Delta n]$
to denote the (local inertial frame) collision term that accounts for thermalization effects due to Compton scattering, double Compton and Bremsstrahlung emission, and the sourcing of distortions by photon injection. These contributions to the collision term will each be treated separately in section~\ref{sec:improvements}. The Newtonian potential perturbations are given by $\Phi$ and $\Psi$, $\tau'$ is the Thomson scattering rate, and ${\rm C}^{(1)}_{\rm T}[\Delta n]$ describes the effect of Thomson scattering on the photon multipoles $\Delta n_{\ell}$ in the presence of baryon velocity perturbations $\beta^{(1)}$. At the background level, the photon distribution is isotropic, such that no Thomson scattering terms appear. 

The background distribution couples to the fluctuations through the local inertial frame transformation of the isotropic thermalization term, $\Psi^{(1)} {\rm C}^{(0)}_{\rm th}[\Delta n]$, and the boosts $\boostO n^{(0)} =G(x)+\boostO \Delta n^{(0)}$. If {\it no} average distortions or non-thermal effects are present, and start with $\Tz=\TCMB$ and initial blackbody radiation at an average temperature $\TCMB$, this means that we only need to consider temperature anisotropies with a thermal spectrum $G(x_{\rm CMB})=x_{\rm CMB}\expf{x_{\rm CMB}}/(\expf{x_{\rm CMB}}-1)^2$ and $x_{\rm CMB}=h\nu/\kB\TCMB$. This automatically yields the well-known CMB temperature hierarchy \citep[e.g.,][]{Ma1995, Seljak1996}
\bealf{
\label{eq:evol_transport_Theta}
&\frac{\partial \Delta n^{(0)}}{\partial \eta}
=0, \qquad \frac{\partial \Theta^{(1)}}{\partial \eta}+\vgh\cdot \nabla \Theta^{(1)}+ \frac{\partial \Phi^{(1)}}{\partial \eta}+ \vgh \cdot \nabla\Psi^{(1)} = \tau'\left[\Theta^{(1)}_0+\frac{1}{10}\,\Theta^{(1)}_2 - \Theta^{(1)} + \beta^{(1)}\chi\right]
}
with temperature fluctuations $\Theta=T/\TCMB-1$ around the average CMB temperature $\TCMB$. This problem can be solved in conjunction with the other perturbation variables \citep[e.g., see][for details]{Ma1995} and then allows us to compute the temperature and polarization anisotropies using the line-of-sight approach \citep{Seljak1996}. We note that we did not explicitly show the contributions from polarization here, as for distortions these effects are currently neglected. 

\subsection{Changing the reference temperature}
\label{sec:change_Tref}
What happens when we {\it choose} to describe the problem with respect to a different reference blackbody temperature, $\Tz\neq \TCMB$? In this case, nothing should change in terms of spectral evolution (we are simply using a different formulation of the problem). Shifting the reference blackbody temperature, we then have the replacement
\begin{align}
\label{eq:shited_ref}
n=\nbb\left(\frac{x_{\rm CMB}}{1+\Theta}\right)\qquad \rightarrow \qquad
\nbb\left(\frac{x (1+\Thetaref)}{(1+\Theta)}\right)
\end{align}
with $x_{\rm CMB}=x \, \Tz/\TCMB=x(1+\Thetaref)$ and $\Thetaref=\Tz/\TCMB-1$. Assuming small variations, this gives
\begin{align}
\label{eq:equilibrium_solution}
\Delta n^{(0)}_{\rm s}&\approx 
-\Thetaref \,G(x), \qquad 
\Delta n^{(1)}_{\rm s}\approx 
\Theta^{(1)}(1-3\Thetaref) \,G(x) - \Theta^{(1)}\Thetaref \,Y(x) = -\oOx{} n_{\rm s}^{(0)} \, \Theta^{(1)}.
\end{align}
up to order $\Thetaref\Theta$ and with $n_{\rm s}^{(0)} = \nbb +  \Delta n^{(0)}_{\rm s}$. We see that the $G$ term has an effective temperature variable $\Theta^{(1), \rm eff}=\Theta^{(1)}(1-3\Thetaref)$ and the $Y$ term an effective $y$-parameter $y^{(1), \rm eff}=-\Thetaref\Theta^{(1)}$.\footnote{We stress here that unless stated otherwise $\Theta^{(1)}$ denotes the temperature solution of the standard perturbed system in Eq.~\eqref{eq:evol_transport_Theta}, which entirely neglects spectral distortions. Corrections to that temperature variable will be denoted as $\delta \Theta^{(1)}$.
}
However, the term $\propto Y(x)=G(x)\left[x\coth(x/2)-4\right]$ in $\Delta n^{(1)}_{\rm s}$ is not a {\it real distortion}, but simply a consequence of shifting the reference temperature by $\Thetaref$ in our computation \citep[as also discussed in][]{chluba_spectro-spatial_2023-II}. After the computation is done, we can obtain the effective power spectra as
\bsub
\label{eq:Cell}
\begin{align}
C_\ell^{TT, \rm eff}&=\TCMB^2(1+\Thetaref)^2\left<\Theta^{(1), \rm eff}_\ell\,\Theta^{(1), \rm eff}_\ell \right>=(1+\Thetaref)^2 (1-3\Thetaref)^2 \, C_\ell^{TT}\approx (1-4\Thetaref)\,C_\ell^{TT}, 
\\
C_\ell^{yT, \rm eff}&=\TCMB^2(1+\Thetaref)^2\left<y^{(1), \rm eff}_\ell\,\Theta^{(1), \rm eff}_\ell \right>=-(1+\Thetaref)^2\Thetaref\,(1-3\Thetaref)\,C_\ell^{TT}
\approx -\Thetaref\,C_\ell^{TT}, 
\end{align}
\esub
where $C_\ell^{TT}$ is the standard power spectrum (computed with $\Tz=\TCMB$), with the extra factor $(1+\Thetaref)^2$ stemming from the conversion to real temperature units. The apparent $yT$ contribution is merely a consequence of our choice for the reference temperature, and the combination 
$C_\ell^{TT, \rm eff}-4 C_\ell^{yT, \rm eff}\approx C_\ell^{TT}$
recovers the standard CMB temperature power spectrum. Below we will see that our improved formulation indeed reproduces this expectation.

\subsection{Changing the initial CMB temperature}
\label{sec:change_TCMB_initial}
What happens when the initial CMB temperature differs from $\TCMB$, i.e., $\TCMB'=\TCMB(1+\Thetabar)$? 
When setting the initial conditions for spectral distortion scenarios this will always be the case, as energy injection/extraction increases/decreases the initially lower/higher photon temperature if fully thermalized \citep[e.g,][]{Chluba2011therm}.
If we choose $\Tz = \TCMB'$, nothing changes for the computation at early times\footnote{The recombination process and also Hubble expansion rate will be modified if the CMB temperature is indeed changed. This directly affects the CMB power spectra \citep[e.g.,][]{Chluba2008T0, Planck2015params}, but for the discussion here, this aspect is not relevant.} (i.e., for variables and initial conditions), but we end up with CMB power spectra that have an amplitude scaled by $(\TCMB'/\TCMB)^2=(1+\Thetabar)^2$, i.e., $C_\ell^{T'T'}=(1+\Thetabar)^2 C_\ell^{TT}$ when converting to real temperature units. What if we now instead use $\Tz = \TCMB=\TCMB'/(1+\Thetabar)$? In that case we have
\begin{align}
\label{eq:shited_initial}
n=\nbb\left(\frac{x'_{\rm CMB}}{1+\Theta}\right)\qquad \rightarrow \qquad
\nbb\left(\frac{x}{(1+\Thetabar)(1+\Theta)}\right)\approx \nbb\left(\frac{x(1-\Thetabar)}{(1+\Theta)}\right),
\end{align}
meaning that the results from the previous section apply with $\Thetaref= -\Thetabar$ in terms of the spectral decomposition, Eq.~\eqref{eq:equilibrium_solution}. For the effective power spectra, we now obtain
\bsub
\label{eq:Cell_initial}
\begin{align}
C_\ell^{TT, \rm eff}&=\frac{{\TCMB'}^2}{(1+\Thetabar)^{2}}\left<\Theta^{(1), \rm eff}_\ell\,\Theta^{(1), \rm eff}_\ell \right>= (1+3\bar{\Theta})^2 \, C_\ell^{TT}\approx (1+6\bar{\Theta})\,C_\ell^{TT}, 
\\
C_\ell^{yT, \rm eff}&=\frac{{\TCMB'}^2}{(1+\Thetabar)^{2}}\left<y^{(1), \rm eff}_\ell\,\Theta^{(1), \rm eff}_\ell \right>=\bar{\Theta}\,(1+3\bar{\Theta})\,C_\ell^{TT}
\approx \bar{\Theta}\,C_\ell^{TT},
\end{align}
\esub
and thus $C_\ell^{T'T'}\approx C_\ell^{TT, \rm eff}-4 C_\ell^{yT, \rm eff}\approx (1+2\bar{\Theta})\,C_\ell^{TT}$ as expected. Again in our computation, we expect this feature to be reproduced.

\subsection{Adding real distortions}
\label{sec:adding_distortions}
What changes when distortions are present? If the average distortion is evolving through average energy or photon injection, this sources time-dependent distortion anisotropies \citep{Chluba2012, chluba_spectro-spatial_2023-II}. In addition, anisotropic energy and photon injection can directly add distortion anisotropies \citep{chluba_spectro-spatial_2023-II, kite_spectro-spatial_2023-III}. If distortion sources stop and thermalization is complete, Eq.~\eqref{eq:equilibrium_solution} should generally solve the problem once the solution for the background spectrum becomes stationary. Below we will show that this is indeed the case; however, tweaks to the way the equations are written are needed to make this more transparent and improve the numerical treatment, as we will explain in the next sections.

\section{Improving the precision of the numerical scheme for the thermalization terms}
\label{sec:improvements}
In this section, we explore how to rewrite the evolution equations for the average and perturbed distortion parts in a more numerically accurate way, treating each physical process of the collision term, ${\rm C}_{\rm th}[\Delta n]$, separately.
This is motivated by the fact that the we want to directly ensure that the correct equilibrium solutions are reached, even if we do not have a perfect (numerical) discretization for the (complicated) Kompaneets operator, $\KompO$, to describe Compton scattering. The main equations describing the Comptonization process in terms of the photon occupation number read \citep{chluba_spectro-spatial_2023-II}
\bsub
\label{eq:evol_Dn}
\bealf{
\label{eq:evol_Dn_a}
\frac{\id \Delta n^{(0)}_0}{\id y_{\rm sc}}\Bigg|_{\rm C}&=\Theta^{(0)}_{\rm e} Y + \KompO \Delta n^{(0)}_0
\\
\label{eq:evol_Dn_b}
\frac{\id \Delta n^{(1)}_0}{\id y_{\rm sc}}\Bigg|_{\rm C}&=
\Theta^{(1)}_{\rm e} Y + \KompO \Delta n^{(1)}_0+\Theta^{(1)}_0\DiffO \Delta n_0^{(0)}+\Theta^{(0)}_{\rm e}\Theta^{(1)}_0 \boostO Y
+ 2\Theta^{(1)}_0\,\DiffO^* \Delta n^{(0)}_0 \boostO \nbb,
}
\esub
where $\id y_{\rm sc}=\tau' \Thz \id \eta$ with $\Thz=\kB\Tz/\me c^2$ is the scattering $y$-parameter, $\Theta_{\rm e}$ is the electron temperature perturbation, $\DiffO=\boostO(\boostO-3)$ is the {\it diffusion operator} and $\DiffstarO =x (4-\boostO)$ is the {\it recoil operator~\citep{Hoey2026Kompaneets}}. We also linearized with respect to the background distortion, keeping only terms at first order in $\Delta n_0^{(0)}$ and no more than one perturbation order, e.g., $\propto \Delta n_0^{(0)} \Theta^{(1)}$.\footnote{This mean we used $\Theta^{(1)}_{\rm e}\DiffO \Delta n_0^{(0)}\rightarrow \Theta^{(1)}_0\DiffO \Delta n_0^{(0)}$,
$\Theta^{(0)}_{\rm e}\DiffO \Delta n_0^{(1)}\rightarrow  \Theta^{(0)}_{\rm e} \Theta_0^{(1)} \DiffO G = \Theta^{(0)}_{\rm e} \Theta_0^{(1)}\DiffO \boostO \nbb = \Theta^{(0)}_{\rm e} \Theta_0^{(1)}\boostO Y$ and $2\,\DiffO^* \Delta n^{(0)}_0 \Delta n^{(1)}_0\rightarrow 2\Theta^{(1)}_0\,\DiffO^* \Delta n^{(0)}_0 \boostO \nbb$ in the more general expression given by \citep{chluba_spectro-spatial_2023-II}.} We also dropped terms relating to perturbed electron scattering (controlled by the {\it baryon density fluctuations} $ \delta_{\rm b}$) and the time transformation into the local inertial frame (related to $\Psi$), which we return to later.

Before discussing any improved numerical treatments, let us rewrite the last term in Eq.~\eqref{eq:evol_Dn_b}, which was previously not treated in \citep{chluba_spectro-spatial_2023-II} but does modify the final equilibrium population. For this, we recognize that the Kompaneets operator can be written as (see Appendix~\ref{app:operator_props} for a summary of important operator properties)
\bealf{
\label{eq:Komp_w}
\KompO&=(\boostO-3)(\boostO-w)=\DiffO+(3-\boostO) w
}
where $w=w_y+4=x(1+2\nbb)$. We can then also use that $2x \nbb = w-x$, implying
\bealf{
2\DiffO^* \Delta n^{(0)}_0\,\boostO \nbb&=
\DiffO^*\Delta n^{(0)}_0 \boostO \frac{w-x}{x} =\DiffO^* \frac{\Delta n^{(0)}_0}{x} \, (1+\boostO) \, (w-x)
=(3-\boostO) \, \Delta n^{(0)}_0 (1+\boostO) w,
}
where we used $\boostO x^{-1}=x^{-1}(1+\boostO)$, $\DiffO^*x^{-1}=(3-\boostO)$ and $(1+\boostO)x=0$. In this expression, we observe an uncomfortable term $\propto w$,\footnote{This term currently has no simple representation in the distortion basis and thus would require new projection integrals.}. However, one can eliminate this term as
\bealf{
\label{eq:important_identity}
(3-\boostO)\Delta n^{(0)}_0 (1+\boostO) w &=(3-\boostO)\left[w\,\Delta n^{(0)}_0 + \boostO w \,\Delta n^{(0)}_0 - w \,\boostO \Delta n^{(0)}_0\right]
\nonumber \\
&=(\KompO-\DiffO)[\Delta n^{(0)}_0-\boostO \Delta n^{(0)}_0]+(3-\boostO)\boostO w \,\Delta n^{(0)}_0 
\nonumber \\
&=(\KompO-\DiffO)[\Delta n^{(0)}_0-\boostO \Delta n^{(0)}_0]+\boostO(\KompO-\DiffO)\Delta n^{(0)}_0
\nonumber \\
&=\KompO\left[\Delta n^{(0)}_0-\boostO \Delta n^{(0)}_0\right]
+\boostO \KompO \Delta n^{(0)}_0
-\DiffO\Delta n^{(0)}_0,
}
where we used $(3-\boostO) w = \KompO-\DiffO$ from Eq.~\eqref{eq:Komp_w} and $(3-\boostO)\boostO=\boostO(3-\boostO)$. Inserting this back into Eq.~\eqref{eq:evol_Dn_b}, we then find
\bealf{
\label{eq:evol_Dn_b_mod}
\frac{\id \Delta n^{(1)}_0}{\id y_{\rm sc}}\Bigg|_{\rm C}&=
\Theta^{(1)}_{\rm e} Y  +\KompO\left[ \Delta n^{(1)}_0-\Theta^{(1)}_0\boostO \Delta n^{(0)}_0\right]+\Theta^{(0)}_{\rm e}\Theta^{(1)}_0 \boostO Y
+\Theta^{(1)}_0(1+\boostO)\KompO\Delta n^{(0)}_0
\nonumber\\[-1mm]
&=\left[\Theta^{(1)}_{\rm e}-\Theta^{(1)}_0(1+\Theta^{(0)}_{\rm e})\right] Y  +\KompO\left[ \Delta n^{(1)}_0-\Theta^{(1)}_0\boostO n^{(0)}_0\right]
+\Theta^{(1)}_0(1+\boostO)\frac{\id \Delta n^{(0)}_0}{\id y_{\rm sc}}\Bigg|_{\rm C},
}
where we used Eq.~\eqref{eq:evol_Dn_a} and set $n^{(0)}_0=\nbb + \Delta n^{(0)}_0$ in the second step. This is a significant improvement since we only encounter operators that we already know how to represent in the spectral basis \citep[e.g., following][]{chluba_spectro-spatial_2023-I,chluba_spectro-spatial_2023-II}.
As we will see below, the expression in Eq.~\eqref{eq:evol_Dn_b_mod} has better numerical properties and also drives the solution to the correct equilibrium distribution even if we do not have a perfect basis representation for the Kompaneets operator. In particular, it allows us to include the effects of the term $2\Theta^{(1)}_0\,\DiffO^* \Delta n^{(0)}_0 \boostO \nbb$, which was previously omitted \citep{chluba_spectro-spatial_2023-II}.

\subsection{Compton equilibrium temperature}
\label{sec:T_eq}
%
Before moving forward, let us consider the Compton equilibrium temperature. It determines the temperature that the electrons would reach in a given radiation field if Compton scattering was extremely efficient \citep[e.g.,][]{Zeldovich1970TCompton, Sazonov2001}. This temperature can be obtained by integrating the Compton scattering terms equation over $x^3 \id x$ and setting the resulting equations to zero. 
In this step, we use 
\bsub
\bealf{
\int \boostO^k f(x) \id x &= \int f(x) \id x
\\
\int x^\alpha\boostO^k f(x) \id x &=  \int \boostO \, x^{\alpha}\,\boostO^{k-1}\,f(x) \id x
-\int (\boostO \, x^{\alpha})\,\boostO^{k-1}\,f(x) \id x
=  [1+\alpha]^k \!\int  x^{\alpha}\,f(x) \id x,}
\esub
where we used the commutator relation $[ x^\alpha,\boostO]=x^\alpha \boostO - \boostO\,x^\alpha= \alpha x^\alpha$. In addition, we can use
\bsub
\bealf{
x^3 \boostO x^{-3}&=\boostO+3,
\qquad
x^3 \DiffO x^{-3}=\boostO x^3 \boostO x^{-3} = \boostO(\boostO+3), \qquad x^3 \DiffstarO x^{-3}
=-\boostO x,
\\
x^3\KompO x^{-3}&= \boostO(\boostO+3) -\boostO w
=\boostO(\boostO-1) -\boostO w_y
}
\esub
to express all the integrals in terms of intensity, $\Delta I = x^3 \Delta n$. For example, this gives
\bealf{
\int x^3 \KompO \Delta n \id x = \int x^3 \KompO x^{-3} \Delta I \id x = \int \left[\boostO(\boostO-1) - \boostO w_y\right]\Delta I \id x
=-\int  w_y \Delta I \id x,
}
where we used that $\int \boostO(\boostO-1) \Delta I \id x=0$. With these identities, from Eq.~\eqref{eq:evol_Dn_a} and \eqref{eq:evol_Dn_b_mod} we can immediately write\footnote{The result does not change when including perturbed electron scattering and the perturbations in the time variable from the local inertial frame transformation \citep[see][]{chluba_spectro-spatial_2023-II}}.
\bealf{
\label{eq:Teq_equations}
0&=4 E_{\nbb} \Theta^{(0)}_{\rm C} -\int x^3 w_y \Delta n^{(0)}_0 \id x
\\ \nonumber 
0&=
4 E_{\nbb}\left[\Theta^{(1)}_{\rm C}-\Theta^{(1)}_0(1+\Theta^{(0)}_{\rm C})\right] -\int x^3 w_y \left[\Delta n^{(1)}_0-\Theta^{(1)}_0\boostO n^{(0)}_0\right] \id x,
}
where $\int x^3 Y \id x = 4 E_{\nbb}$. We then obtain the Compton equilibrium temperatures
\bealf{
\label{eq:Teq}
\Theta^{(0)}_{\rm C}&=\frac{\int x^3 w_y \Delta n^{(0)}_0 \id x}{4 E_{\nbb}}, \qquad \Theta^{(1)}_{\rm C} \approx 
\Theta^{(1)}_0 \left[1+\Theta^{(0)}_{\rm C}\right]+ \frac{\int x^3 w_y \left[\Delta n^{(1)}_0-\Theta^{(1)}_0 \boostO n^{(0)}_0\right] \id x}{4 E_{\nbb}}
}
with $\boostO n^{(0)}_0= G+\boostO \Delta n^{(0)}_0$. This shows a very important property of the Comptonization problem: If $\Delta n^{(1)}_0\rightarrow \Theta^{(1)}_0\,\boostO n^{(0)}_0$, then the perturbed Compton equilibrium temperature is
\bealf{
\Theta^{(1)}_{\rm C}
\approx\Theta^{(1)}_0[1+\Theta^{(0)}_{\rm C}].
}
If the average spectrum is in addition undistorted, we have $\Theta^{(0)}_{\rm C}=\Thetabar$ and hence $\Theta^{(1)}_{\rm C}
\approx\Theta^{(1)}_0[1+\bar{\Theta}]$. This is consistent with fluctuations at a higher average temperature $(1+\bar{\Theta})(1+\Theta_0)\approx 1+\bar{\Theta}+ \Theta^{(1)}_0(1+\bar{\Theta})$, which shows that the operator combination in Eq.~\eqref{eq:evol_Dn_b_mod} indeed drives the solution towards the correct equilibrium if only a shift in the blackbody temperature is present.
Without the term $2\DiffO^* \Delta n^{(0)}_0 \Delta n^{(1)}_0$ this would not be the case showing that the arguments of \citep{chluba_spectro-spatial_2023-II} do not apply in this situation.

We note that the Compton equilibrium temperature can be consistently computed using the expressions above even if the change is simply due to the choice of reference temperature. In this case, the timescale on which equilibrium is reached has to be modified by $\Tz/\TCMB=(1+\Thetaref)$, but this does not modify the equilibrium temperature perturbation. If we start with $n(x_{\rm CMB})$ and use $\Tz=\TCMB(1+\Thetaref)$, we have $\Theta^{(0)}_{\rm C}=-\Thetaref$. The physical electron temperature is then 
\bealf{
\Te^{(0)}=\Tz(1+\Theta^{(0)}_{\rm C})\approx \TCMB(1+\Thetaref)(1-\Thetaref)\approx \TCMB
}
as expected. In the calculations, the absolute temperature only matters for the Comptonization timescale, as already mentioned before.

\subsection{Quasi-stationary solutions}
\label{sec:QS_sol}
In the absence of any heating or photon sources, we know that the spectrum reaches a kinetic equilibrium with the electrons. For the average distortion this is $\Delta n^{(0)}_{\rm qs}=\Theta^{(0)}_0 G + \mu^{(0)}_0 M$, where $\Theta^{(0)}_0$ and $\mu^{(0)}_0$ are fixed by the photon number and energy densities. In this case the Compton equilibrium temperature at the background level is $\Theta^{(0)}_{\rm qs}\approx \Theta^{(0)}_0 + \eta_M \mu^{(0)}_0$. 

In a similar way, the quasi-stationary solution at the perturbed level is $\Delta n^{(1)}_{\rm qs}=\Theta^{(1)}_0 (G+\boostO \Delta n^{(0)}_{\rm qs})$, such that $\Theta^{(1)}_{\rm qs}
\approx\Theta^{(1)}_0[1+\Theta^{(0)}_{\rm qs}]$. Since the Kompaneets operator gives $\KompO G=-Y$ and $\KompO M=-\eta_M Y$, it can be extremely helpful to use the quasi-stationary solutions as references, thereby eliminating these fast modes from the photon energy diffusion problem. This then implies
\bsub
\label{eq:def_Dm}
\bealf{
\Delta m^{(0)}_0&=\Delta n^{(0)}_0-\Delta n^{(0)}_{\rm qs}=\Delta n^{(0)}_0-\left(\Theta^{(0)}_0\, G+\mu^{(0)}_0\, M\right)
\\[1mm]
\Delta m^{(1)}_0&=\Delta n^{(1)}_0-\Delta n^{(1)}_{\rm qs}=\Delta n^{(1)}_0-\Theta_0^{(1)}\left[ G + \boostO \left(\Theta^{(0)}_0\, G+\mu^{(0)}_0\, M\right)\right]
}
\esub
as better variables in combination with the Kompaneets operator. 
With these variables, from Eq.~\eqref{eq:Teq} one also has
\bealf{
\label{eq:Teq_qs}
\Theta^{(0)}_{\rm C}&=\Theta^{(0)}_{\rm qs}+\frac{\int x^3 w_y \,\Delta m^{(0)}_0 {\rm d} x}{4 E_{\nbb}}, \quad
\Theta^{(1)}_{\rm C} \approx 
\Theta^{(1)}_0 \left[1+\Theta^{(0)}_{\rm C}\right]+\frac{\int x^3 w_y \left[\Delta m^{(1)}_0-\Theta^{(1)}_0 \boostO \Delta m^{(0)}_0\right] {\rm d} x}{4 E_{\nbb}}
}
for convenience. Here we used 
\bealf{
\label{eq:Dn1_rewrite}
\Delta n^{(1)}_0-\Theta^{(1)}_0 \boostO n^{(0)}_0=\Delta m^{(1)}_0+\Delta n^{(1)}_{\rm qs}-\Theta^{(1)}_0 (G+\boostO \Delta n^{(0)}_0)=\Delta m^{(1)}_0-\Theta^{(1)}_0 \boostO \Delta m^{(0)}_0,
}
given that $\Delta n^{(1)}_{\rm qs}=\Theta^{(1)}_0 (G+\boostO \Delta n^{(0)}_{\rm qs})$.
We will see in the next sections that these modifications are indeed very useful. Here, we note that $\Delta m^{(0)}_0$ has {\it no} contributions from $G$ and $M$ once the spectral basis is chosen. However, for $\Delta m^{(1)}_0$ we generally have a correction $\Theta^{(1), \rm full}_0-\Theta^{(1)}_0(1+3\Theta^{(0)}_0)\neq 0$ for $G$ and $\mu^{(1)}_0-\mu^{(1)}_{\rm qs}\neq 0$ for $M$ until a quasi-stationary state is reached. However, as we will see below this correction is identically canceled by the conversion of $G$ and $M$ into $-Y$ when the Kompaneets operator is applied to these terms. We also highlight that once a spectral basis is chosen, the energy exchange integrals can be computed using the simple scalar product in Eq.~\eqref{eq:Compton_exchange_E}.

\subsection{Rewriting the Compton terms}
\label{sec:num_sol}
To numerically apply the Eqns.~\eqref{eq:evol_Dn_a} and \eqref{eq:evol_Dn_b_mod}, we require a few key features. We want to numerically conserve the number of photons, as scattering does not change photon number. To ensure this it is beneficial to use distortion basis functions that carry vanishing photon number, $\int x^2 f(x) \id x=0$. As proposed by \citep{chluba_spectro-spatial_2023-I}, the basis $Y$ and $M$ extended by $Y_k=(1/4)^k \boostO^k Y$ has this property (see Sect.~\ref{sec:prelim}).
Second, we need to ensure that the correct equilibrium solution is reached. The latter is numerically achieved better by using the quasi-stationary solutions as explained in Sect.~\ref{sec:QS_sol}. With this we find
\bealf{
\label{eq:evol_Dn_a_mod}
\frac{\id \Delta n^{(0)}_0}{\id y_{\rm sc}}\Bigg|_{\rm C}&=\Theta^{(0)}_{\rm e} Y + \KompO \Delta n^{(0)}_0
=\left[\Theta^{(0)}_{\rm e}-\Theta^{(0)}_{\rm qs}\right] Y + \KompO \Delta m^{(0)}_0
}
at the background level. In this case, $\KompO \Delta m^{(0)}_0$ only couples distortion modes, with no terms affecting the temperature term $\propto G$, and $M$ only being sourced by the conversion of $Y\rightarrow Y_1 \rightarrow \ldots \rightarrow Y_N \rightarrow M$. Enforcing energy conservation for this system with the basis 
extended by $Y_k$ up to $Y_{15}$ then leads to the treatment described in \citep{chluba_spectro-spatial_2023-I} for the average distortion evolution. Importantly, for $\Delta m^{(0)}_0\rightarrow 0$ but $\Theta^{(0)}_{\rm e}\neq \Theta^{(0)}_{\rm qs}=\Theta^{(0)}_0+\eta_M \mu^{(0)}_0$, this means that $y$-type distortions are being sourced and then reprocessed through the Kompaneets operator until $\Theta^{(0)}_{\rm e}\rightarrow \Theta^{(0)}_{\rm qs}=\Theta^{(0)}_0+\eta_M \mu^{(0)}_0$, ensuring exact convergence of the scheme to full kinetic equilibrium under Compton scattering. 
If in addition $\mu^{(0)}_0$ vanishes, one naturally obtains full thermal equilibrium with $\Delta n^{(0)}_0 = \Theta^{(0)}_0 G$ and $\Theta^{(0)}_{\rm e}=\Theta^{(0)}_0$.

How do things change at the perturbed level? To obtain a numerically more stable scheme, we again need to ensure that all terms $\propto G$ and $\propto M$ are propagated correctly through the Kompaneets operator. Using the variables $\Delta m^{(0)}_0 $ and $\Delta m^{(1)}_0$ as defined in Eq.~\eqref{eq:def_Dm} with Eq.~\eqref{eq:evol_Dn_b_mod} and \eqref{eq:evol_Dn_a_mod} then yields the replacements
\bealf{
\KompO\left[ \Delta n^{(1)}_0-\Theta^{(1)}_0\boostO n^{(0)}_0\right]
&\rightarrow \KompO \Delta m^{(1)}_0 - \Theta^{(1)}_0 \KompO \boostO \Delta m^{(0)}_0 
\\
\nonumber
\Theta^{(1)}_0(1+\boostO)\frac{\id \Delta n^{(0)}_0}{\id y_{\rm sc}}\Bigg|_{\rm C}
&\rightarrow \Theta^{(1)}_0\left\{\left[\Theta^{(0)}_{\rm e}-\Theta^{(0)}_{\rm qs}\right] Y + \KompO \Delta m^{(0)}_0
+4\left[\Theta^{(0)}_{\rm e}-\Theta^{(0)}_{\rm qs}\right] Y_1 + \boostO \KompO \Delta m^{(0)}_0
\right\}
}
where we used Eq.~\eqref{eq:Dn1_rewrite} and $\boostO Y=4Y_1$ to simplify the expression. Collecting all terms in Eq.~\eqref{eq:evol_Dn_b_mod}, we then obtain
\bealf{
\label{eq:evol_Dn_b_mod_II}
\frac{\id \Delta n^{(1)}_0}{\id y_{\rm sc}}\Bigg|_{\rm C}
&=\left[\Theta^{(1)}_{\rm e}-\Theta^{(1)}_0(1+\Theta^{(0)}_{\rm e})+\Theta^{(1)}_0\left(\Theta^{(0)}_{\rm e}-\Theta^{(0)}_{\rm qs}\right)\right] Y  + \KompO \Delta m^{(1)}_0 - \Theta^{(1)}_0 \KompO \boostO \Delta m^{(0)}_0
\nonumber\\[-1mm]
&\qquad 
+ \Theta^{(1)}_0 \KompO \Delta m^{(0)}_0 
+ 4\Theta^{(1)}_0\left[\Theta^{(0)}_{\rm e}-\Theta^{(0)}_{\rm qs}\right]Y_1
+\Theta^{(1)}_0 \boostO \KompO \Delta m^{(0)}_0
\nonumber\\[2mm]
&=\left[\Theta^{(1)}_{\rm e}-\Theta^{(1)}_0\left(1+\Theta^{(0)}_{\rm qs}\right)\right] Y  + 4 \Theta^{(1)}_0
\left\{\Theta^{(0)}_{\rm e} 
-\Theta^{(0)}_{\rm qs} \right\} Y_1
\\[1.5mm] \nonumber
&\qquad\qquad + \KompO \Delta m^{(1)}_0 + \Theta^{(1)}_0 \KompO \Delta m^{(0)}_0 +\Theta^{(1)}_0 \left\{\boostO \KompO -\KompO \boostO \right\}\Delta m^{(0)}_0.
}
Written in this form, it is evident that we indeed reach kinetic equilibrium with respect to Compton scattering for $\Delta m^{(0)}_0 \rightarrow 0$, $\Delta m^{(1)}_0 \rightarrow 0$, $\Theta^{(0)}_{\rm e}\rightarrow \Theta^{(0)}_{\rm qs}$ and $\Theta^{(1)}_{\rm e}\rightarrow \Theta^{(1)}_0(1+\Theta^{(0)}_{\rm qs})$ . We note that within the distortion basis $\{Y, Y_1, \ldots, Y_{15}, M\}$ \citep{chluba_spectro-spatial_2023-I, chluba_spectro-spatial_2023-II}, all the operators appearing in this expression are represented sufficiently well, meaning that the expression in Eq.~\eqref{eq:evol_Dn_b_mod_II} provides a good starting point for numerical applications. 

\subsection{Accounting for energy injection}
\label{sec:num_sol_energy}
We can now also add energy injection to the problem to find the correct electron temperature given a heating term and the radiation field \citep[see Appendix~C of][for details]{chluba_spectro-spatial_2023-II}. We can assume that the electron temperature evolves along a sequence of quasi-stationary stages and is always rapidly driven towards Compton equilibrium.
Starting with the evolution equation for the electron temperature and setting $\partial_t \Te\approx 0$, at the background level this yields \citep{Chluba2011therm, chluba_spectro-spatial_2023-II}
\bsub
\bealf{
\Lambda_{\rm C}^{(0)}
&=\kappa 
\left\{-\frac{1}{4 E_{\nbb}}\int x^3 \frac{\id \Delta n^{(0)}_0}{\id y_{\rm sc}}\Bigg|_{\rm C} \id x
\right\}
=\kappa 
\left\{ 
\Theta_{\rm C}^{(0)}-\Theta^{(0)}_{\rm e}
\right\}
\\
\label{eq:Theta_e_zero}
&\Lambda_{\rm C}^{(0)}+\dot{Q}^{(0)}_{\rm c}\approx 0\qquad \rightarrow \qquad
\Theta^{(0)}_{\rm e}\approx \Theta^{(0)}_{\rm C}+\frac{\dot{Q}^{(0)}_{\rm c}}{\kappa}
}
\esub
with $\kappa=4\rho_z \dot{\tau} \, \Thz $, where $\rho_z$ is the photon energy density. Here $\Lambda_{\rm C}^{(0)}$ is the Compton energy exchange integral between photons and electrons at the background level and $\dot{Q}^{(0)}_{\rm c}$ is the energy injection term into the electrons caused by the interactions through collisions. We neglected the small corrections from the heat capacity of the non-relativistic electrons and their adiabatic cooling effect, which causes very small average distortions $\mu\simeq -\pot{3}{-9}$ \citep{Chluba2005, Chluba2011therm, Khatri2011BE} and hence tiny distortion anisotropies.

At first order in perturbations, we also need to take into account the modified electron scattering rate and transformation of the time variable into the local inertial frame. The latter drops out of the problem when Hubble cooling corrections are neglected \citep{chluba_spectro-spatial_2023-II}, leaving the condition $\Lambda_{\rm C}^{(1)}+\dot{Q}^{(1)}_{\rm c}\approx 0$, where the Compton energy exchange integral with the electrons is
\bealf{
\Lambda_{\rm C}^{(1)}
&=\kappa 
\left\{-\frac{1}{4 E_{\nbb}}\int x^3 \frac{\id \Delta n^{(1)}_0}{\id y_{\rm sc}}\Bigg|_{\rm C} \id x 
+\delta^{(1)}_{\rm b} 
\left[\Theta_{\rm C}^{(0)} - \Theta_{\rm e}^{(0)}\right]
\right\}
\nonumber \\
&=\kappa 
\left\{ 
\Theta^{(1)}_0(1+\Theta^{(0)}_{\rm e})  +
\frac{\int x^3 w_y \left[\Delta n^{(1)}_0-\Theta^{(1)}_0\boostO n^{(0)}_0\right] \id x}{4 E_{\nbb}}
-\Theta^{(1)}_{\rm e} +(5\Theta^{(1)}_0+\delta^{(1)}_{\rm b})\left[\Theta_{\rm C}^{(0)} - \Theta_{\rm e}^{(0)}\right]
\right\}
\nonumber \\
&=\kappa 
\left\{ 
\Theta_{\rm C}^{(1)}-\Theta^{(1)}_{\rm e}
+\left(\delta^{(1)}_{\rm b} + 4\Theta^{(1)}_0\right)
\left[\Theta_{\rm C}^{(0)} - \Theta_{\rm e}^{(0)}\right]
\right\}.
}
Here, we started from Eq.~\eqref{eq:evol_Dn_b_mod} and used Eq.~\eqref{eq:Teq} to eliminate the intermediate integrals, but the same result is obtained from Eq.~\eqref{eq:evol_Dn_b_mod_II} with Eq.~\eqref{eq:Teq_qs}, showing consistency of the expressions. 
Again solving the condition $\Lambda_{\rm C}^{(1)}+\dot{Q}^{(1)}_{\rm c}\approx 0$ for $\Theta_{\rm e}^{(1)}$, we then find
\bealf{
\label{eq:Theta_e_first}
\Theta^{(1)}_{\rm e}\approx \Theta^{(1)}_{\rm C}+\frac{\dot{Q}^{(1)}_{\rm c}}
{\kappa}
-\left[\delta^{(1)}_{\rm b} + 4\Theta^{(1)}_0\right] \frac{\dot{Q}^{(0)}_{\rm c}}
{\kappa}.
}
This is consistent with \citep{chluba_spectro-spatial_2023-II} but at a slightly modified Compton equilibrium temperature $\Theta^{(1)}_{\rm C}$, due to the additional stimulated scattering corrections taken into account here. 

Putting things together, from Eq.~\eqref{eq:evol_Dn_a_mod} and \eqref{eq:evol_Dn_b_mod_II} with Eq.~\eqref{eq:Theta_e_zero} and \eqref{eq:Theta_e_first}, we then have the final expressions describing all effects of Compton scattering with electrons in the presence of perturbations in the CMB spectrum:
\bsub
\label{eq:evol_Dn_final}
\bealf{
\label{eq:evol_Dn_final_a}
\frac{\id \Delta n^{(0)}_0}{\id y_{\rm sc}}\Bigg|_{\rm C}&=\left[\Theta^{(0)}_{\rm C}-\Theta^{(0)}_{\rm qs}\right] Y + \frac{\dot{Q}^{(0)}_{\rm c}}{\kappa} Y + \KompO \Delta m^{(0)}_0 
\\
\label{eq:evol_Dn_final_b}
\frac{\id \Delta n^{(1)}_0}{\id y_{\rm sc}}\Bigg|_{\rm C}&=
(\delta^{(1)}_{\rm b} + \Psi^{(1)})\frac{\id \Delta n^{(0)}_0}{\id y_{\rm sc}}\Bigg|_{\rm {C}}+\left[\Theta^{(1)}_{\rm C}+\frac{\dot{Q}^{(1)}_{\rm c}}{\kappa} 
-\left[\delta^{(1)}_{\rm b} + 4\Theta^{(1)}_0\right] \frac{\dot{Q}^{(0)}_{\rm c}}{\kappa}-\Theta^{(1)}_0\left(1+\Theta^{(0)}_{\rm qs}\right)\right] Y 
\nonumber \\ 
&\;\,+ 4 \Theta^{(1)}_0\left\{\Theta^{(0)}_{\rm C}+ \frac{\dot{Q}^{(0)}_{\rm c}}{\kappa}-\Theta^{(0)}_{\rm qs} \right\}Y_1 
+ \KompO \Delta m^{(1)}_0 + \Theta^{(1)}_0 \KompO \Delta m^{(0)}_0 
+\Theta^{(1)}_0 \left\{\boostO \KompO -\KompO \boostO \right\}\Delta m^{(0)}_0
\nonumber\\
&=
(\delta^{(1)}_{\rm b} + \Theta^{(1)}_0 + \Psi^{(1)})\left[\Theta^{(0)}_{\rm C}-\Theta^{(0)}_{\rm qs}\right] Y+\left[\Theta^{(1)}_{\rm C}-\Theta^{(1)}_0\left(1+\Theta^{(0)}_{\rm C}\right)\right] Y  + \left[\frac{\dot{Q}^{(1)}_{\rm c}}{\kappa} 
+\Psi^{(1)} \frac{\dot{Q}^{(0)}_{\rm c}}{\kappa} \right]Y 
\nonumber\\ 
&\qquad + 4 \Theta^{(1)}_0\left\{\Theta^{(0)}_{\rm C}-\Theta^{(0)}_{\rm qs} \right\}Y_1+ 4 \Theta^{(1)}_0\frac{\dot{Q}^{(0)}_{\rm c}}{\kappa} (Y_1-Y)
\\[2mm] \nonumber
&\qquad\qquad + \KompO \Delta m^{(1)}_0 + (\delta^{(1)}_{\rm b} +\Theta^{(1)}_0+ \Psi^{(1)})\,\KompO \Delta m^{(0)}_0 
+\Theta^{(1)}_0 \left\{\boostO \KompO -\KompO \boostO \right\}\Delta m^{(0)}_0
\nonumber\\
&=
\Delta \Theta^{(1)}_{\rm C} Y + \KompO \Delta m^{(1)}_0  + \left[\frac{\dot{Q}^{(1)}_{\rm c}}{\kappa} 
+\Psi^{(1)} \frac{\dot{Q}^{(0)}_{\rm c}}{\kappa} \right]Y 
+(\delta^{(1)}_{\rm b} + \Theta^{(1)}_0 + \Psi^{(1)})\left[ \Delta \Theta^{(0)}_{\rm C} Y+\KompO \Delta m^{(0)}_0\right]
\nonumber\\ 
&\qquad 
+ 4 \Theta^{(1)}_0 \left[ \Delta \Theta^{(0)}_{\rm C}Y_1+ \frac{\dot{Q}^{(0)}_{\rm c}}{\kappa} (Y_1-Y)\right]
+\Theta^{(1)}_0 \left\{\boostO \KompO -\KompO \boostO \right\}\Delta m^{(0)}_0.
}
\esub
{In} the last line we introduced $\Delta \Theta^{(0)}_{\rm C}=\Theta^{(0)}_{\rm C}-\Theta^{(0)}_{\rm qs}$ and $\Delta \Theta^{(1)}_{\rm C}=\Theta^{(1)}_{\rm C}-\Theta^{(1)}_0\left(1+\Theta^{(0)}_{\rm C}\right)$ and also grouped terms conveniently.
Note that at first order in perturbations we added the term from changes in the time-coordinate {and perturbed scattering}, i.e., $(\delta^{(1)}_{\rm b} + \Psi^{(1)}) \id \Delta n^{(0)}_0/\id y_{\rm sc}|_{\rm {C}}$.

We note that $\kappa\propto \Thz^5$. {In Eq.~\eqref{eq:evol_Dn_final} this term only appear in the combination} $\Thz/\kappa$ as part of the heating rate, meaning that we have a modified injection rate when changing {the reference temperature} $\Tz$, i.e., $\dot{Q}^{(0)}_{\rm c}/\kappa\rightarrow \dot{Q}^{(0)}_{\rm c}\,(1-4\Thetaref)/\kappa_{\rm CMB}$ and $\dot{Q}^{(1)}_{\rm c}/
\kappa\rightarrow \dot{Q}^{(1)}_{\rm c}/
\kappa_{\rm CMB}$ (with corrections that are higher order in the background distortion), where $\kappa_{\rm CMB}$ is $\kappa$ {evaluated} for $\Tz=\TCMB$. However, this effect can be captured by simply rescaling the energy release rate (and thus reinterpreting the related constraint), should this level of precision be required. For most cases, we can omit this distinction, which becomes important for {\it large distortion} scenarios \citep[e.g.,][]{Chluba2020large, Acharya2022large}.

Written in the form in Eq.~\eqref{eq:evol_Dn_final}, we can directly see that in the absence of energy injection kinetic equilibrium with respect to Compton scattering at the background level is reached when $\Delta m^{(0)}\rightarrow 0$, which implies $\Theta^{(0)}_{\rm C}\rightarrow \Theta^{(0)}_{\rm qs}$ and $\Delta n^{(0)}_0\rightarrow \Delta n^{(0)}_{\rm qs}=\Theta^{(0)}_{0} G+\mu_0^{(0)} M$. 
Full thermal equilibrium at the background level furthermore means $\mu_0^{(0)}=0$.
This eliminates most of the terms in the perturbed equation, leaving only 
\label{eq:evol_Dn_eq_background}
\bealf{
\frac{\id \Delta n^{(1)}_0}{\id y_{\rm sc}}\Bigg|^{\rm eq}_{\rm {C}}
&=
\left[\Theta^{(1)}_{\rm C}-\Theta^{(1)}_0\left(1+\Theta^{(0)}_{\rm qs}\right)\right] Y 
+ \KompO \Delta m^{(1)}_0,
}
which vanishes when $\Delta n^{(1)}_0\rightarrow \Theta^{(1)}_{0} \boostO n^{(0)}_0=\Theta^{(1)}_{0} (G + \boostO \Delta n^{(0)}_{\rm qs})={\Delta n^{(1)}_{\rm qs}}$. {For $\Delta n^{(0)}_{\rm qs}=\Theta^{(0)}_{0} G$} this produces full equilibrium distributions, thus improving the initial formulation of \citep{chluba_spectro-spatial_2023-II} with the differences stemming from the term $2\Theta^{(1)}_0\,\DiffO^* \Delta n^{(0)}_0 \boostO \nbb$ at the perturbed level and a significant regrouping of terms to achieve a better numerical representation and physical transparency. 

\subsection{Accounting for double Compton and Bremsstrahlung}
\label{sec:photon_prod}
To complete the formulation, one has to add the effects of photon emission and absorption by double Compton and Bremsstrahlung, which can be accurately modeled using {\tt DCpack} \citep{Ravenni2020DC} and {\tt BRpack}~\citep{Chluba2020BRpack}. However, to include these processes in the frequency hierarchy, we have to define how quickly $\mu$ converts to $T$, which can be achieved using quasi-stationary approximations \citep{Sunyaev1970mu, Danese1982, Burigana1991, Hu1995PhD, Chluba2014}.
This calculation was carried out in Appendix C.3 of \citep{chluba_spectro-spatial_2023-II}, which we briefly recap. 

We start by stating the emission terms at background and perturbed order. The full terms are given in Eq.~(A.10) of \citep{chluba_spectro-spatial_2023-II}; however, here we only need the leading contributions in the limit $x\ll 1$, which read [see Appendix~\ref{sec:simplified_pert_em} for a simplified derivation]
\begin{align}
\label{eq:emission_terms}
\frac{\id  n^{(0)}_0}{\id y_{\rm sc}}\Bigg|_{\rm em}
&\approx
-\frac{\Lambda(x, \Thz)}{\Thz x^2}\left[\Delta n^{(0)}_0-
\frac{\Theta^{(0)}_{\rm e}}{x}\right]
\\\nonumber
\frac{\id  n^{(1)}_0}{\id y_{\rm sc}}\Bigg|_{\rm em}
&\approx 
-\frac{\Lambda(x, \Thz)}{\Thz x^2}\Bigg\{ \Delta n^{(1)}_0-
\frac{\Theta^{(1)}_{\rm e}}{x} 
+\left(\delta^{(1)}_{\rm b}+\Psi^{(1)}+\Theta^{(1)}_0\left[\frac{\partial \ln \Lambda}{\partial\ln\Thg}\Bigg|_{\Thz}+\frac{\partial \ln \Lambda}{\partial\ln\The}\Bigg|_{\Thz}-1 \right]\right) \left[\Delta n^{(0)}_0
-\frac{\Theta^{(0)}_{\rm e}}{x} \right]\Bigg\}.
\end{align}
Here, {the subscript 'em' denotes that the terms relate to emission processes;} $\Lambda(x, \The, \Thg)$ is the double Compton and Bremsstrahlung emission coefficient, which we approximated as $\Lambda(x, \The, \Thg)\propto  \Thg^2$ (assuming double Compton dominates {and temperature corrections \citep{Chluba2007a} can be omitted}). 

To continue, we also need the low frequency limits of Eq.~\eqref{eq:evol_Dn_final}. All terms $\propto Y_k$ can be dropped as they only scale as $\propto 1/x$, while the leading terms scale as $\propto 1/x^2$. These will lead to small frequency-dependent corrections to the photon production rate and can in principle be included using perturbative methods as in \citep{Chluba2014}. We will also assume {external} heating has stopped, leaving only terms involving the Kompaneets operator. In the low-frequency limit, from Eq.~\eqref{eq:Komp_w} with $w\rightarrow 2$, we then have $\KompO^{\rm low}=(\boostO-3)(\boostO-2)$, which means that $\boostO \KompO -\KompO \boostO\approx 0$. {Using Eq.~\eqref{eq:evol_Dn_final} this} then gives 
\bsub
\label{eq:evol_Dn_low}
\bealf{
\frac{\id \Delta n^{(0)}_0}{\id y_{\rm sc}}\Bigg|^{\rm low}_{\rm {C}}&=\KompO^{\rm low} \Delta m^{(0)}_0 \approx \KompO^{\rm low} \left[\Delta n^{(0)}_0-\frac{\Theta^{(0)}_{\rm e}}{x}\right]
\\
\frac{\id \Delta n^{(1)}_0}{\id y_{\rm sc}}\Bigg|^{\rm low}_{\rm {C}}
&=
\KompO^{\rm low} \Delta m^{(1)}_0 + (\delta^{(1)}_{\rm b} +\Theta^{(1)}_0+ \Psi^{(1)})\,\KompO^{\rm low} \Delta m^{(0)}_0 
\nonumber\\[-1mm]
&\approx 
\KompO^{\rm low} \left[\Delta n^{(1)}_0-\frac{\Theta^{(1)}_{\rm e}}{x}\right] + (\delta^{(1)}_{\rm b} +\Theta^{(1)}_0+ \Psi^{(1)})\,\KompO^{\rm low} \left[\Delta n^{(0)}_0-\frac{\Theta^{(0)}_{\rm e}}{x}\right], 
}
\esub
where we assume that the Compton equilibrium temperature is reached. {Adding the emission terms from Eq.~\eqref{eq:emission_terms} and assuming quasi-stationary conditions, we} then find the solutions
\bsub
\label{eq:evol_Dn_low_II}
\bealf{
0&\approx \KompO^{\rm low} \left[\Delta n^{(0)}_0-\frac{\Theta^{(0)}_{\rm e}}{x}\right]-\frac{\Lambda(\Thz)}{\Thz x^2}\left[\Delta n^{(0)}_0-
\frac{\Theta^{(0)}_{\rm e}}{x}\right]\quad \rightarrow \quad 
\Delta n^{(0)}_0 = \frac{\Theta^{(0)}_{\rm e}}{x} + \frac{\mu_{\infty}^{(0)} \expf{-\xc/x}}{x^2}
\\
0&\approx 
\KompO^{\rm low} \left[\Delta n^{(1)}_0-\frac{\Theta^{(1)}_{\rm e}}{x}\right] + (\delta^{(1)}_{\rm b} +\Theta^{(1)}_0+ \Psi^{(1)})\,\KompO^{\rm low} \left[\Delta n^{(0)}_0-\frac{\Theta^{(0)}_{\rm e}}{x}\right] 
\nonumber\\
&\qquad \qquad -\frac{\Lambda(x, \Thz)}{x^2}\Bigg\{ \Delta n^{(1)}_0-
\frac{\Theta^{(1)}_{\rm e}}{x} 
+\left(\delta^{(1)}_{\rm b}+\Psi^{(1)}+\Theta^{(1)}_0\right) \left[\Delta n^{(0)}_0
-\frac{\Theta^{(0)}_{\rm e}}{x} \right]\Bigg\}
\nonumber\\
&= \KompO^{\rm low} \left[\Delta n^{(1)}_0-\frac{\Theta^{(1)}_{\rm e}}{x}\right]-\frac{\Lambda(\Thz)}{\Thz x^2}\left[\Delta n^{(1)}_0-
\frac{\Theta^{(1)}_{\rm e}}{x}\right]\quad \rightarrow \quad 
\Delta n^{(1)}_0 = \frac{\Theta^{(1)}_{\rm e}}{x} + \frac{\mu_{\infty}^{(1)} \expf{-\xc/x}}{x^2}}
\esub
with critical frequency $\xc=\sqrt{\Lambda/\Thz}$ and where we used $\partial \ln \Lambda/\partial\ln\Thg|_{\Thz}+\partial \ln \Lambda/\partial\ln\The|_{\Thz}-1 \equiv 1$. 
The solutions $\Delta n^{(i)}_0$ allow us to compute the photon production rate at background and perturbed order, which in turn can be used to compute the conversion rate of $\mu\rightarrow \Theta$ \citep[see Eq.~(C.20) of][]{chluba_spectro-spatial_2023-II}, confirming the effective description to account for the effects of photon production using \citep{chluba_spectro-spatial_2023-II}
\bsub
\label{eq:effective_em_abs}
\begin{align}
\frac{\id  n^{(0)}_0}{\id y_{\rm sc}}\Bigg|_{\rm em}^{\rm eff}
&\approx \gamma_T\,\xc\,\mu_{0} ^{(0)} G -\gamma_N\,\xc\,\mu_{0} ^{(0)} M
\\
\frac{\id  n^{(1)}_0}{\id y_{\rm sc}}\Bigg|_{\rm em}^{\rm eff}
&\approx\gamma_T\,\xc\,\mu_{0} ^{(1)} G -\gamma_N\,\xc\,\mu_{0}^{(1)} M + \left(\delta^{(1)}_{\rm b}+\Psi^{(1)}+\Theta^{(1)}_0
\right) \left[\gamma_T\,\xc\,\mu_{0} ^{(0)} G -\gamma_N\,\xc\,\mu_{0} ^{(0)} M\right].
\end{align}
\esub
Here $\gamma_T\approx 0.1387$ and $\gamma_N \approx 0.7769$ ensure energy conservation when converting from $M$ to $G$ \citep[see][for additional details]{Chluba2014, chluba_spectro-spatial_2023-II}. 

We mention that a naive treatment using $\Thg \xc \propto \Thg^{3/2}=\Thz^{3/2}(1+\Theta^{(1)}_0)^{3/2}$ would {suggest} a term $\propto (3/2) \Theta^{(1)}_0$ instead of $\propto \Theta^{(1)}_0$. The local Compton and double Compton scattering rates adjust the spectrum in such a way that the coefficient indeed reduces to unity \citep{chluba_spectro-spatial_2023-II}.
Can we recover this result in another way? Let us look at how the critical frequency is determined. The equation to derive the frequency dependence of the average chemical potential is given by \citep{chluba_spectro-spatial_2023-II}
\bealf{
\label{eq:mu_equation}
0\approx \The \hat{\mathcal{K}}_x \Delta n_0 + \Thz \Theta_{\rm e} Y(x)- \frac{\Lambda(x, \The,\Thg)\,(1-\expf{-x\,\Thz/\The})}{x^3}\left[\Delta n_0-G(x)\,\Theta_{\rm e} \right],
}
where we have only linearized in terms of the photon occupation number and used the Kompaneets operator $\hat{\mathcal{K}}_x=x^{-2} \partial_x x^4 [\partial_x + 1+2 \nbb]$, leaving all temperature variables, $\theta_i=kT_i/\me c^2$ {or} explicitly $\The = \Thz (1+\Theta_{\rm e})$, $\Thg = \Thz (1+\Theta_0)$, as they are according to their physical origin. 

In the low frequency limit , $G\rightarrow 1/x$, $\hat{\mathcal{K}}_x\rightarrow x^{-2} \partial_x x^2 \partial_x x^2$, $Y\rightarrow -2/x$, such that 
\bealf{
\label{eq:mu_equation_low}
0\approx \frac{\The}{x^2}\partial_x x^2 \partial_x x^2 \Delta n_0 - \frac{2\Thz \Theta_{\rm e}}{x}- \frac{\Thz\,\Lambda(x, \The,\Thg)}{\The x^2}\left[\Delta n_0-\frac{\Theta_{\rm e}}{x} \right].
}
Replacing $\Delta n_0 = \Delta n_{0,\rm d}+\frac{\Theta_{\rm e}}{x}$ and dividing by $\The/x^2$, we then have
\bealf{
\label{eq:mu_equation_low_dist}
0\approx \partial_x x^2 \partial_x x^2 \Delta n_{0,\rm d}- \frac{\Thz\,\Lambda(x, \The,\Thg)}{\The^2}\Delta n_{0,\rm d}
+ 2x \left(1- \frac{\Thz}{\The}\right) \Theta_{\rm e}.
}
The last term takes the form $\left(1- \frac{\Thz}{\The}\right) \Theta_{\rm e}=\Theta^2_{\rm e}/(1+\Theta_{\rm e})$. With the Ansatz $\Delta n_{0,\rm d}=\mu_\infty \expf{-\xc/x}/x^2$, {for the distortion (indicated by the subscript 'd')} we can write $\partial_x x^2 \partial_x x^2 \Delta n_{0,\rm d}=\mu_\infty\,\partial_x x^2 \partial_x \expf{-\xc/x} = \mu_\infty\,\xc \partial_x \expf{-\xc/x} = \mu_\infty\,\xc^2\expf{-\xc/x}/x^2 \equiv \xc^2 \Delta n_{0,\rm d}$, meaning
\bealf{
\label{eq:mu_equation_low_dist_final}
0\approx \left[\xc^2- \frac{\Thz\,\Lambda(x, \The,\Thg)}{\The^2}\right]\Delta n_{0,\rm d}
+ \frac{2x \, \Theta^2_{\rm e}}{1+\Theta_{\rm e}}
}
from Eq.~\eqref{eq:mu_equation_low_dist}.
For this equation to have solutions, one has to assume that around $x\simeq \xc$ the last term is negligible -- an approximation that is well-justified, even if at higher frequencies time-dependent corrections can appear \citep{Chluba2014} -- and we also need to determine the critical frequency such that the brackets vanish. We shall be using $\Lambda \approx  \frac{4\alpha}{3\pi}\,\Thg^2\,\mathcal{I}_4$ for double Compton emission\footnote{The procedure can be augmented to include slow frequency variations \citep{Chluba2014}.}, neglecting any temperature or frequency corrections \citep{Chluba2014}. Here, we defined the {double Compton} emission integral $\mathcal{I}_4 = \int x^4 \nbb(1+\nbb) \id x=4\pi^4/15$, which then implies 
\bealf{
\label{eq:x_crit_fin}
\xc = \sqrt{\frac{4\alpha}{3\pi}\,\mathcal{I}_4\,\Thz}\times \frac{1+\Theta_0}{1+\Theta_{\rm e}}\approx \sqrt{\frac{4\alpha}{3\pi}\,\mathcal{I}_4\,\Thz}.
}
where in the last step we assumed $\Theta_{\rm e}\approx \Theta_0$. This means that in the treatment the critical frequency variable itself should {\it not} be considered as a perturbed quantity, as the dynamics of double Compton and Compton scattering will establish a new balance that leaves the solution invariant. However, the time-scale of the problem is given by $\id y_{\rm e} = \The \id \tau\approx  \Thg \id \tau$, which is also how we defined the quasi-stationary solution, meaning that $\Thg \xc \rightarrow \Thz (1+\Theta_0) \xc$. This then gives the desired result.

\subsection{Final expression for the thermalization terms}
\label{sec:therm_final}
We can now summarize the effective description of the Compton, double Compton and Bremsstrahlung processes. Collecting terms from Eq.~\eqref{eq:evol_Dn_final} and Eq.~\eqref{eq:effective_em_abs}, we then have the final set of distortion thermalization terms
\bsub
\label{eq:evol_Dn_final_all}
\bealf{
\frac{\id \Delta n^{(0)}_0}{\id y_{\rm sc}}\Bigg|_{\rm d}&\approx \Delta \Theta^{(0)}_{\rm C} Y + \KompO \Delta m^{(0)}_0 +\gamma_T\,\xc\,\mu_{0} ^{(0)} G -\gamma_N\,\xc\,\mu_{0} ^{(0)} M+ \frac{\dot{Q}^{(0)}_{\rm c}}{\kappa} Y 
\\
\frac{\id \Delta n^{(1)}_0}{\id y_{\rm sc}}\Bigg|_{\rm d}
&\approx 
\Delta \Theta^{(1)}_{\rm C} Y + \KompO \Delta m^{(1)}_0  +\gamma_T\,\xc\,\mu_{0}^{(1)} G -\gamma_N\,\xc\,\mu_{0}^{(1)} M + \left[\frac{\dot{Q}^{(1)}_{\rm c}}{\kappa} 
+\Psi^{(1)} \frac{\dot{Q}^{(0)}_{\rm c}}{\kappa} \right]Y
\nonumber\\
&\qquad 
+(\delta^{(1)}_{\rm b} + \Theta^{(1)}_0 + \Psi^{(1)})\left[ \Delta \Theta^{(0)}_{\rm C} Y+\KompO \Delta m^{(0)}_0 + \gamma_T\,\xc\,\mu_{0} ^{(0)} G -\gamma_N\,\xc\,\mu_{0} ^{(0)} M\right]
\nonumber\\ 
&\qquad \qquad 
+ 4 \Theta^{(1)}_0 \left[ \Delta \Theta^{(0)}_{\rm C}Y_1+ \frac{\dot{Q}^{(0)}_{\rm c}}{\kappa} (Y_1-Y)\right]
+\Theta^{(1)}_0 \left\{\boostO \KompO -\KompO \boostO \right\}\Delta m^{(0)}_0.
}
\esub
We will next convert these equations into evolution equations in the distortion basis for efficient numerical {treatment. Before, we mention} that because $\id y_{\rm sc}=\Thz\id \tau=(\kB\Tz/\me c^2)\id \tau$ with $\Tz=\TCMB(1+\Thetaref)$, we generally have to scale the terms by an factor of $(1+\Thetaref)$ when changing the reference temperature even if the CMB temperature is $\TCMB$. However, this factor enters at higher order in the average distortion unless large $\Thetaref$ is being considered, and thus can be dropped.

\subsection{Numerical implementation for thermalization terms}
\label{sec:num_setup}
We now have all the required equations to discuss the final numerical implementation of the problem. Following \citep{chluba_spectro-spatial_2023-I, chluba_spectro-spatial_2023-II}, we use the spectral basis $\{G, Y_0, Y_1, \ldots, Y_{15}, M\}$ with $Y=Y_0$ and $Y_k=(1/4)^k \boostO^k Y$. 
{We can then} convert all differential operators into matrix multiplications. Similarly, integrals over the photon distribution can {be converted} into scalar products.
{We use} the following mappings
\begin{align}
\label{eq:mappings_num}
&\Delta n^{(i)}_\ell 
\rightarrow 
\vek{y}^{(i)}_\ell,
\quad  
\Delta m^{(i)}_\ell 
\rightarrow 
\Delta \vek{y}^{(i)}_\ell = \vek{y}^{(i)}_\ell - \vek{y}^{(i), \rm qs}_\ell,
\quad 
\boostO \Delta n^{(i)}_\ell
\rightarrow 
M_{\rm B}\,\vek{y}^{(i)}_\ell,
\quad
\boostO \Delta m^{(i)}_\ell
\rightarrow 
M_{\rm B}\,\Delta \vek{y}^{(i)}_\ell,
\nonumber\\[2mm]
&\Delta \Theta^{(0)}_{\rm C}=\Theta^{(0)}_{\rm C}-\Theta^{(0)}_{\rm qs}=\frac{\int x^3 w_y \,\Delta m^{(0)}_0 {\rm d} x}{4 E_{\nbb}} = \vek{b}_{\rm \Theta} \cdot \Delta \vek{y}^{(0)}_0,
\nonumber\\
&\Delta \Theta^{(1)}_{\rm C}=\Theta^{(1)}_{\rm C}-\Theta^{(1)}_0 \left[1+\Theta^{(0)}_{\rm C}\right] \approx 
\frac{\int x^3 w_y \left[\Delta m^{(1)}_0-\Theta^{(1)}_0 \boostO \Delta m^{(0)}_0\right] {\rm d} x}{4 E_{\nbb}}=\vek{b}_{\rm \Theta} \cdot\left[\Delta \vek{y}^{(1)}_0 - \Theta^{(1)}_0 M_{\rm B}\, \Delta \vek{y}^{(0)}_0\right],
\nonumber\\
&\KompO \Delta m^{(i)}_0 = M_{\rm C} \Delta \vek{y}^{(i)}_0, \quad \left\{\boostO \KompO -\KompO \boostO \right\}\Delta m^{(0)}_0 = M_{\rm BC} \Delta \vek{y}^{(0)}_0
\end{align}
where $\ell$ is the multipole index. Here, we defined the boost matrix, $M_{\rm B}$, the Compton matrix, $M_{\rm C}$, and the matrix $M_{\rm BC}$ to describe the effect of the commutator $[\boostO, \KompO]$. All these matrices can be pre-computed and remain independent of the solution~\citep[see][for details]{chluba_spectro-spatial_2023-I, chluba_spectro-spatial_2023-II}. 

Before considering the full evolution equation, we clarify an important property of the Kompaneets operator. For this, we first define the matrix $M_\Theta=\vek{e}_Y\otimes \vek{b}_{\rm \Theta}$. This matrix simply has $\vek{b}_{\rm \Theta}^T$ as the row of the $y$-variable, and otherwise vanishes, meaning $M_\Theta \vek{y}_0=(\vek{b}_{\rm \Theta} \cdot \vek{y}_0)\,\vek{e}_Y$.
{This means} $$\Delta \Theta^{(0)}_{\rm C} Y+\KompO \Delta m^{(0)}_0 \rightarrow M_\Theta \Delta \vek{y}^{(0)}_0 + M_{\rm C} \Delta \vek{y}^{(0)}_0 \equiv M_\Theta \vek{y}^{(0)}_0 + M_{\rm C} \vek{y}^{(0)}_0.$$ Let us first clarify the last step: although $\Delta \vek{y}^{(0)}_0=\vek{y}^{(0)}_0-\vek{y}^{(0)}_{\rm qs}$, we have $M_\Theta \vek{y}^{(0)}_{\rm qs} + M_{\rm C} \vek{y}^{(0)}_{\rm qs}=0$. This is because the background quasi-stationary solution is $\vek{y}^{(0)}_{\rm qs}=\Theta^{(0)}_0 \vek{e}_G + \mu^{(0)}_0 \vek{e}_M$ and $\vek{b}_{\rm \Theta} \cdot \vek{y}^{(0)}_{\rm qs}=\Theta^{(0)}_0+\eta_M \mu^{(0)}_0$ while $M_{\rm C}  \vek{y}^{(0)}_{\rm qs}=-(\Theta^{(0)}_0+\eta_M \mu^{(0)}_0)\,\vek{e}_Y$. This then means that in $(\vek{b}_{\rm \Theta} \cdot \vek{y}^{(0)}_0)\,\vek{e}_Y + M_{\rm C} \vek{y}^{(0)}_0$, the effects of $\Theta^{(0)}_0$ and $\mu^{(0)}_0$ cancel identically. We can therefore define the {\it Kompaneets matrix} $M_{\rm K}=M_\Theta + M_{\rm C}$, which is a block matrix that has vanishing columns for the $\Theta$ and $\mu$ variables and {\it only couples} to the $y_k$ variables of $\vek{y}$, leading to a cascading rotation\footnote{{Compton scattering conserves the number of photons and slowly converts $Y$ into $M$ while conserving energy in the process. This can be thought of as a rotation of the basis in an energy cascade towards $M$ \citep{chluba_spectro-spatial_2023-I}.}} of $Y\rightarrow Y_1 \rightarrow \ldots \rightarrow Y_{15} \rightarrow M$ without sourcing $G$ (i.e., the row for $\Theta$ also vanishes).
Looking at Eq.~\eqref{eq:evol_Dn_final_all} and Eq.~\eqref{eq:mappings_num}, in a similar manner we have
\bealf{
\nonumber
\Delta \Theta^{(1)}_{\rm C} Y + \KompO \Delta m^{(1)}_0
\; \rightarrow \;
M_{\Theta} \left[\Delta \vek{y}^{(1)}_0 - \Theta^{(1)}_0 M_{\rm B}\, \Delta \vek{y}^{(0)}_0\right]+ M_{\rm C} \Delta \vek{y}^{(1)}_0=M_{\rm K} \Delta \vek{y}^{(1)}_0 - \Theta^{(1)}_0\,M_{\Theta} M_{\rm B}\, \Delta \vek{y}^{(0)}_0.
}
The last term can be combined with $4 \Theta^{(1)}_0 \Delta \Theta^{(0)}_{\rm C}Y_1
+\Theta^{(1)}_0 \left\{\boostO \KompO -\KompO \boostO \right\}\Delta m^{(0)}_0$ to give
\bealf{
\nonumber
&\Delta \Theta^{(1)}_{\rm C} Y + \KompO \Delta m^{(1)}_0
+4 \Theta^{(1)}_0 \Delta \Theta^{(0)}_{\rm C}Y_1
+\Theta^{(1)}_0 \left\{\boostO \KompO -\KompO \boostO \right\}\Delta m^{(0)}_0
\\
&\qquad \rightarrow \;
M_{\rm K} \Delta \vek{y}^{(1)}_0 - \Theta^{(1)}_0\,M_{\Theta} M_{\rm B}\, \Delta \vek{y}^{(0)}_0 + 
\Theta^{(1)}_0 M_{\rm B} M_\Theta \Delta \vek{y}^{(0)}_0
+
\Theta^{(1)}_0\left[ M_{\rm B} M_{\rm C} - M_{\rm C} M_{\rm B} \right] \Delta \vek{y}^{(0)}_0
\nonumber\\
&\qquad \qquad  =M_{\rm K} \Delta \vek{y}^{(1)}_0 + \Theta^{(1)}_0\left[ M_{\rm B} M_{\rm K} - M_{\rm K} M_{\rm B} \right] \Delta \vek{y}^{(0)}_0,
\nonumber\\
&\qquad \qquad  =M_{\rm K} \vek{y}^{(1)}_0 - M_{\rm K} \vek{y}^{(1)}_{\rm qs} + \Theta^{(1)}_0\left[ M_{\rm B} M_{\rm K} - M_{\rm K} M_{\rm B} \right] \Delta \vek{y}^{(0)}_0,
}
where we note that $4 \Theta^{(1)}_0 \Delta \Theta^{(0)}_{\rm C}Y_1\rightarrow 4 \Theta^{(1)}_0 \vek{e}_{Y_1} (\vek{b}_{\Theta} \cdot \Delta \vek{y}^{(0)}_0)=\Theta^{(1)}_0 M_{\rm B} \,\vek{e}_{Y} (\vek{b}_{\Theta} \cdot \Delta \vek{y}_0^{(0)})\equiv \Theta^{(1)}_0 M_{\rm B} M_\Theta\,\Delta \vek{y}_0^{(0)}$.

From Eq.~\eqref{eq:evol_transport} and Eq.~\eqref{eq:evol_Dn_final_all}, which have to be considered together with the equations for the {other perturbation variables, $\Phi, \Psi$, etc. \citep[e.g.,][]{Ma1995}}, we then have the final system to account for the thermalization terms in the spectral distortion evolution
\bsub
\label{eq:evol_transport_yn}
\bealf{
&\frac{\partial \vek{y}^{(0)}_0}{\partial \eta}
=\tau'\Thz \left[ M_{\rm K} \vek{y}^{(0)}_0 + \vek{D}^{(0)}_0 \right] + \frac{{\vek{Q}'}^{(0)}}{4} 
\\[2mm]
&\frac{\partial \vek{y}^{(1)}}{\partial \eta}+\vgh\cdot \nabla \vek{y}^{(1)}+
\vek{b}^{(0)}_0 \left(\frac{\partial \Phi^{(1)}}{\partial \eta}+ \vgh \cdot \nabla\Psi^{(1)} \right)= 
{\rm C}^{(1)}_{\rm T}[\vek{y}]+{\rm C}^{(1)}_{\rm d}[\vek{y}]
\\
&{\rm C}^{(1)}_{\rm T}[\vek{y}]=\tau'\left[\vek{y}^{(1)}_0+\frac{1}{10}\,\vek{y}^{(1)}_2 - \vek{y}^{(1)} + \vek{b}^{(0)}_0 \beta^{(1)}\chi\,\right], 
\\[2mm]
&{\rm C}^{(1)}_{\rm d}[\vek{y}]={\rm C}^{(1)}_{\rm th}[\vek{y}]+\Psi^{(1)} {\rm C}^{(0)}_{\rm th}[\vek{y}]
\\ \nonumber
&\qquad\;\,\,\,= \tau' \Thz
\left[M_{\rm K} \vek{y}^{(1)}_0 + \vek{D}^{(1)}_0 - \Theta^{(1)}_0 M_{\rm K} M_{\rm B} \,\vek{y}^{(0)}_{\rm qs}\right] + \frac{{\vek{Q}'}^{(1)}}{4} + \Theta^{(1)}_0\,\vek{C}^{(0)}_0 
\\ \nonumber
&\qquad\qquad
+\tau' \Thz \Bigg\{(\delta^{(1)}_{\rm b} +\Theta^{(1)}_0+ \Psi^{(1)})\, \left[M_{\rm K} \vek{y}^{(0)}_0 + \vek{D}^{(0)}_0 \right]+\Theta^{(1)}_0 M_{\rm BK} \Delta \vek{y}^{(0)}_0\Bigg\}
\\[2mm]
&\vek{b}^{(0)}_0=\boostO n^{(0)}= \vek{e}_G + M_{\rm B} \vek{y}^{(0)}_0,
\qquad 
{\vek{y}^{(0)}_{\rm qs}=\left(\Theta^{(0)}_0,0, \ldots, 0, \mu^{(0)}_0\right)^T}
\\
&\vek{D}^{(i)}_0 =\xc \,\mu_{0}^{(i)} \left(\gamma_T, 0, \ldots, 0, -\gamma_N\right)^T, \qquad
\vek{C}^{(0)}_0= \frac{{Q'_{\rm c}}^{(0)}}{ \rho_z}\left(0,-1, 1, 0,\ldots, 0\right)^T
\\
&{\vek{Q}'}^{(0)}= \left(0,\frac{{Q'_{\rm c}}^{(0)}}{\rho_z},\ldots, 0\right)^T, \qquad 
{\vek{Q}'}^{(1)}= \left(0,\frac{{Q'_{\rm c}}^{(1)}}{\rho_z}+\Psi^{(1)}\frac{{Q'_{\rm c}}^{(0)}}{\rho_z},\ldots, 0\right)^T.
}
\esub
Here, we introduced the matrix $M_{\rm BK}=M_{\rm B} M_{\rm K} - M_{\rm K} M_{\rm B}$.
As part of the derivations, we also used that $\vek{y}^{(1)}_{\rm qs}=\Theta^{(1)}_0 \left[\vek{e}_G+M_{\rm B} \vek{y}^{(0)}_{\rm qs}\right]={\Theta^{(1)}_0 \vek{b}^{(0)}_0}$ and thus $M_{\rm K} \vek{y}^{(1)}_{\rm qs} = \Theta^{(1)}_0\,M_{\rm K} M_{\rm B} \vek{y}^{(0)}_{\rm qs}$.
Written in this form, the tendency towards the correct equilibrium is directly ensured and thus improves the form given in \citep{chluba_spectro-spatial_2023-II}. We note that even if the form of the equation appears quite different, we confirmed all terms given in \citep{chluba_spectro-spatial_2023-II} but added terms stemming from stimulated scattering effects that were previously neglected. We also note that here $\beta^{(1)}$, $\Phi^{(1)}$ and $\Psi^{(1)}$ require the {\it full solutions} including distortion corrections. We will carefully separate those corrections in Sect.~\ref{sec:num_setup_Theta_corr}.

We can further simplify matters by defining the {\it thermalization matrix} $M_{\rm T}$, where we replace the $\mu$ column of $M_{\rm K}$ with the $\vek{d}_0=\xc \left(\gamma_T, 0, \ldots, 0, -\gamma_N\right)^T$ to directly capture emission effects, {i.e., $M_{\rm T}\,\vek{y}^{(0)}_0=M_{\rm K} \vek{y}^{(0)}_0 + \vek{D}^{(0)}_0$}. Similarly, {because $M_{K}\vek{y}^{(0)}_{\rm qs}=0$, we find $M_{\rm BK} \Delta \vek{y}^{(0)}_0=M_{\rm BK} \vek{y}^{(0)}_0+M_{K}M_{\rm B} \,\vek{y}^{(0)}_{\rm qs}$, which cancels the explicit appearances of $\vek{y}^{(0)}_{\rm qs}$ in ${\rm C}^{(1)}_{\rm d}[\vek{y}]$, finally yielding} 
\bsub
\label{eq:evol_transport_yn_improved}
\bealf{
&\frac{\partial \vek{y}^{(0)}_0}{\partial \eta}
=\tau'\Thz M_{\rm T}  \vek{y}^{(0)}_0  + \frac{{\vek{Q}'}^{(0)}}{4},
\\
&{\rm C}^{(1)}_{\rm d}[\vek{y}]= \tau' \Thz 
M_{\rm T} \vek{y}^{(1)}_0  + \frac{{\vek{Q}'}^{(1)}}{4} + \Theta^{(1)}_0\,\vek{C}^{(0)}_0
+\tau' \Thz \left[(\delta^{(1)}_{\rm b} +\Theta^{(1)}_0+ \Psi^{(1)})\, M_{\rm T} +{\Theta^{(1)}_0 M_{\rm BK}}\right]\vek{y}^{(0)}_0,
}
\esub
{after} we grouped terms relating to {\it perturbed heating}, i.e., $\frac{1}{4}{\vek{Q}'}^{(1)} + \Theta^{(1)}_0\,\vek{C}^{(0)}_0$, and {\it perturbed thermalization} effects (the terms in the brackets). 

\newpage

\subsubsection{Consistency checks for background temperature shift}
\label{sec:check_T_ref}
Let us check if the system defined above gives the correct result when a simple change in the background blackbody temperature is being considered. This can also be required when initializing the problem with a non-zero difference in the CMB temperature, as for example needed for dark-photon to photon conversion problems, which have to start with blackbody radiation at $\Tin>\TCMB$ \citep{Chluba2024DP}.
In this case, $\vek{y}^{(0)}_{0, \rm s}=(\bar{\Theta}, 0, \ldots, 0)^T$ and $\vek{y}^{(1)}_{0, \rm s}=\Theta^{(1)}_0\,(1+3\bar{\Theta}, \bar{\Theta}, 0, \ldots, 0)^T=\Theta^{(1)}_0\,(1, 0, \ldots, 0)^T + \Theta^{(1)}_0 M_{\rm B} \vek{y}^{(0)}_{0, \rm s}$ which follows from Eq.~\eqref{eq:shited_initial}.
Inserting this into the system and assuming no external heating (${\vek{Q}'}^{(0)}={\vek{Q}'}^{(1)}=0$), we find that there is {\it no change} to the background evolution, i.e., $\partial_\eta \vek{y}^{(0)}_{0}=0$, as expected. 
At the perturbed level, we have $\vek{y}^{(1)}_{0, \rm s}=\Theta^{(1)}_0\,\vek{b}^{(0)}_0 =\Theta^{(1)}_0 M_{\rm B} \,\vek{y}^{(0)}_{0, \rm s}$ and from Eq.~\eqref{eq:evol_transport_yn_improved}
\bealf{
\label{eq:background_fixed_check}
&\frac{{\rm C}^{(1)}_{\rm d}[\vek{y}]}{\tau' \Thz}= M_{\rm T} \vek{y}^{(1)}_{0, \rm s}  
{+\Theta^{(1)}_0 M_{\rm BK}\,\vek{y}^{(0)}_{0, \rm s}}
= M_{\rm K} \vek{y}^{(1)}_{0, \rm s} - \Theta^{(1)}_0 M_{\rm K} M_{\rm B} \vek{y}^{(0)}_{0,\rm s} \equiv 0
}
{Here we used $M_{\rm T} \vek{y}^{(0)}_{0, \rm s}=0$ and $M_{\rm BK} \vek{y}^{(0)}_{0, \rm s}=-M_{\rm K} M_{\rm B} \vek{y}^{(0)}_{0, \rm s}$ together with $M_{\rm K} \vek{y}^{(1)}_{0, \rm s}=\Theta^{(1)}_0 M_{\rm K} M_{\rm B} \vek{y}^{(0)}_{0, \rm s}$. The solution to Eq.~\eqref{eq:evol_transport_yn} is then $\vek{y}^{(1)}_{\ell, \rm s}=\Theta^{(1)}_0\,\vek{b}^{(0)}_0$, where $\vek{b}^{(0)}_0$ factors out. This} implies that we only have to solve for the standard temperature perturbations, $\Theta^{(1)}$, {without extra sources for distortion anisotropies or temperature corrections}, demonstrating the consistency of the improved system.

\section{Kinematic corrections}
\label{sec:k_corrections}
We now also include first order kinematic corrections into the description of the scattering problem. This {causes} new source terms in the dipolar part of the spectrum that ensure the full equilibrium is reached. In the rest frame of the moving thermal electron distribution, we shall assume that (aside from the effect of Thomson scattering) only the monopole part of the photon distribution scatters at leading order in the electron temperature, while dipole, quadrupole and octupole scattering effects \citep[see][for these terms]{Chluba2012} are negligible. In the rest frame of the moving cloud we can therefore consider the scattering process using the Kompaneets equation:
\bealf{
\label{eq:evol_Dn_rest}
\frac{\id \Delta n'_0}{\id y'_{\rm sc}}&=\Theta'_{\rm e} Y(x') + \KompOp \Delta n'_0 + \DiffstarOp (\Delta n'_0)^2.
}
where here the prime means that the quantity is evaluated in the moving cloud's rest frame. 
We are after terms of order $\beta^{(1)} \Theta_0^{(0)}$, {while neglecting higher order velocity terms that are relevant to the dissipation of acoustic modes \citep[e.g.,][]{Chluba2012, Khatri2012mix, Ota2014, Ota2017}}. At this order, there are no kinematic corrections to the electron temperature, i.e., $\Theta'_{\rm e}\rightarrow \Theta_{\rm e}$. Similarly, we can directly neglect the last term, which would be of higher order. For the transformation of $\Delta n$ into the cloud frame, we can immediately use $\Delta n'_0\approx \Delta n_0(x')$. Then all that is left is to perform a Lorentz boost back into the CMB frame. This then gives the final result \citep[see Appendix~A2.1 of][for detailed intermediate steps]{chluba_spectro-spatial_2023-II}
\begin{align}
\label{eq:Main_Eq_lab_lin}
\frac{\partial \Delta n^{(1)}}{\partial y_{\rm sc}}
\Bigg|^{\rm sc}_{\rm kin}
&\approx 
-\beta^{(1)} \chi\left\{\Theta_{\rm e}^{(0)}\,\left[ \Yspec -4\Ynspec{1}\right]
+\left[
\KompO - \KompO \boostO 
+
\DiffO^*\,\left(A-2G
\right)
\right] \Delta n^{(0)}_0\right\}
\end{align}
for the kinematic corrections, where $A=1+2\nbb$ and $\chi=\vbetah\cdot\vgh$ is the direction cosine. This can again be rewritten to eliminate the terms $\propto \DiffO^*$. First we can use $\DiffO^*\,A=\KompO - \DiffO$. We also have $2\DiffO^* G \Delta n^{(0)}_0 = 2\DiffO^* \Delta n^{(0)}_0 \boostO \nbb$, meaning that we can apply Eq.~\eqref{eq:important_identity} to find
\begin{align}
\label{eq:Main_Eq_lab_lin_final}
\frac{\partial \Delta n^{(1)}}{\partial y_{\rm sc}}
\Bigg|^{\rm sc}_{\rm kin}
&\approx 
-\beta^{(1)} \chi\left\{\Theta_{\rm e}^{(0)} \Yspec + \KompO \Delta n^{(0)}_0
-\Theta_{\rm e}^{(0)}\,\boostO Y
+\left[- \KompO \boostO 
+\KompO - \DiffO-2 \DiffO^* G
\right] \Delta n^{(0)}_0\right\}
\nonumber \\
&=-\beta^{(1)} \chi\left\{\Theta_{\rm e}^{(0)} \Yspec + \KompO \Delta n^{(0)}_0
-\Theta_{\rm e}^{(0)}\,\boostO Y
-\boostO\KompO \Delta n^{(0)}_0\right\}
\equiv -\beta^{(1)} \chi \,(1-\boostO)\left\{\Theta_{\rm e}^{(0)} \Yspec + \KompO \Delta n^{(0)}_0\right\}
\nonumber \\
&\approx - \beta^{(1)} \chi \,(1-\boostO)\left\{\Delta \Theta_{\rm C}^{(0)} \Yspec + \KompO \Delta m^{(0)}_0 + \frac{\dot{Q}^{(0)}_{\rm c}}{\kappa} Y \right\}
%
\end{align}
where in the last step we used \eqref{eq:Theta_e_zero}. This is a significant simplification, that allows us to easily include these terms into the frequency hierarchy.

To add kinematic corrections due to double Compton and Bremsstrahlung, we will use the effective description in Eq.~\eqref{eq:effective_em_abs} for the moving frame and then readily obtain
\begin{align}
\label{eq:effective_em_abs_kin}
\frac{\id  
{\Delta n^{(1)}}}{\id y_{\rm sc}}\Bigg|_{\rm em}^{\rm eff, kin}
&\approx -\beta^{(1)} \chi\,(1-\boostO)\left\{\gamma_T\,\xc\,\mu_{0} ^{(0)} G -\gamma_N\,\xc\,\mu_{0} ^{(0)} M\right\}
\end{align}
for the related (approximate) kinematic corrections. This means that in our spectral basis the kinematic corrections to the thermalization terms take the final form
\bealf{
\label{eq:evol_transport_yn_improved_kin}
&\frac{\partial \vek{y}^{(1)}_1}{\partial \eta}\Bigg|_{\rm kin}
=- {\beta^{(1)} \chi} \, (1-M_{\rm B})\left\{\tau'\Thz M_{\rm T}  \vek{y}^{(0)}_0  + \frac{{\vek{Q}'}^{(0)}}{4}\right\}= - {\beta^{(1)} \chi} \,(1-M_{\rm B})\, \frac{\partial \vek{y}^{(0)}_0}{\partial \eta}
}
as new contribution to the dipolar distortion anisotropies. In hindsight, this has the obvious interpretation that the background source has to be boosted into the local frame with $-{\beta^{(1)}\chi\,}(1-M_{\rm B})$.
When the background solution becomes stationary, this term vanishes. Also, the source term is independent of the value of $\Theta^{(0)}_0$, as this does not couple to the thermalization matrix as explained earlier. Again, these kinematic correction terms have to be included to obtain a consistent formulation of the problem when considering temperature corrections. 

\subsection{Changes to the baryon velocity equation}
\label{sec:baryon_velocity}
As already highlighted in \citep{chluba_spectro-spatial_2023-II}, the presence of distortions also affects the momentum exchange between electrons and photons and thus the baryons. Here, we are interested in including all corrections to also allow us to compute the correct CMB temperature corrections, which were previously omitted. This will also capture the aforementioned kinematic corrections. 

Following \citep{chluba_spectro-spatial_2023-II}, the change of the baryon momentum by scattering with photons at first order in perturbations with distortions is 
\bealf{
\label{eq:evol_1_beta}
\frac{\partial \beta^{(1)}}{\partial \eta}\bigg|_{\rm sc}
&\approx - \tau'\,\frac{\rho_z}{\rho_{\rm b}} \,\frac{1}{2} \int \frac{x^3 \chi}{E_{\nbb}}
\left[n^{(1)}_0+\frac{1}{10}\,n^{(1)}_2 - n^{(1)} + \beta^{(1)}\,\chi\,\boostO n^{(0)}_0-\beta^{(1)} \chi\,(1-\boostO) \frac{\partial \Delta n^{(0)}_0}{\partial \tau}\Bigg|_{\rm ex}\right] \id x \id \chi
\nonumber\\
&=\tau'\,\frac{\rho_z}{\rho_{\rm b}} \,\int \frac{\chi^2}{2}  \id \chi \int \frac{x^3 }{E_{\nbb}}
\left[3\tilde{n}_1^{(1)} - \beta^{(1)}\,\boostO n^{(0)}_0+\beta^{(1)} (1-\boostO) \frac{\partial \Delta n^{(0)}_0}{\partial \tau}\Bigg|_{\rm ex}\right] \id x 
\nonumber\\
&=\tau'\,\frac{4\rho_z}{3\rho_{\rm b}} \int \frac{x^3 }{4 E_{\nbb}}
\left[3\tilde{n}_1^{(1)} - \beta^{(1)}\,\boostO n^{(0)}_0+\beta^{(1)} (1-\boostO) \frac{\partial \Delta n^{(0)}_0}{\partial \tau}\Bigg|_{\rm ex}\right] \id x,
}
where we inserted the Legendre polynomial expansion, $n^{(1)}=\sum_\ell (2\ell+1) \,\tilde{n}_\ell^{(1)}P_\ell(\chi)$, and also included the contributions arising from Compton scattering kinematic corrections and corrections to the emission process, with the momentum exchange being mediated by 
\begin{align}
\label{eq:ex}
\frac{\partial \Delta n^{(0)}_0}{\partial \tau}
\Bigg|_{\rm ex}
&\approx \Thz\left[\Theta_{\rm e}^{(0)} \Yspec + \KompO \Delta n^{(0)}_0  +\gamma_T\,\xc\,\mu_{0} ^{(0)} G -\gamma_N\,\xc\,\mu_{0} ^{(0)} M\right]
\end{align}
according to Eq.~\eqref{eq:Main_Eq_lab_lin_final} and Eq.~\eqref{eq:effective_em_abs_kin}. We now use $\tilde{n}_1^{(1)}=\Theta^{(1)}_1 G +\vek{B}(x)\cdot\vek{y}^{(1)}_1$ for the Legendre coefficient of the perturbed photon dipole spectrum, where $\vek{y}^{(1)}_1=(\delta \Theta^{(1)}, y^{(1)}_{0}, y^{(1)}_1, \ldots, y^{(1)}_{15}, \mu^{(1)})^T_1$ is the dipole distortion vector and $\vek{B}(x)=(G, Y_0, Y_1, \ldots, Y_{15}, M)^T$ is the spectral basis vector. 
Note that we cleanly separated the temperature correction, $\delta\Theta^{(1)}=\Theta^{(1), \rm full}-\Theta^{(1)}$, in the distortion vector. Here, $\Theta^{(1), \rm full}$ is the solution of the full problem with distortions included, while $\Theta^{(1)}$ is the solution for the standard temperature hierarchy {which neglects any distortion effects}. 
We then have
\bsub
\bealf{
\int \frac{x^3 }{4 E_{\nbb}}\, \tilde{n}_1^{(1)}\id x 
&= \Theta^{(1)}_1 + \delta \Theta^{(1)}_1 +\sum_{\rm k} y^{(1)}_{k, 1}+\frac{\epsilon_M}{4}\,\mu^{(1)}_1=\Theta^{(1)}_1+\frac{\epsilon^{(1)}_{1}}{4}
\\
\int \frac{x^3 }{4 E_{\nbb}}\, \boostO n^{(0)}_0 \id x 
&= 1 + 4\Theta^{(0)}_0 + \sum_{\rm k} 4y^{(0)}_{k, 0}+\epsilon_M \,\mu^{(0)}_0=1+\epsilon^{(0)}_{0}
\\
\int \frac{x^3 }{4 E_{\nbb}}\, (1-\boostO) \frac{\partial \Delta n^{(0)}}{\partial \tau}\Bigg|_{\rm ex} \id x 
&= - \int \frac{ 3 x^3}{4 E_{\nbb}}\, \frac{\partial \Delta n^{(0)}}{\partial \tau}\Bigg|_{\rm ex} \id x
=-3 \Thz\left[\Theta^{(0)}_{\rm e}- \Theta^{(0)}_{\rm C}\right]
\approx  -\frac{3}{4} \frac{{Q'_{\rm c}}^{(0)}}{\tau' \rho_z}
}
\esub
with $\epsilon_f=\int x^3 f \id x/E_{\nbb}$ and thus $\epsilon_M=1/1.4007$. In the last line, we used Eq.~\eqref{eq:evol_Dn_a_mod} and Eq.~\eqref{eq:Teq_qs}, and also that by construction the emission and absorption process does not change the photon energy density, as we neglected the small energy exchange with the electrons. 

We can now write the final equation for the baryon velocity including the effects of distortions. However, in Eq.~\eqref{eq:evol_1_beta} we {\it implicitly} used the full solution at first order in perturbation theory, $\beta^{(1), \rm full}$. We now also more carefully split the velocity variable, $\beta^{(1), \rm full}=\beta^{(1)}+\delta \beta^{(1)}$, into the part without distortions, $\beta^{(1)}$, and the correction, $\delta\beta^{(1)}$, due to the presence of distortions. Putting everything together, we then finally have the two evolution equations
\bealf{
\label{eq:evol_1_beta_final}
\frac{\partial \beta^{(1)}}{\partial \eta}\bigg|_{\rm sc}
&\approx 
\frac{3\tau'}{R} 
\left[
\Theta^{(1)}_1-\frac{\beta^{(1)}}{3}\right],
\quad
\frac{\partial \delta \beta^{(1)}}{\partial \eta}\bigg|_{\rm sc}
\approx 
\frac{3\tau'}{R} 
\left[\frac{\epsilon^{(1)}_{1}}{4}-\frac{\epsilon^{(0)}_{0}\,\beta^{(1)}}{3}-\frac{\delta \beta^{(1)}}{3}
-\frac{\beta^{(1)}}{\tau'}\,\frac{{\mathcal{Q}'}^{(0)}}{4}\right],
}
where $R=\frac{3\rho_{\rm b}}{4\rho_z}$ describes the baryon loading and ${\mathcal{Q}'}^{(0)}={Q'_{\rm c}}^{(0)}/\rho_z$. The terms $\propto \epsilon^{(1)}_{1}$ and $\propto \epsilon^{(0)}_{0}$ were also given in \citep{chluba_spectro-spatial_2023-II} and describe changes to the photon energy and momentum densities. The new term $\propto {\mathcal{Q}'}^{(0)}$ appears from the sourcing of distortions due to the heating of the baryons. 

There is {another} subtle aspect. If we chose $\Tz=\TCMB$, but initially start with {a blackbody field at a temperature} $\TCMB'=\TCMB(1+\bar{\Theta})$, this means that effectively $\Tz=\TCMB'/(1+\bar{\Theta})\approx \TCMB'(1-\bar{\Theta})$. Because $R=\frac{3\rho_{\rm b}}{4\rho_z}\approx \frac{3\rho_{\rm b}}{4\rho_{\rm CMB}}(1+4\bar\Theta)$, an additional modification to the baryon velocity equation appears that is captured by
\label{eq:evol_1_beta_final_slip}
\bealf{
\frac{\partial \delta \beta^{(1)}}{\partial \eta}\bigg|^{*}_{\rm sc}
\approx 
-4\bar{\Theta}\,\frac{3\tau'}{R_{\rm CMB}} 
\left[
\Theta^{(1)}_1-\frac{\beta^{(1)}}{3}\right]
+
\frac{3\tau'}{R_{\rm CMB}} 
\left[\frac{\epsilon^{(1)}_{1}}{4}-\frac{\epsilon^{(0)}_{0}\,\beta^{(1)}}{3}-\frac{\delta \beta^{(1)}}{3}
-\frac{\beta^{(1)}}{\tau'}\,\frac{{\mathcal{Q}'}^{(0)}}{4}\right]
}
with $R_{\rm CMB}=\frac{3\rho_{\rm b}}{4\rho_{\rm CMB}}$. This correction has to be added when starting with modified initial temperature. However, in most calculations we will use $\Tz=\TCMB$ (i.e., $\Thetaref=0$) and start with a blackbody at this temperature (or $\bar\Theta=0$), so that this term does not appear.

\section{Photon source terms}
\label{sec:Photon sources}
As an extension of the formulation of \citep{chluba_spectro-spatial_2023-II}, we now include photon sources. We shall only consider cases for which the source function can be easily expressed in terms of the basis functions and is independent of the distortions itself. For example, this will preclude us from modeling scenarios with narrow line emission processes \citep[e.g.,][]{Chluba2015GreensII, Bolliet2020dp} or where the spectrum itself affects the emission processes. However, processes with photon conversion \citep{Chluba2024DP, Cyr2024Axions} can be captured this way. We will assume that the source term is isotropic in the restframe of the particles, meaning we can write
\bealf{
\label{eq:sources}
\frac{\partial S_0(x, \eta)}{\partial \eta}
&\approx 
G(x)\,\partial_\eta S_G (\eta)+Y(x)\,\partial_\eta S_Y (\eta)+\ldots +M(x)\,\partial_\eta S_M (\eta) \rightarrow \vek{S}'_0(x, \eta),
}
where $\vek{S}'_0(x, \eta)$ is the component source term vector (with prime denoting conformal time derivative {as before}).
The physical picture is that the sources are due to some local collision of particles. As such, we have to take into account corrections from the transformation into the local inertia frame, i.e., those from $\Psi$ and the source velocity, $\beta_S$. This then means that we have to add the terms
\bealf{
\label{eq:sources_orders_monopole}
\vek{S}'^{(0)}(x, \eta), \qquad {\rm and} \qquad  
{\vek{S}'}^{(1)}_0(x, \eta)+\Psi^{(1)}\,\vek{S}'^{(0)}(x, \eta)
}
to the monopole equations at background and perturbed order, and
\bealf{
\label{eq:sources_orders_dipole}
\frac{\partial \vek{y}^{(1)}_1}{\partial \eta}\Bigg|_{\rm S}
=-{\beta_S^{(1)} \chi_S}\,(1-M_{\rm B})\,\vek{S}'^{(0)}(x, \eta)
}
to the dipole equations to account for kinematic corrections. The boost matrix appears from the transformation of the source spectra while the other term arises from kinematic corrections of the time-variable, similar to how this appeared for the Compton scattering corrections. We note here that if the emitting particle is moving with the baryons, we have $\beta_S^{(1)}=\beta_{\rm b}^{(1)}$, which we will assume below. Alternatively, we could include the effect of emission by particles comoving with the dark matter, i.e., $\beta_S^{(1)}=\beta_{\rm cdm}^{(1)}$, or not moving at all, $\beta_S^{(1)}=0$, depending on the scenario; however, we shall not consider these options in the present work. We also note that ${\vek{S}'}^{(1)}_0(x, \eta)$ generally also depends on the (matter) density fluctuations, such that the relevant perturbed variables also have to be considered.

\subsection{Correction to the velocity equation of the source particles}
\label{sec:corr_v_source}
Like for the effect of scattering on the baryon velocity, {we} obtain a correction term 
\bealf{
\label{eq:evol_1_beta_S}
\frac{\partial \beta^{(1)}_S}{\partial \eta}\bigg|_{\rm S}
&\approx 
-\frac{3\beta^{(1)}_S}{R_S}\,\frac{\vek{b}_{\rm \epsilon} \cdot \vek{S}'^{(0)}(\eta)}{4}
=-\frac{3\tau'}{R_S}\,\frac{\beta^{(1)}_S}{\tau'}\,\frac{{\mathcal{S}'}^{(0)}}{4}
}
to the velocity equation of the source with ${\mathcal{S}'}^{(0)}=\vek{b}_{\rm \epsilon} \cdot \vek{S}'^{(0)}(\eta)$ describing the effective relative energy injection of the source. Here $R_S=\frac{3\rho_S}{4\rho_z}$ is the photon loading due to the source particle. We expect this term to usually be negligible. Below we will only consider baryons as the source.

\section{Checking the consistency of the formulation}
\label{sec:consistency}
We have already shown that the new form for the thermalization terms drives the solution towards the correct equilibrium (see discussion around Eq.~\eqref{eq:evol_Dn_eq_background} and also Sect.~\ref{sec:check_T_ref} for our discretized version). To setup problems, {we also} need the correct initial conditions. For applications, we may want to start with a different initial temperature $\TCMB'=\TCMB(1+\bar{\Theta})$ and various choices of the reference temperature $\Tz=\TCMB(1+\Thetaref)$. In the absence of any energy injection this should not alter the final result for the CMB power spectra, as we have already explained in Sect.~\ref{sec:eq_solutions_general}.

\subsection{Initial condition for the baryon velocity with modified blackbody temperature}
\label{sec:v_b_initial}
To explore this point, we start by asking how the initial condition for the baryon velocity changes when we start with a modified CMB average temperature $\TCMB'=\TCMB(1+\bar{\Theta})$, while neglecting any energy or photon injection, {i.e., assume that there are no real spectral distortions. When choosing the reference temperature for the computation as} $\Tz=\TCMB'$, nothing changes with the initial conditions and we can carry our computations out as usual (i.e., no distortion terms), with the only difference that our final power spectra are scaled by $(\TCMB'/\TCMB)^2=(1+\bar{\Theta})^2$ relative to the standard case with $\bar{\Theta}=0$ (see Sect.~\ref{sec:change_TCMB_initial}).  
What if we instead want to use $\Tz=\TCMB$ in our calculation? The final result should {\it not} change. We know that in this case we have $\Theta^{(1), \rm full}=(1+3\bar{\Theta})\Theta^{(1)}$ and $y^{(1)}_{0}= \bar{\Theta}\Theta^{(1)}$ initially. For adiabatic initial conditions we also have $\Theta^{(1)}_1=\beta^{(1)}/3$ for the standard variables independent of the choice for $\Tz$. What do we get for {$\delta \Theta^{(1)}_1$ and} $\delta\beta^{(1)}$? We have $\delta \Theta^{(1)}_1=3\bar{\Theta}\Theta^{(1)}_1$ and  $y^{(1)}_{0,1}= \bar{\Theta}\Theta^{(1)}_1$, implying $\epsilon^{(1)}_{1}/4=3\bar{\Theta}\Theta^{(1)}_1+\bar{\Theta}\Theta^{(1)}_1=4\bar{\Theta}\Theta^{(1)}_1$. Similarly, we have $\epsilon^{(0)}_{0}=4 \bar{\Theta}$, such that initially
\label{eq:del_beta_initial}
\bealf{
\frac{\epsilon^{(1)}_{1}}{4}-\frac{\epsilon^{(0)}_{0}\,\beta^{(1)}}{3} = 0 \qquad \rightarrow \qquad  \delta\beta^{(1)}=0
}
{from Eq.~\eqref{eq:evol_1_beta_final}.}
Does this make sense? Indeed this seems to be correct, since the velocity field should not be affected by our choice of $\Tz$ in our computation; only our spectral decomposition depends on $\Tz$. Assuming that our system consistently conserves the spectral decomposition, we will have the photon distribution in terms of {$\Theta^{(1), \rm full}_\ell=(1+3\bar{\Theta})\Theta^{(1)}_\ell$ and $y^{(1)}_{0, \ell}=\bar{\Theta}\Theta^{(1)}_\ell$} throughout the computation, in the end giving
\label{eq:Dn_change_Tz}
\bealf{
\Delta n^{(1)} = (1+3\bar{\Theta})\Theta^{(1)}\,G(x)+ \bar{\Theta}\Theta^{(1)}_{0}\,Y(x)\approx  \nbb \left(\frac{x}{(1+\bar\Theta)(1+\Theta^{(1)})}\right)
}
as the correct final result.
In a similar manner, all the other standard perturbation variables remain unaffected by our choice of reference temperature in the spectral decomposition, as we will see below.

\subsection{Consistently solving for the corrections to the temperature perturbations}
\label{sec:num_setup_Theta_corr}
In this paper, we are interested in also obtaining the corrections to the temperature equations caused by thermalization effects and distortion physics. Our system [e.g., in Eq.~\eqref{eq:evol_transport_yn}] can be further modified to {\it separately} capture the corrections caused by thermalization effects to the CMB temperature perturbations. For this, we simply use the standard {Boltzmann hierarchy, Eq.~\eqref{eq:evol_transport_Theta} complemented by those for the other perturbation variables \citep[e.g.,][]{Ma1995}}, to obtain $\Theta^{(1)}$ {(i.e., without distortion effects)} and then only consider the correction, $\delta\Theta^{(1)}=\Theta^{(1), \rm full}-\Theta^{(1)}$ in the variable $\vek{y}^{(1)}_0$, i.e., $\vek{y}^{(1)}_0 = \left(\delta \Theta^{(1)}, y^{(1)}_{0}, y^{(1)}_{1}, \ldots, y^{(1)}_{15}, \mu^{(1)}\right)_0^T$. In addition, we have to replace $\vek{b}^{(0)}_0\rightarrow M_{\rm B} \vek{y}^{(0)}_0$ in all the equations for $\vek{y}^{(0)}_0$ {to avoid double counting contributions}. 

However, as we have seen in Eq.~\eqref{eq:evol_1_beta_final}, the presence of average distortions causes corrections to the baryon velocity. In a similar way, all other perturbation variables ($\Psi, \Phi, \delta_{\rm b}$, etc.) {are} slightly modified. {We shall denote the corrections with an additional '$\delta$' in front of the perturbation variable, e.g., $\Phi^{\rm full}=\Phi^{(1)}+\delta \Phi$, $\Psi^{\rm full}=\Psi^{(1)}+\delta \Psi$, $\delta_{\rm m}^{\rm full}=\delta_{\rm m}^{(1)}+\delta\delta_{\rm m}$, etc. Aside from the modification to the baryon velocity equation, the only additional equations} that are {\it directly} affected by the distortions are those for the Newtonian potentials. These depend on the matter {and} radiation (i.e., photons and neutrinos) energy density perturbations, $\delta \rho_{\rm m}$ and $\delta \rho_{\rm r, \ell}$, respectively \citep{Ma1995}
\begin{align}
\label{eq:Potentials}
k^2 \Phi + 3 \mathcal{H}(\partial_\eta \Phi-\mathcal{H} \Psi)&=4 \pi G a^2 [\delta \rho_{\rm m}+\delta \rho_{\rm r, 0}],\qquad
k^2 (\Phi+\Psi)=-8 \pi G a^2 \delta \rho_{\rm r, 2},
\end{align}
where $\mathcal{H}=a H = a^{-1} \partial_\eta a$ is the conformal time Hubble parameter. In these equations, the chosen reference temperature again has to be taken into account. Assuming that the CMB energy density initially is $\rho_{\rm CMB}$ with $T=\TCMB$, but that we chose $\Tz =\TCMB(1+\Thetaref)$, one has the photon contribution
\begin{align}
\label{eq:Potentials_energy_density}
\delta \rho^{(1)}_{\gamma, 
\ell}
=4\rho_z \left[\Theta^{(1)}_\ell+\frac{\epsilon^{(1)}_\ell}{4}\right]
=4\rho_{\rm CMB}(1+\Thetaref)^4
\left[\Theta^{(1)}_\ell+\frac{\epsilon^{(1)}_\ell}{4}\right]\approx 
4\rho_{\rm CMB}\left\{
\Theta^{(1)}_\ell
+
\left[\frac{\epsilon^{(1)}_\ell}{4}+4\Thetaref\Theta^{(1)}_\ell\right]
\right\}
\end{align}
noting that $\epsilon^{(1)}_\ell$ just depends on the distortion parameters and $\delta \Theta^{(1)}_\ell$. From Eq.~\eqref{eq:equilibrium_solution}, we then have $\epsilon^{(1)}_\ell/4=-4\Thetaref\Theta^{(1)}_\ell$, which cancels the term from the shift in the reference temperature. This means that also for the potentials initially one has $\delta \Phi=\delta\Psi=0$; however, as in Eq.~\eqref{eq:Potentials_energy_density}, the term $16\rho_{\rm CMB}\,\Thetaref\Theta^{(1)}_\ell$ has to be added {when setting up the corresponding evolution equations}.

On the other hand, if the initial energy density of the CMB is $\TCMB'=\TCMB(1+\bar\Theta)$ but we use $\Tz=\TCMB=\TCMB'/(1+\bar\Theta)$ {to formulate the problem}, then this case is equivalent to the above, but with $\Thetaref=-\bar\Theta$, again leaving the potential initial conditions unaffected.
In summary, one can apply Eq.~\eqref{eq:Potentials} for both $\Phi^{(1)}, \Psi^{(1)}$ and $\delta\Phi^{(1)}, \delta\Psi^{(1)}$, where for the latter two one has to use
\begin{align}
\delta\delta \rho^{(1)}_{\gamma, \ell}=4\rho_{\rm CMB}
\left[\frac{\epsilon^{(1)}_\ell}{4}+4(\Thetaref-\Thetabar)\Theta^{(1)}_\ell\right]
\end{align}
in the evaluation of the right hand sides.
Similarly, the modified baryon velocity, temperature monopole and dipole equations explicitly read
\bsub
\begin{align}
\label{eq:vb_and_Theta_0_1_equations}
\partial_\eta \delta \beta^{(1)}&= \kB\delta \Psi^{(1)}-\mathcal{H} \delta \beta^{(1)}
+\frac{3 \tau'}{R}\left[\delta\Theta^{(1)}_1-\frac{\delta\beta^{(1)}}{3}\right]
\\
&\qquad \qquad 
+4(\Thetaref-\Thetabar)\frac{3 \tau'}{R}\left[\Theta^{(1)}_1-\frac{\beta^{(1)}}{3}\right]
+\frac{3 \tau'}{R}\left[\frac{\epsilon^{(1)}_{1}}{4}
-\frac{\epsilon^{(0)}_{0}\beta^{(1)}}{3}
-\delta\Theta^{(1)}_1
-\frac{\beta^{(1)}}{4\tau'}\,\left({\mathcal{Q}'}^{(0)}
+{\mathcal{S}'}^{(0)}\right)\right]
\nonumber\\
\partial_\eta \delta \Theta^{(1)}_0&=-k \delta \Theta^{(1)}_1-\partial_\eta\delta \Phi^{(1)}-3\Theta^{(0)}_0 \partial_\eta \Phi^{(1)}
\nonumber\\[2mm]
&\qquad\qquad +
\tau' \Thz \,\gamma_T \xc\left\{\mu_{0} ^{(1)} + \left(\delta^{(1)}_{\rm b}+\Psi^{(1)}+\Theta^{(1)}_0
\right) \mu_{0} ^{(0)} \right\} + {\Theta'}^{(1)}_{S,0}+\Psi^{(1)} {\Theta'}^{(0)}_{S,0},
\\[2mm] \nonumber
\partial_\eta \delta \Theta^{(1)}_1&=\frac{k}{3}\delta \Theta^{(1)}_0-\frac{2 k}{3}\delta \Theta^{(1)}_2+\frac{k}{3}\delta \Psi^{(1)}
-\tau'\left[\delta \Theta^{(1)}_1-\frac{\delta \beta^{(1)}}{3}\right]
\nonumber\\
&\qquad \qquad 
+k \Theta^{(0)}_0 \Psi^{(1)}+\tau'\Theta^{(0)}_0\beta^{(1)}
+\frac{2\beta^{(1)}}{3} \left[\tau' \Thz \gamma_T \xc \mu_{0} ^{(0)}+{\Theta'}^{(0)}_{S,0}\right],
\end{align}
\esub
where ${\Theta'}^{(i)}_{S,0}$ is the temperature source term (should it be present), which here we assume to be comoving with the baryons.
For each equation, we separated the new terms {in} the second line of the equation.
Here, we can see that the only direct connection to the thermalization effects comes from the conversion of $\mu$ to $\Theta$ by photon emission and absorption processes, the terms $\propto \xc$, which are only effective at $z\gtrsim \pot{3}{5}$ \citep[e.g.,][]{Burigana1991, Hu1993}. This causes a source in the monopole temperature equation and kinematic correction to the dipole term, where we used $-(1-\boostO) G=2 G$. For the remaining distortion equations, we can then use Eq.~\eqref{eq:evol_transport_yn} with Eq.~\eqref{eq:evol_transport_yn_improved}, Eq.~\eqref{eq:evol_transport_yn_improved_kin} and the photon sources Eq.~\eqref{eq:sources_orders_monopole} and Eq.~\eqref{eq:sources_orders_dipole}. 
This new approach is extremely helpful for small distortion scenarios, as otherwise the temperature corrections become numerically difficult to separate. This setup is now implemented in {\tt CosmoTherm} \citep{Chluba2011therm} and will be used below.

\section{Summary of the improved frequency hierarchy equations} 
\label{sec:FH_eq}
We are now in the position to write down the final frequency hierarchy (FH) for the $\vek{y}^{(i)}$. For convenience, here we give a summary of all the required equations and shall also link to the line-of-sight approach for the computation of the signal power spectra.

Collecting all terms from electron scattering and emission processes, Eq.~\eqref{eq:evol_transport_yn} and Eq.~\eqref{eq:evol_transport_yn_improved}, kinematic corrections, Eq.~\eqref{eq:evol_transport_yn_improved_kin}, and external photon source terms, Eq.~\eqref{eq:sources_orders_monopole} and Eq.~\eqref{eq:sources_orders_dipole}, then gives the system
\bsub
\label{eq:evol_transport_yn_final}
\bealf{
\label{eq:evol_transport_yn_final_bg}
&\frac{\partial \vek{y}^{(0)}_0}{\partial \eta}
=\tau'\Thz M_{\rm T} \,\vek{y}^{(0)}_0 + \frac{{\vek{Q}'}^{(0)}}{4} + \vek{S}'^{(0)}
\\
&{\vek{Q}'}^{(0)}= \left(0,\frac{{Q'_{\rm c}}^{(0)}}{\rho_z},\ldots, 0\right)^T,
\\
&\vek{S}'^{(0)} = \left({S'_G}^{(0)},{S'_Y}^{(0)},\ldots, {S'_{Y_N}}^{(0)}, {S'_M}^{(0)}\right)^T,
\\[4mm]
&\frac{\partial \vek{y}^{(1)}}{\partial \eta}+\vgh\cdot \nabla \vek{y}^{(1)}+
\vek{b}^{(0)}_0 \left(\frac{\partial \Phi^{(1)}}{\partial \eta}+ \vgh \cdot \nabla\Psi^{(1)} \right)= 
{\rm C}^{(1)}_{\rm T}[\vek{y}]+{\rm C}^{(1)}_{\rm all}[\vek{y}]
\\
&{\rm C}^{(1)}_{\rm T}[\vek{y}]=\tau'\left[\vek{y}^{(1)}_0+\frac{1}{10}\,\vek{y}^{(1)}_2 - \vek{y}^{(1)} + \vek{b}^{(0)}_0 \beta^{(1)}\chi\,\right], 
\\[2mm]
&{\rm C}^{(1)}_{\rm all}[\vek{y}]
= \tau' \Thz \left\{
M_{\rm T} \vek{y}^{(1)}_0  + \left[(\delta^{(1)}_{\rm b} +\Theta^{(1)}_0+ \Psi^{(1)})\, M_{\rm T} + \Theta^{(1)}_0 M_{\rm BK}\right]\vek{y}^{(0)}_0
- \beta^{(1)} \chi \,(1-M_{\rm B}) M_{\rm T} \,\vek{y}^{(0)}_0\right\}
\nonumber\\
&\qquad\qquad
+ \frac{{\vek{Q}'}^{(1)}}{4} + \Theta^{(1)}_0\,\vek{C}^{(0)}_0- \beta^{(1)} \chi \,(1-M_{\rm B}) \,\frac{{\vek{Q}'}^{(0)}}{4}
+ {\vek{S}'}^{(1)} - \beta^{(1)}_S \chi_S (1-M_{\rm B})\, \vek{S}'^{(0)}
\\
&\vek{b}^{(0)}_0=\boostO n^{(0)}= \vek{e}_G + M_{\rm B} \vek{y}^{(0)}_0,
\qquad 
\vek{C}^{(0)}_0 = \frac{{Q'_{\rm c}}^{(0)}}{ \rho_z}\left(0,-1, 1, 0,\ldots, 0\right)^T,
\\
&{\vek{Q}'}^{(1)}= \left(0,\frac{{Q'_{\rm c}}^{(1)}}{\rho_z}+\Psi^{(1)}\frac{{Q'_{\rm c}}^{(0)}}{\rho_z},\ldots, 0\right)^T, 
\\
&{\vek{S}'}^{(1)} 
= \left({S'_G}^{(1)},{S'_Y}^{(1)},\ldots, {S'_{Y_N}}^{(1)}, {S'_M}^{(1)}\right)^T+\Psi^{(1)}\,\vek{S}'^{(0)}.
}
\esub
We mention here that the photon source term $\vek{S}'^{(1)}$ can in principle have an additional angular dependence. If this was indeed the case at the level of the collision process, it means that we would also have to add light aberration effects \citep{Challinor2002, Chluba2011ab, Dai2014}. However, here we shall not consider this case and assume that the source term is isotropic in the rest frame of the moving source. 
We also stress that to isolate the corrections to the temperature variables one has to consider the procedure in Sect.~\ref{sec:num_setup_Theta_corr}, with modified equations for $\delta \Phi, \delta \Psi, \delta \beta, \delta \Theta_0$ and $\delta \Theta_1$.

The next step is to go to Fourier space and perform a Legendre expansion of the first order system for $\vek{y}^{(1)}(\eta, \chi, \vek{r})$, assuming that we are dealing only with scalar perturbations (i.e., only $m=0$ matters in terms of the angular dependence). For the Legendre expansion in Fourier space, we use the common convention 
\bealf{
X(\eta, \chi, k)=\sum_\ell (2\ell+1)\,(-\i)^\ell X_\ell(\eta, k)\,P_\ell(\chi),
\qquad
X_\ell(\eta, k)=\int \frac{\i^\ell P_\ell(\chi)\, X(\eta, \chi, k) \id \chi}{2}
}
where $P_\ell(\chi)$ are the Legendre polynomials. We note that we do not explicitly distinguish Fourier space variables and assume that it is clear from the context (i.e., when $k$ is part of the argument).

Carrying out all the transformations for the perturbed variables \citep[e.g., see][for some additional intermediate steps]{chluba_spectro-spatial_2023-II}, we then obtain the final FH:
\bsub
\label{eq:evol_1_final_hierarchy}
\bealf{
\label{eq:evol_1_final_hierarchy_0}
\frac{\partial \vek{y}^{(1)}_0}{\partial \eta}
&=-\kB\,\vek{y}^{(1)}_1\!-\!
\frac{\partial \Phi^{(1)}}{\partial \eta}
\vek{b}^{(0)}_0
+\frac{{\vek{Q}'^{(1)}}}{4}+ \Theta^{(1)}_0\,\vek{C}^{(0)}_0+{\vek{S}'^{(1)}_0}
\\
\nonumber
&\qquad
+\tau' \Thz \Bigg\{M_{\rm T} \vek{y}^{(1)}_0+\left[(\delta^{(1)}_{\rm b} +\Theta^{(1)}_0+ \Psi^{(1)})\, M_{\rm T} +\Theta^{(1)}_0 M_{\rm BK}\right]\vek{y}^{(0)}_0 \Bigg\},
\\
\label{eq:evol_1_final_hierarchy_1}
\frac{\partial \vek{y}^{(1)}_1}{\partial \eta}
&=k \,
\left(\frac{1}{3}
\vek{y}^{(1)}_{0}-\frac{2}{3}
\vek{y}^{(1)}_{2}\right)
+\frac{k}{3}\Psi^{(1)} \,\vek{b}^{(0)}_0
-\tau'\left[
\vek{y}^{(1)}_1-\frac{\beta^{(1)}}{3}\vek{b}^{(0)}_0\right] + {\vek{S}'^{(1)}_1}
\\
\nonumber
&\qquad
- \frac{\beta^{(1)}}{3}\,(1-M_{\rm B})\, \left[\tau'\Thz M_{\rm T} \vek{y}^{(0)}_0 + \frac{{\vek{Q}'}^{(0)}}{4} + \vek{S}'^{(0)}_0 \right] ,
\\
\frac{\partial \vek{y}^{(1)}_2}{\partial \eta}
&=
k \,
\left(\frac{2}{5}\vek{y}^{(1)}_1-\frac{3}{5}
\vek{y}^{(1)}_3
\right)
-\frac{9}{10}\,\tau'\,\vek{y}^{(1)}_2 +{\vek{S}'^{(1)}_2},
\\
\frac{\partial \vek{y}^{(1)}_{\ell\geq 3}}{\partial \eta}
&=
k \,
\left(\frac{\ell}{2\ell+1}
\vek{y}^{(1)}_{\ell-1}-
\frac{\ell+1}{2\ell+1}
\vek{y}^{(1)}_{\ell+1}\right)
-\tau'\vek{y}^{(1)}_\ell + {\vek{S}'^{(1)}_\ell}.
}
\esub
We note that here we redefined $\i \beta\rightarrow \beta$ as part of the Fourier and Legendre transformations.
We also generalized the photon source term $\vek{S}'_{\ell}$, allowing it to contribute to all multipoles, a generalization that for instance appears for photon to dark photon conversion scenarios \citep{Evangelista2026DP}. This system can then be solved using {\tt CosmoTherm}.

\subsection{Line of sight solution} \label{sec:line-of-sight}
Using the solutions to FH problem, Eq.~\eqref{eq:evol_1_final_hierarchy}, to obtain the transfer functions for various observables, we can then compute the power spectra and cross-power spectra \citep[see][for explicit expression]{kite_spectro-spatial_2023-III}. In \cite{chluba_spectro-spatial_2023-II}, the well-known line of sight approach \citep{CMBFAST} was generalized to include the effects of spectral distortions. 
This gives the integral solutions at the final conformal time, $\eta_f$, as
\bsub
\label{eq:formal_sol_Leg_fin}
\begin{align}
\vek{y}^{(1)}_\ell(\eta_f, k) &= \int_0^{\eta_f} \id \eta \,g(\eta)\, \left[\mathcal{\vek{S}}^{\rm T}_\ell(\eta, \eta_f, k)+\mathcal{\vek{S}}^{\rm th}_\ell(\eta, \eta_f, k)
+\mathcal{\vek{S}}^{\rm ex}_\ell(\eta, \eta_f, k)\right]
\\
\mathcal{\vek{S}}^{\rm T}_\ell(\eta, \eta_f, k)&=
\left[
 \vek{y}_0^{(1)}+\Psi^{(1)} \vek{b}^{(0)}_0
 +
 \left(\frac{\partial \Psi^{(1)}}{\partial \eta}
 -\frac{\partial \Phi^{(1)}}{\partial \eta}
 \right)\frac{\vek{b}^{(0)}_0}{\tau'}
 \right]\,j_\ell
+\beta^{(1)} \vek{b}^{(0)}_0\,j^{(1,0)}_\ell 
+ \frac{\vek{y}_2^{(1)}}{2} \,j^{(2,0)}_\ell
\\
\mathcal{\vek{S}}^{\rm th}_\ell(\eta, \eta_f, k)&=
\Thz \Bigg\{M_{\rm T} \vek{y}^{(1)}_0+\left[(\delta^{(1)}_{\rm b} +\Theta^{(1)}_0+ \Psi^{(1)})\, M_{\rm T} +\Theta^{(1)}_0 M_{\rm BK}+ \Psi^{(1)} M_{\rm B} M_{\rm T}\right]\vek{y}^{(0)}_0 \Bigg\} j_\ell
\nonumber\\
&\qquad\qquad
-\Thz \,\beta^{(1)} (1-M_{\rm B})\, M_{\rm T} \vek{y}^{(0)}_0  \,j^{(1,0)}_\ell
\\[2mm]
\mathcal{\vek{S}}^{\rm ex}_\ell(\eta, \eta_f, k)&=
\left[\Psi^{(1)}M_{\rm B}\left\{\frac{{\vek{Q}'^{(0)}}}{4\tau'}+\frac{\vek{S}'^{(0)}_0}{\tau'}\right\}+\frac{{\vek{Q}'^{(1)}}}{4\tau'}+\Theta^{(1)}_0\,\frac{\vek{C}^{(0)}_0}{\tau'}+\frac{\vek{S}'^{(1)}_0}{\tau'}\right]
j_\ell
\\ \nonumber
&\qquad +\Bigg[\frac{3 \vek{S}'^{(1)}_1}{\tau'} - \beta^{(1)} (1-M_{\rm B})\, \left\{\frac{{\vek{Q}'}^{(0)}}{4\tau'} + \frac{\vek{S}'^{(0)}_0}{\tau'} \right\}\Bigg]\,j^{(1,0)}_\ell
+\frac{5 \vek{S}'^{(1)}_2}{\tau'}\,j^{(2,0)}_\ell
\end{align}
\esub
Here, $g(\eta)=\tau'\,\expf{-\tau_{\rm b}}=\partial_\eta \expf{-\tau_{\rm b}}$ is the Thomson visibility function and where $\tau_{\rm b}=\tau(\eta_f)-\tau(\eta)$ is the Thomson optical depth between $\eta$ and $\eta_f$. The sources were split into those from Thomson scattering, $\mathcal{\vek{S}}^{\rm T}_\ell$, thermalization effects, $\mathcal{\vek{S}}^{\rm th}_\ell$, and external sources of heat and photons, $\mathcal{\vek{S}}^{\rm ex}_\ell$, assuming that the latter become negligible at $\ell>2$. 
The $j^{(a,b)}_\ell(x)$ are defined as in \citep{Hu1997, chluba_spectro-spatial_2023-II} and all have the argument $x=\kB\Delta\eta$ with $\Delta \eta=\eta_f-\eta$.
We also added previously neglected terms relating to the time-dependence of $\vek{b}^{(0)}_0$ [see Appendix~\eqref{app:new_term}] that may become important when external sources of heat or photons are active around or after last scattering as we explain in Appendix~\ref{eq:add_LOS_term}.

We note again that our treatment does not include all effects from polarization in the photon field. For this we would also need to consider the thermalization of polarized distortion anisotropies \citep{Pitrou2009}, which we leave for future work. We also note that for a consistent treatment of the corrections to the photon temperature field one would further need to consider the Boltzmann hierarchy of the photon sources. This becomes apparent when for example thinking of photon to dark photon conversion processes, which implies that perturbations from the photon fluid source perturbations in the dark photon fluid. However, these aspects have to be considered for each scenario separately.

\section{Illustrations for various scenarios}
\label{sec:illustrations}
In this section, we consider a few examples to highlight the properties of the improved set of equations. A more extensive application to various distortion scenarios will be left to the future, while here we focus on conceptually new aspects and confirmation of the treatment. To compute transfer functions and signal power spectra we follow the methods described in \citep{chluba_spectro-spatial_2023-II} and \citep{kite_spectro-spatial_2023-III}, with the {main change being in} the new treatment of the thermalization effects in the photon transport. We shall assume $\Thetaref=0$ unless stated otherwise. We will also use the {FH} setup including the basis up to $Y_{15}$. 
The transfer function examples will furthermore all be given in the {\it scattering basis}, which ensures orthogonality to the temperature perturbations, and thus models the transition between $\mu$ and $y$ more smoothly. Power spectra will be presented in the {\it observational basis}, which is closest to what one would obtain when analyzing data using a $T$, $y$ and $\mu$ decomposition \citep[see][for more details]{chluba_spectro-spatial_2023-I}.

\subsection{Basic numerical tests}
\label{sec:numerical_tests}
Before we apply the new treatment to specific scenarios, we numerically confirmed a few general aspects using {\tt CosmoTherm}. Adding a temperature shift either as initial condition or by continuously sourcing $G$-terms does not create any background level distortions. Similarly, starting\footnote{We initialize our anisotropic setup with adiabatic initial conditions at redshift $z=10^7$, which means that modes with $\kB\leq 1\,\Mpc^{-1}$ are {\it well} outside of the horizon.} the anisotropic distortion with $\delta\Theta^{(1)}_\ell \approx 3 \bar{\Theta} \Theta^{(1)}_\ell$ and $y^{(1)}_{0,\ell} \approx \bar{\Theta} \Theta^{(1)}_\ell$ and then leaving $\Thetabar$ unchanged {(i.e., starting with a higher temperature blackbody spectrum, $\TCMB'=\TCMB(1+\bar{\Theta})$ while using the reference temperature $\Tz=\TCMB$)} conserves the solution throughout the computation. For example, starting with $\Thetabar=0.01$ (a very large modification), we find tiny distortion creation at the level of $\mu \lesssim 10^{-10}$, which are caused by the imperfect discretization of the Kompaneets operator, causing some small $Y\rightarrow M$ conversion. On the other hand, when switching off the perturbed scattering terms [e.g., specifically those {$\propto \tau' \Thz$}] or kinematic corrections {[those $\propto \beta^{(1)} (1-M_{\rm B})$] in Eq.~\eqref{eq:evol_1_final_hierarchy}}, we find noticeable departures from the expected equilibrium, showing that the new formulation is required to consistently conserve the spectrum of the anisotropies in the absence of out-of-equilibrium processes. 

\subsection{Single change in the background temperature}
\label{sec:Tz-changes-delta}
As another illustrative example, we consider a single change in the background temperature at some redshift $z_{\rm s}$. Comparing to the situation when we start with a non-zero temperature shift in the initial condition, this allows us to ask how long it takes for modes to equilibrate. At $z_{\rm s}\gtrsim 10^4$, we also expect some level of spectral evolution towards this equilibrium {due to} thermalization terms. 

\begin{figure}[H]
    \centering
\includegraphics[width=0.46\columnwidth]{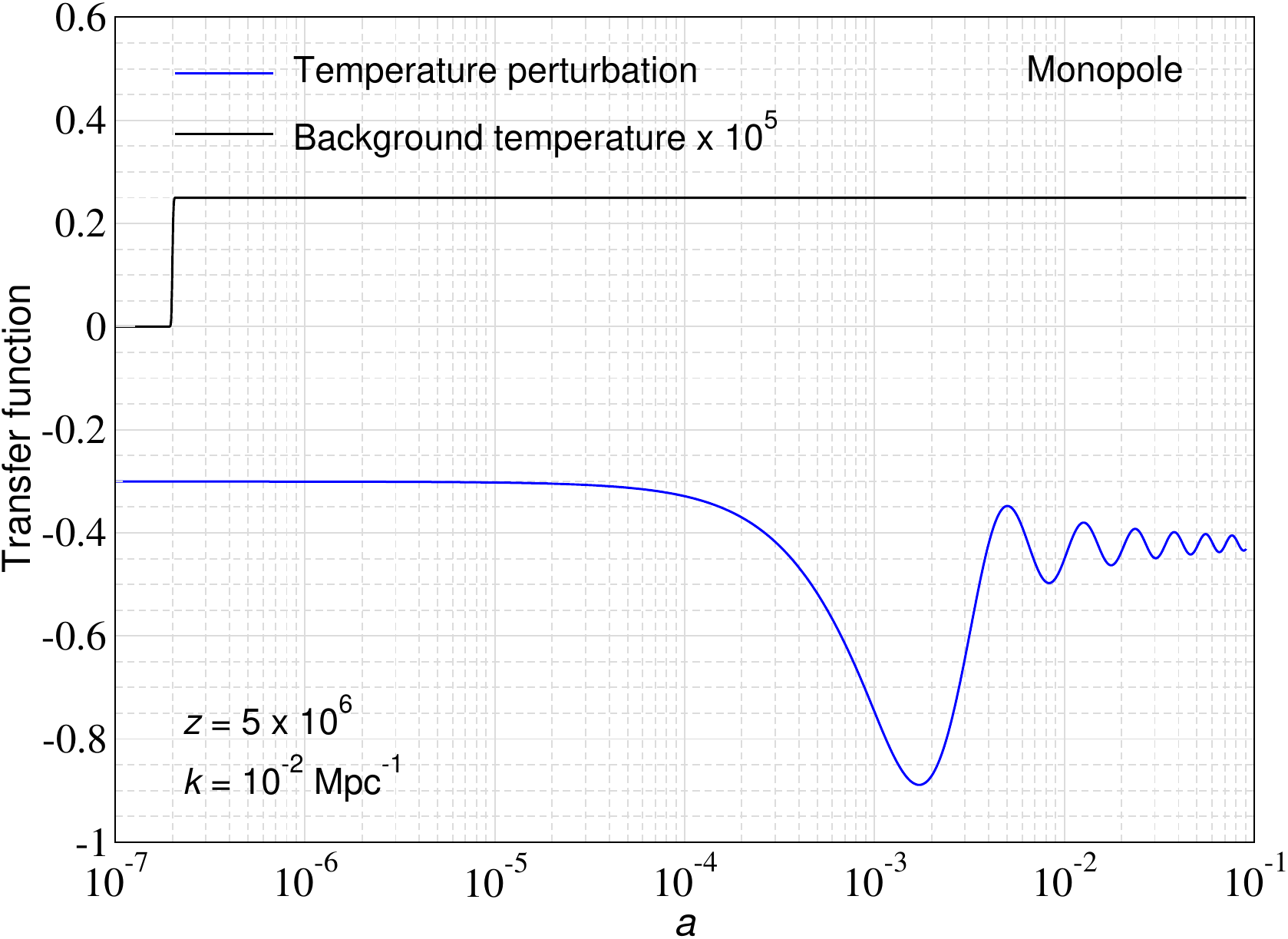}
\hspace{3mm}
\includegraphics[width=0.46\columnwidth]{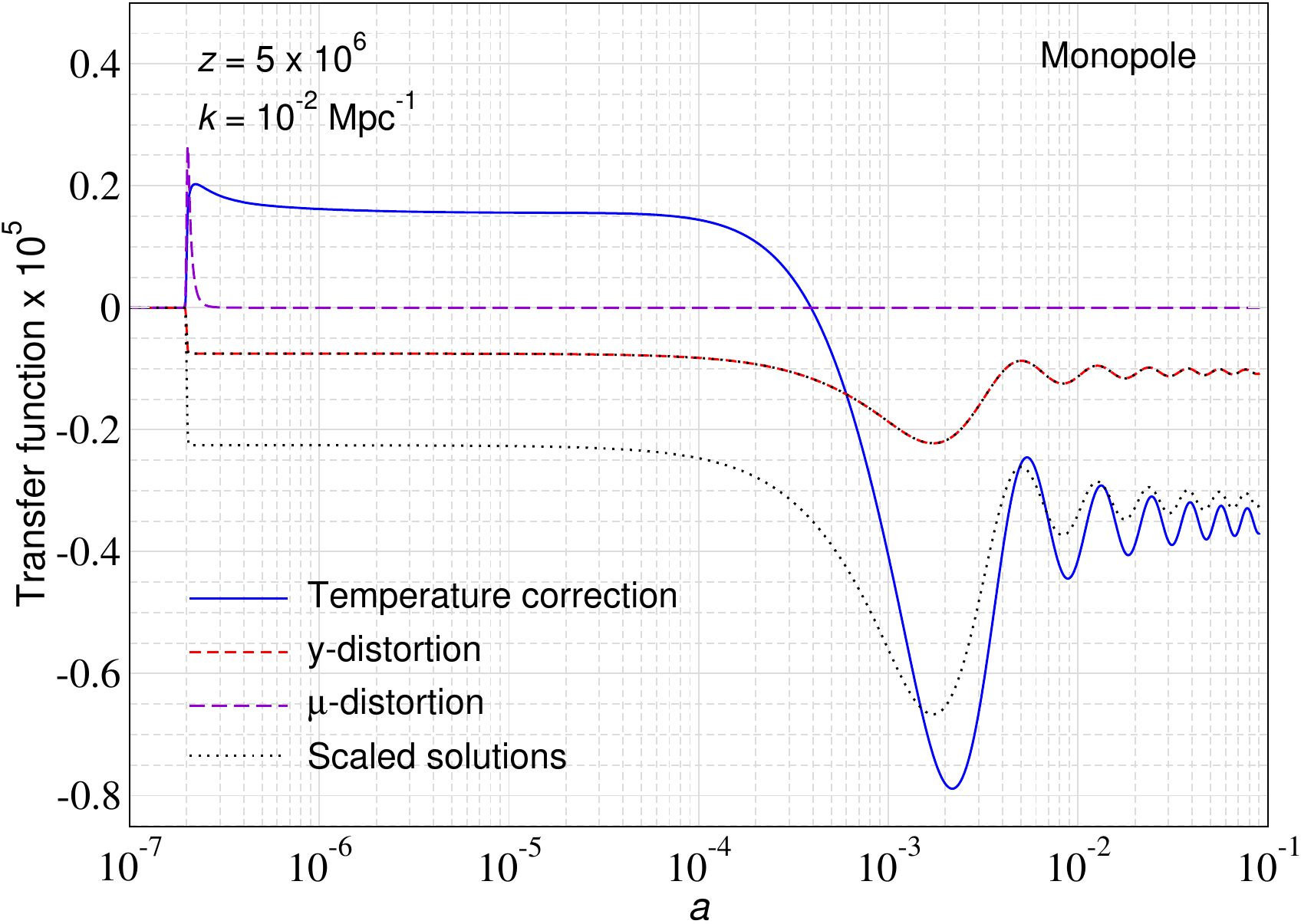}
\\
\includegraphics[width=0.46\columnwidth]{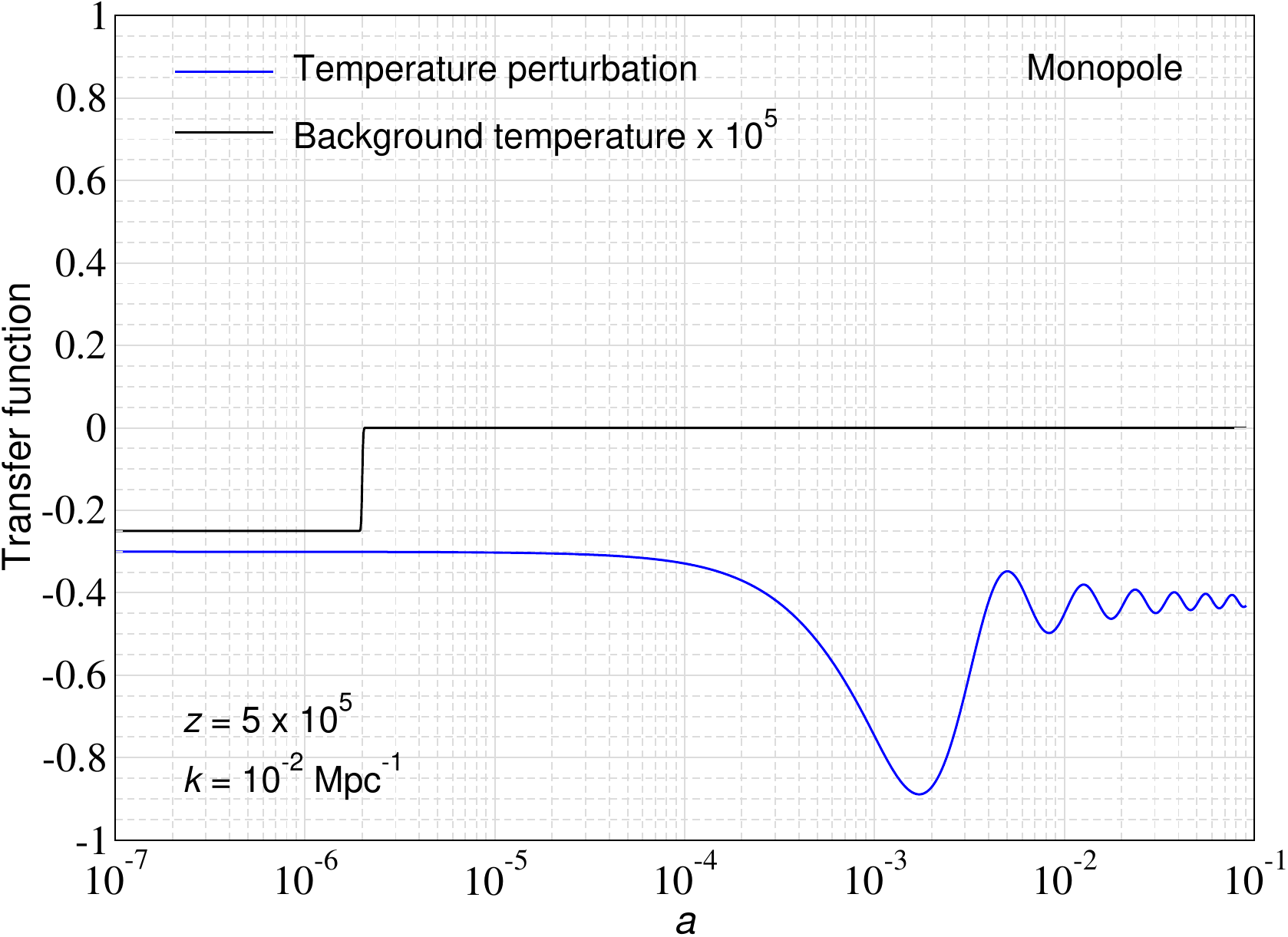}
\hspace{3mm}
\includegraphics[width=0.46\columnwidth]{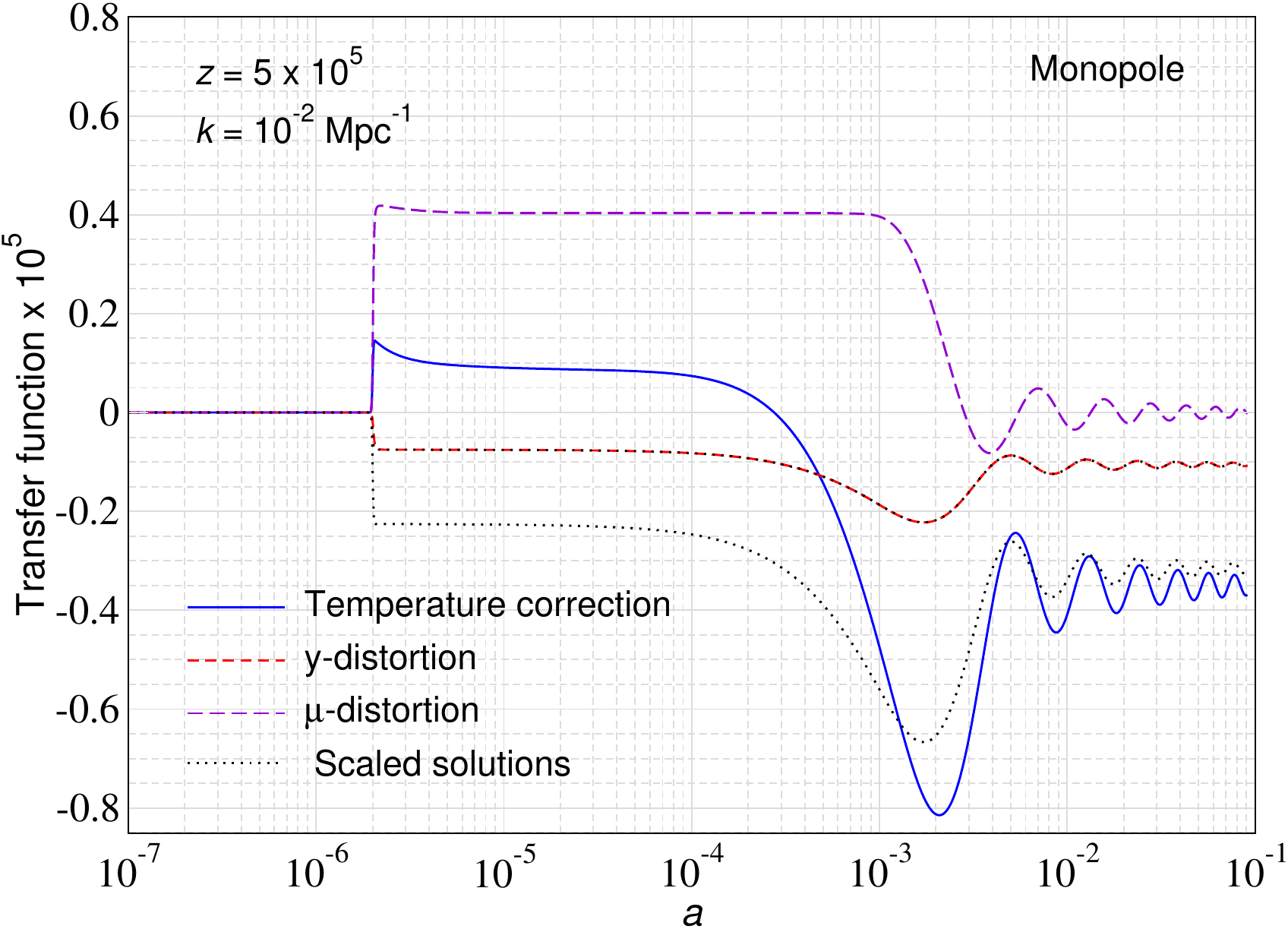}
\\
\includegraphics[width=0.46\columnwidth]{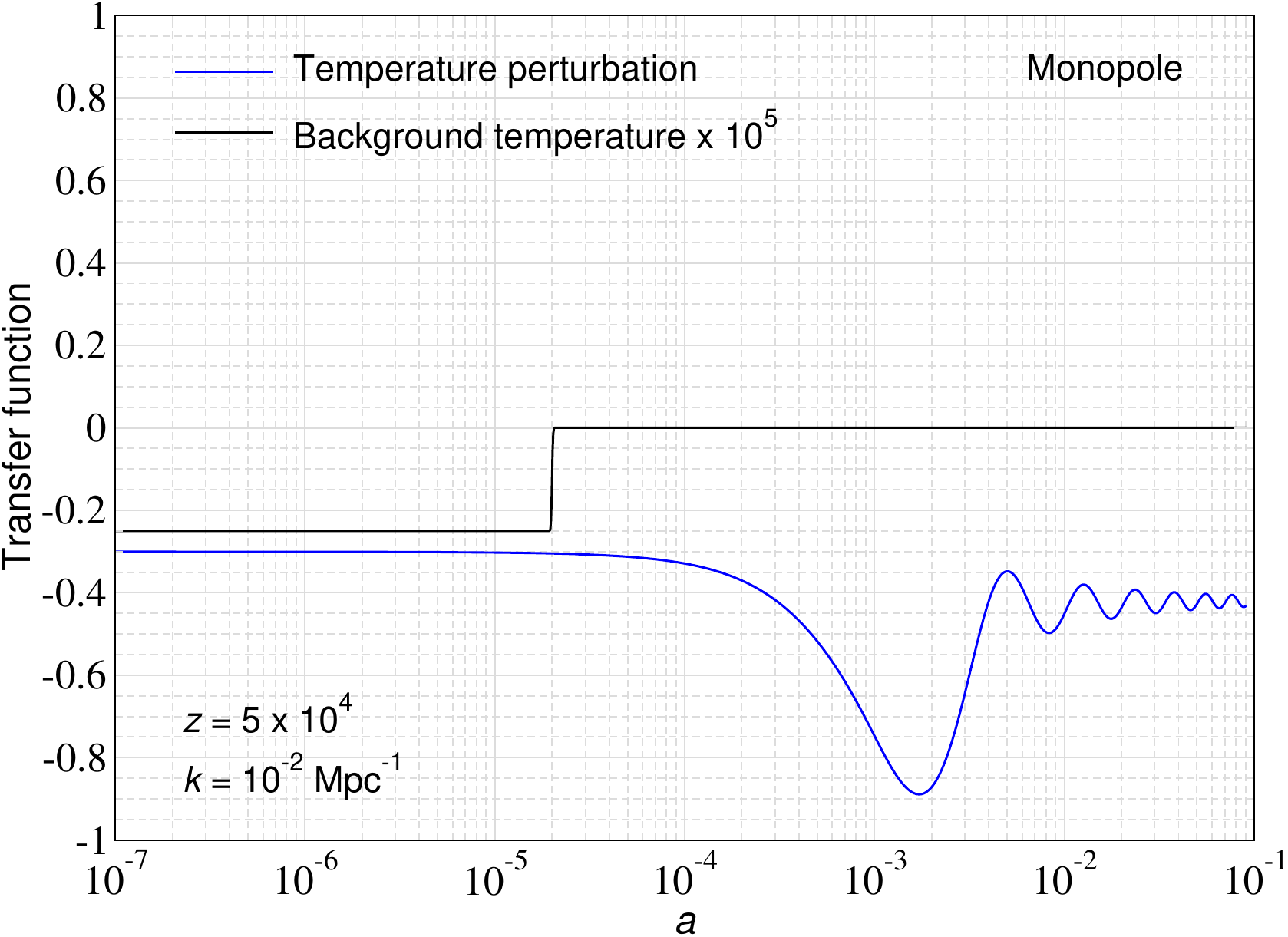}
\hspace{3mm}
\includegraphics[width=0.46\columnwidth]{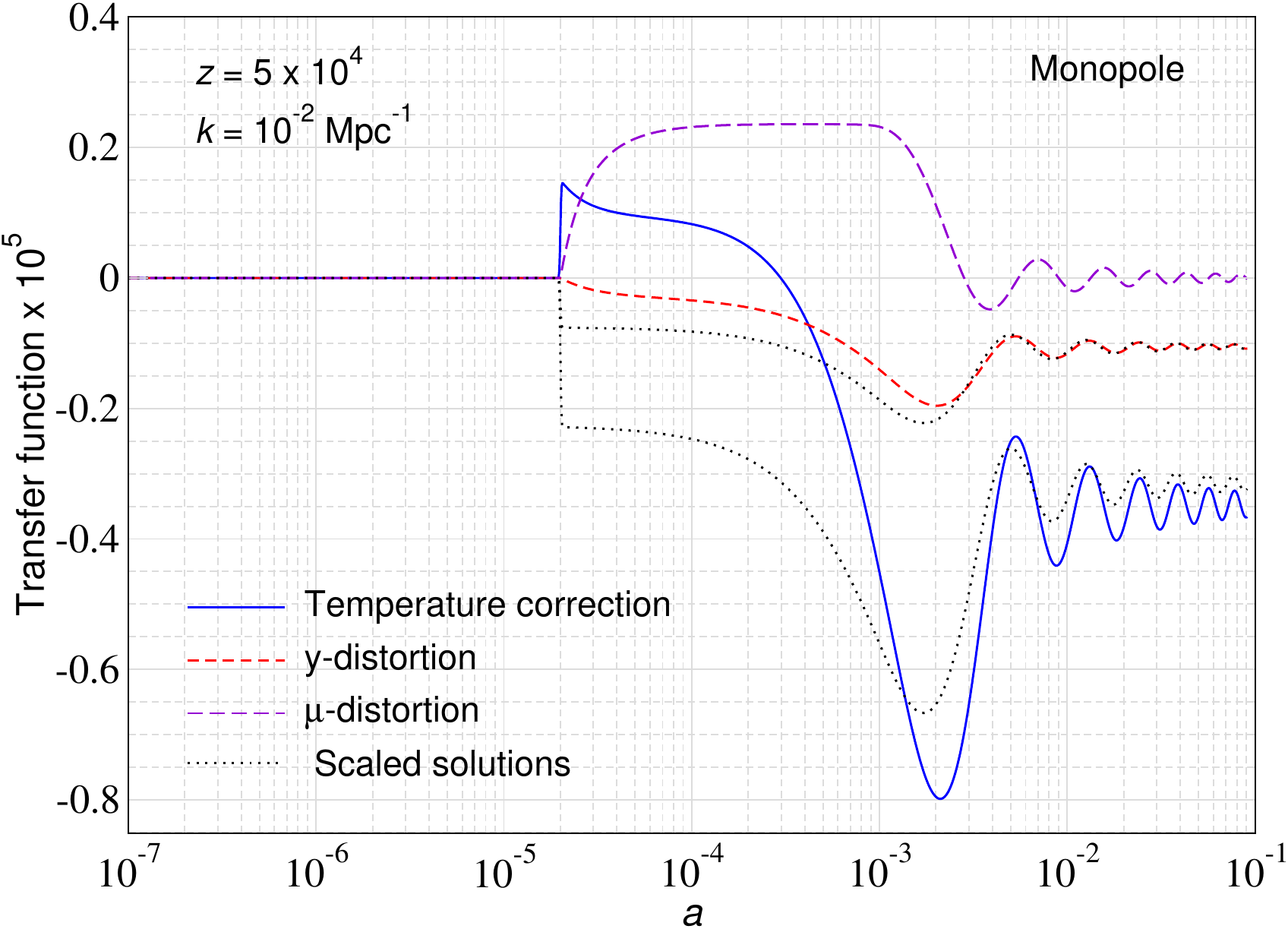}
\\
\includegraphics[width=0.46\columnwidth]{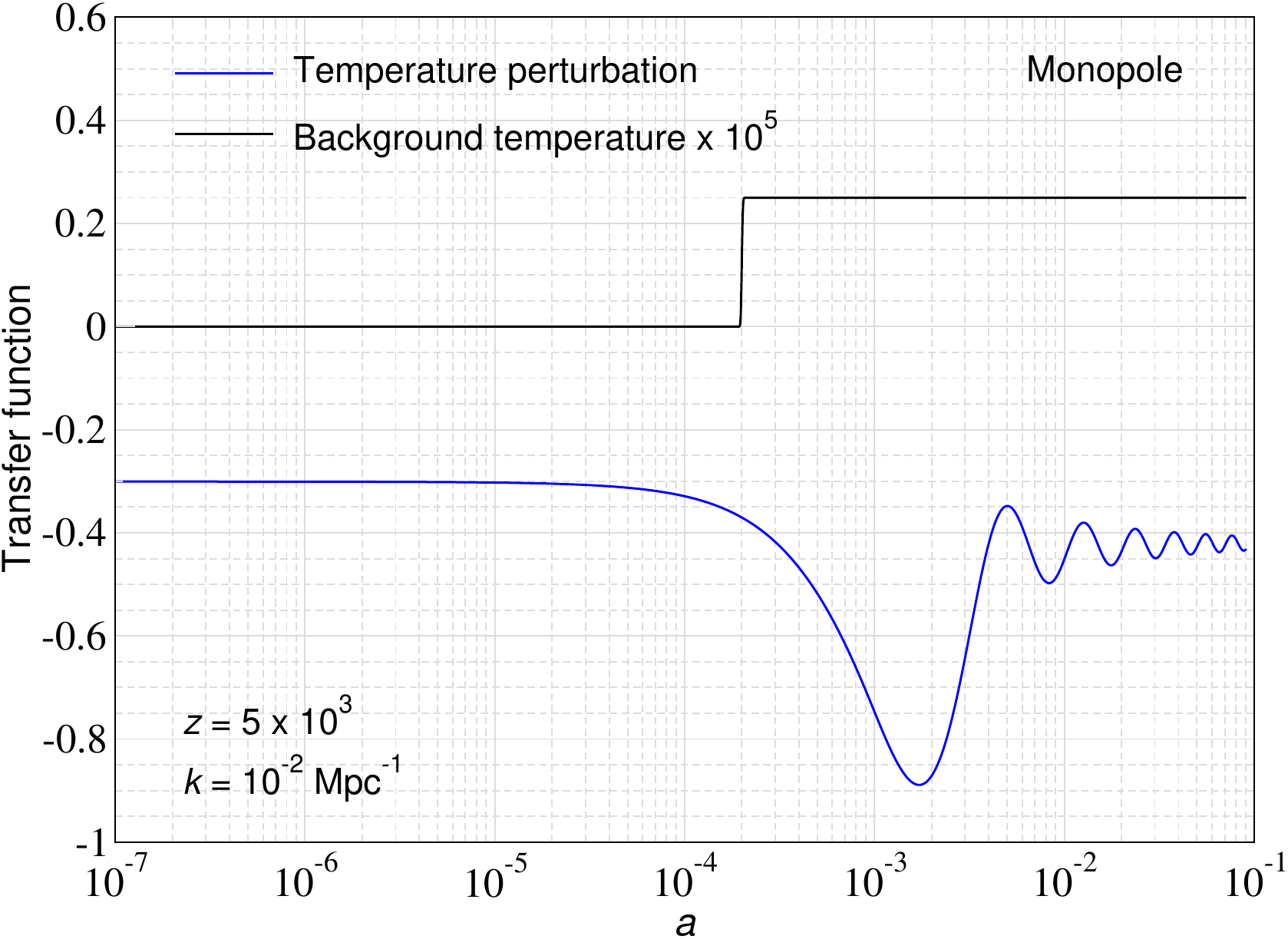}
\hspace{3mm}
\includegraphics[width=0.46\columnwidth]{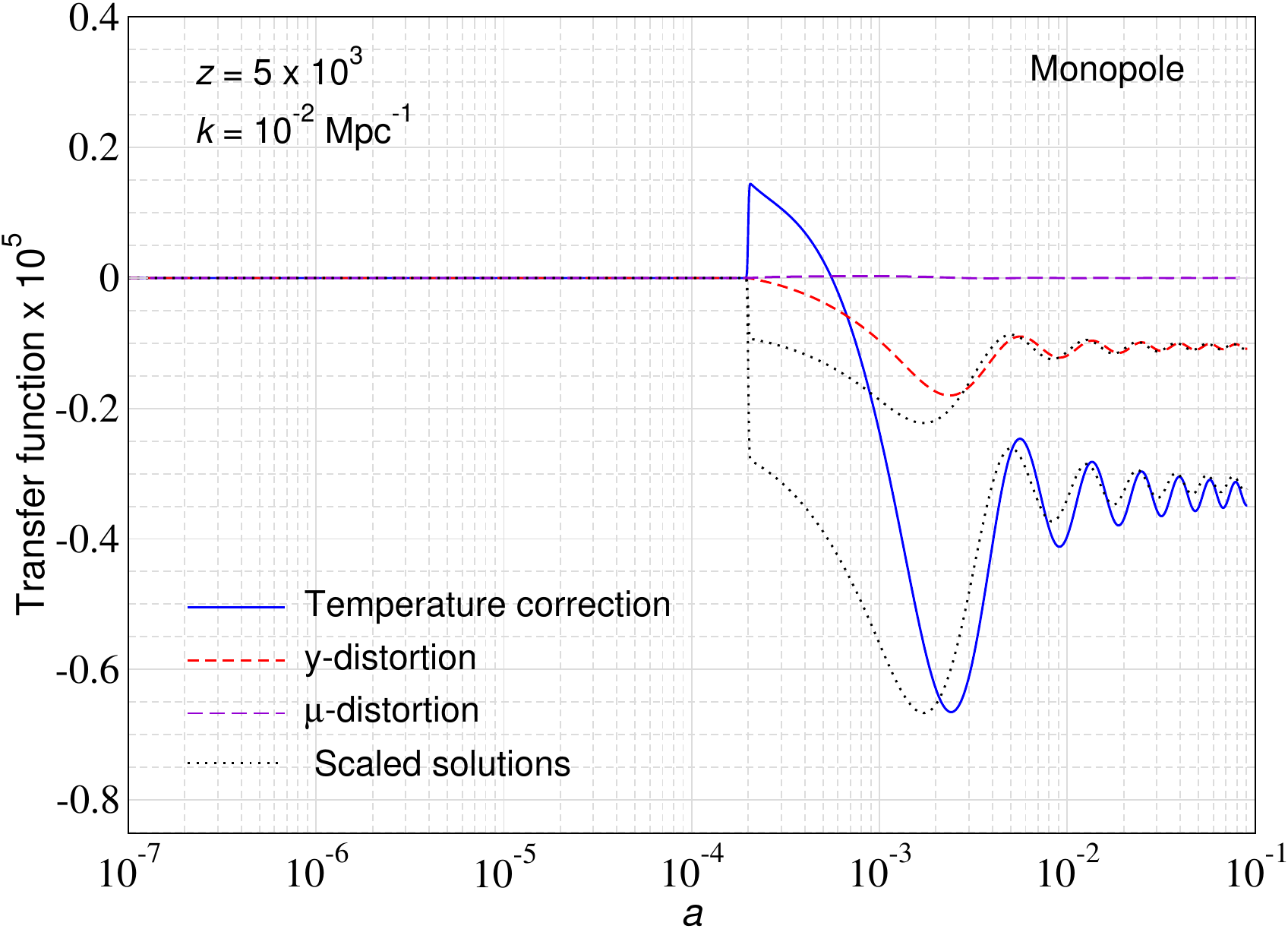}
\\
\caption{Transfer functions for the monopole perturbations as a function of scale factor $a=1/(1+z)$ and assuming a single injection of a temperature shift at $z_{\rm s}=z$ as annotated. The left column shows the evolution of the background temperature (no background distortions are sourced by construction) and temperature perturbations. The right column shows the induced temperature correction and $y$ and $\mu$-parameters.
For comparison we show the scaled solutions $\delta\Theta^{(1)}_0 \approx 3 \bar{\Theta} \Theta^{(1)}_0$ and $y^{(1)}_{0,0} \approx \bar{\Theta} \Theta^{(1)}_0$.
We note the changing scale on the $y$-axis.}
\label{fig:T-shift-delta-II}
\vspace{-4mm}
\end{figure}

To setup scenarios, we first consider initial conditions without any distortions or temperature shift and then start the sourcing of an isotropic temperature shift at $z_{\rm s}$. We also include the changes in $\Psi^{(1)}$ for anisotropic sourcing to account for changes in the local cosmic time, meaning\footnote{In practice we use a narrow Gaussian to model the $\delta$-function in redshift. The differential equation solver automatically threads carefully across this $\delta$-function around the injection redshift.}
\bealf{
\frac{\id \mathcal{S}^{(0)}(z, x)}{\id z}&=\bar{\Theta} \, \delta(z-z_{\rm s}) \,G(x), 
\qquad \frac{\id \mathcal{S}^{(1)}(z, x)}{\id z}= \Psi^{(1)}\,\frac{\id \mathcal{S}^{(0)}(z, x)}{\id z}.
}
This explicit `photon source' can be readily translated into the numerical setup, for which we also include kinematic corrections (see Sect.~\ref{sec:Photon sources}). We note that this source term is not expected to leave the spectrum undistorted, as the local spectrum does not automatically receive the correct number of photons to remain a blackbody. 
We furthermore expect this to lead to contributions from {\it isocurvature} modes, since we are locally changing the mixture of particles. 

In Fig.~\ref{fig:T-shift-delta-II}, we show the results for a run with $\bar\Theta=\pot{2.5}{-6}$ (energy injection $\delta\rho/\rho=10^{-5}$) and varying redshifts $z_{\rm s}$ for a mode with wavenumber $k=10^{-2}\,\Mpc^{-1}$. The left panel demonstrates that the background spectrum correctly shifts towards the new blackbody spectrum with new average temperature $\TCMB(1+\bar\Theta)$. No average distortions are being produced by construction. However, for the monopole perturbations the picture is a little different. For $z_{\rm s}=\pot{5}{5}$ and $z_{\rm s}=\pot{5}{4}$, we notice a $\mu$-distortion anisotropy. For $z_{\rm s}>10^6$, $\mu$-distortions are sourced but quickly thermalize, while at $z_{\rm s}<10^4$ conversion of $y$ into $\mu$ becomes inefficient. The $y$-distortion monopole is consistent with the scaled solution, $y^{(1)}_{0,0} \approx \bar{\Theta} \Theta^{(1)}_0$ for injection at $z_{\rm s}=\pot{5}{6}$ and $z_{\rm s}=\pot{5}{5}$, while departures become noticeable for the other cases. For the transfer functions of $\delta \Theta^{(1)}_{0}$, we observe departures from the scaled solution $\delta\Theta^{(1)}_0 \approx 3 \bar{\Theta} \Theta^{(1)}_0$ in all cases, although at late times the solution always becomes very similar. A small phase shift furthermore indicates presence of an isocurvature-{type} mode.

Let us try to understand the appearance of the $\mu$-distortion anisotropy for early injections. At superhorizon scales, we have $\Psi^{(1)}=-2\Theta^{(1)}_0$ for adiabatic initial conditions, meaning that right after the injection we have $\delta \Theta^{(1)}_0= -2\Theta^{(1)}_0 \Thetabar$ or $\delta \Theta^{(1)}_0 \approx \pot{1.5}{-6}$ for our case, which is in agreement with our computations especially for $z_{\rm s}=\pot{5}{5}$ and $z_{\rm s}=\pot{5}{4}$. Since this is a direct temperature source, no further distortion anisotropy is being caused by this contribution. However, the scattering term $-\Theta^{(1)}_0 M_{\rm BK} \,\vek{y}^{(0)}_{\rm qs}$ in Eq.~\eqref{eq:evol_transport_yn_improved}, which arises from Compton energy exchange with the average spectrum, {does source} distortion anisotropies. {To understand this, let} us first compute $M_{\rm B} \vek{y}^{(0)}_{\rm qs}=3\Thetabar\,\vek{e}_G +\Thetabar\,\vek{e}_Y$ meaning a $y$-source with energy $\epsilon^{(1)}_0=-4\Theta^{(1)}_0 \Thetabar \approx \pot{3.0}{-6}$ at superhorizon scales. For $z_{\rm s}>10^5$, the Kompaneets operator quickly converts this source into a $\mu$-distortion by repeated scattering with $\mu^{(1)}_0\approx 1.4 \epsilon^{(1)}_0 \approx \pot{4.2}{-6}$ initially, as clearly visible for the solution with $z_{\rm s}=\pot{5}{5}$. For this case, we can also observe a small evolution of $\mu^{(1)}_0$ even before horizon-crossing due to local conversions of $\mu\rightarrow T$. This effect becomes more dramatic if we inject at $z_{\rm s}= \pot{2}{6}$, leading to an exponential suppression of $\mu$ and related conversion to $T$, boosting the amplitude of $\delta \Theta^{(1)}_0 $ by about $\epsilon^{(1)}_0/4\simeq \pot{7.5}{-7}$.

We mention that the anisotropic photon source term also causes changes in the local potential, leading to part of the evolution of $\delta\Theta_0^{(1)}$ seen in Fig.~\ref{fig:T-shift-delta-II} while the mode is at superhorizon scales. In physical scenarios, additional gravitational effects may have to be considered, however, without a specific scenario in mind, this is difficult to assess. {We also stress that the superhorizon evolution seen in our solutions is not acausal: the average CMB spectrum is changed by local thermalization physics occurring the same way in all patches and these affect the perturbation variables at all scales. This effect is absent when the average spectrum does not evolve.}

\subsection{Changes in the temperature-redshift relation}
\label{sec:Tz-changes}
Next we consider possible changes to the temperature redshift relation (TRR). This scenario was motivated as {\it adiabatic photon production} process in cosmologies with decaying vacuum energy density \citep[e.g.,][]{Freese1987, Overduin1993, Lima1996, Lima2000}, leading to a departure of the temperature-redshift relation from the $1+z$ scaling. Although it is clear that these kind of scenarios are hard to imagine without creating CMB spectral distortions \citep{Caldwell2008, Caldwell2013, Chluba2014TRR}, a consistent demonstration also of the effects on the CMB temperature anisotropies was not carried out so far. 

As a concrete example, we shall consider a temperature redshift-relation of the form.
\bealf{
T(z)&=T_0(1+z)^{1-\beta}\qquad \rightarrow \qquad \bar{\Theta}=\frac{T-\Tz}{\Tz}=\frac{1}{(1+z)^{\beta}}-1
}
for the average CMB background spectrum where $\Tz=T_0(1+z)$. This then implies the temperature injection or source term
\bealf{
\frac{\id \mathcal{S}(z, x)}{\id z}&\approx  G(x) \frac{\id \bar{\Theta}}{\id z} = -\frac{\beta\, (1+\bar{\Theta})}{1+z} G(x).
}
We neglected any effects on the background expansion rate in our computations, which applies when the source of photons is related to another relativistic species. Otherwise, additional redshifting effects have to be taken into account, but we shall leave a detailed discussion for future work. 

To setup scenarios, we consider initial conditions without any distortions and then start the sourcing of isotropic temperature shifts below some specified redshift\footnote{The transition is modeled using a simple tanh-switch with relative width $\Delta z/z = 0.1$.}, $z_{\rm s}$. We also include the changes in $\Psi^{(1)}$ for anisotropic sourcing to account for changes in the local cosmic time, meaning
\bealf{
\frac{\id \mathcal{S}^{(0)}(z, x)}{\id z}&= -\frac{\beta\, (1+\bar{\Theta})}{1+z} G(x), 
\qquad \frac{\id \mathcal{S}^{(1)}(z, x)}{\id z}= \Psi^{(1)}\,\frac{\id \mathcal{S}^{(0)}(z, x)}{\id z}.
}
This explicit `photon source' can again be readily translated into the numerical setup {by adding a source term to the temperature variable}.

\begin{figure}
    \centering
\includegraphics[width=0.495\columnwidth]{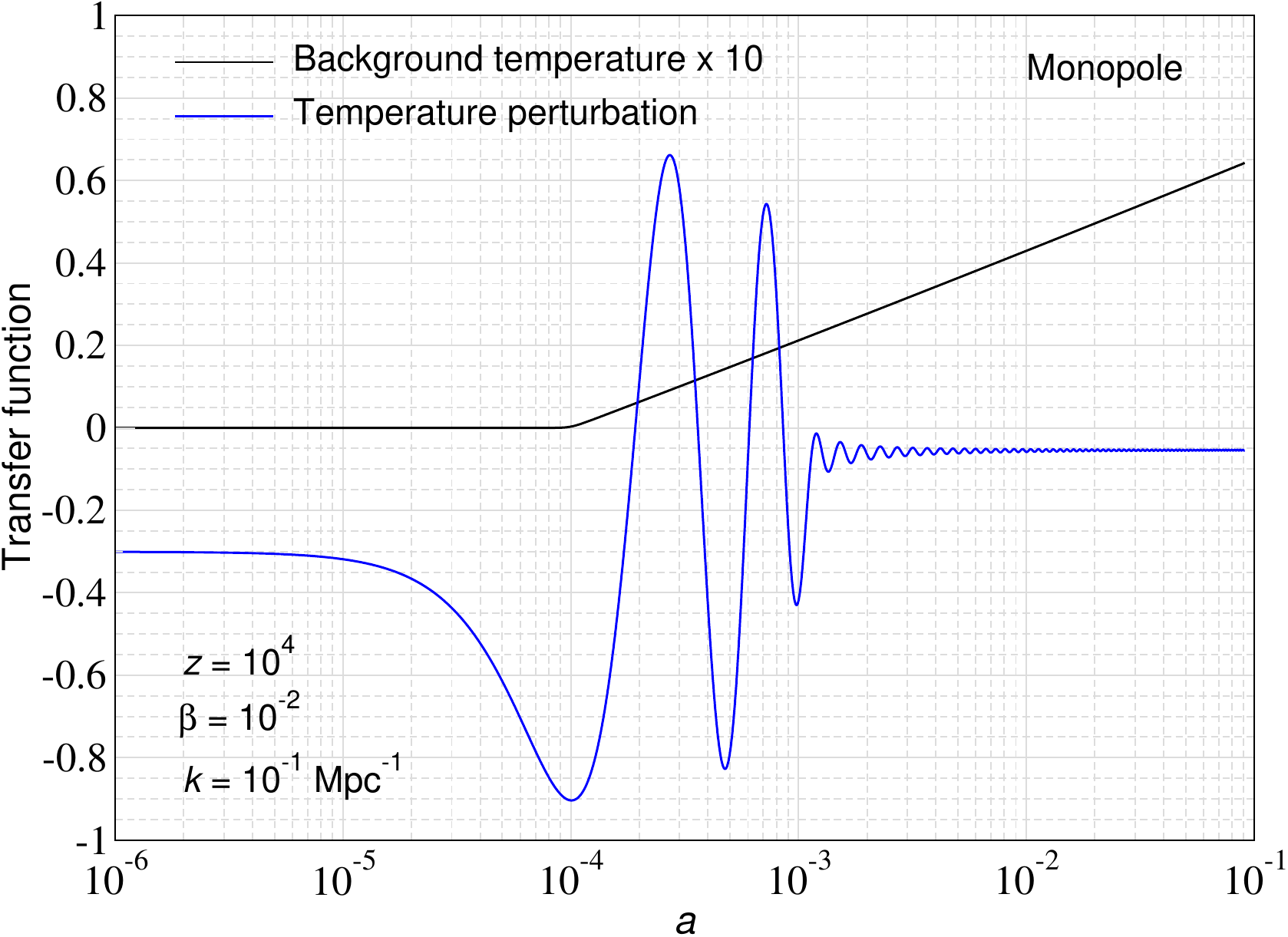}
\includegraphics[width=0.495\columnwidth]{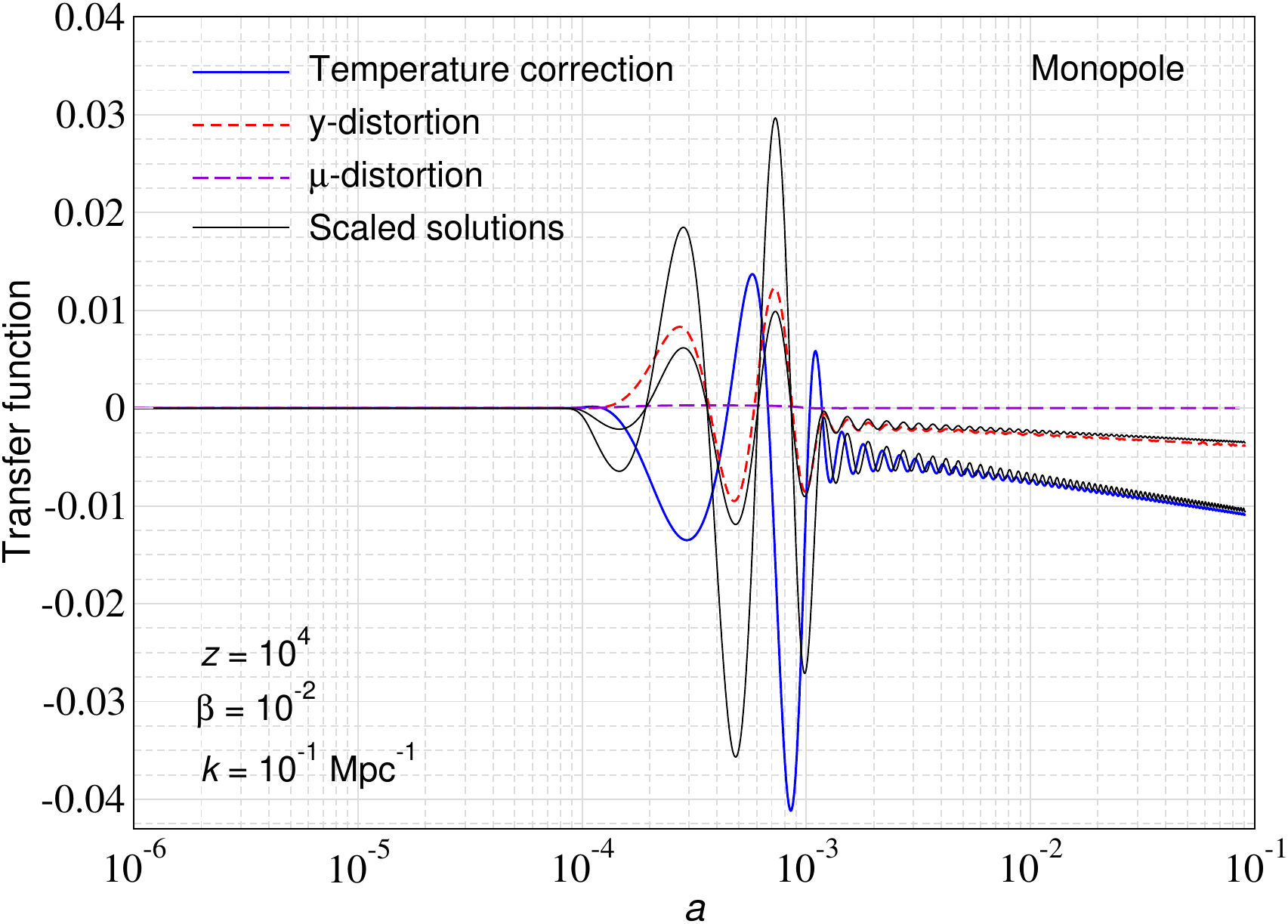}
\\
\includegraphics[width=0.495\columnwidth]{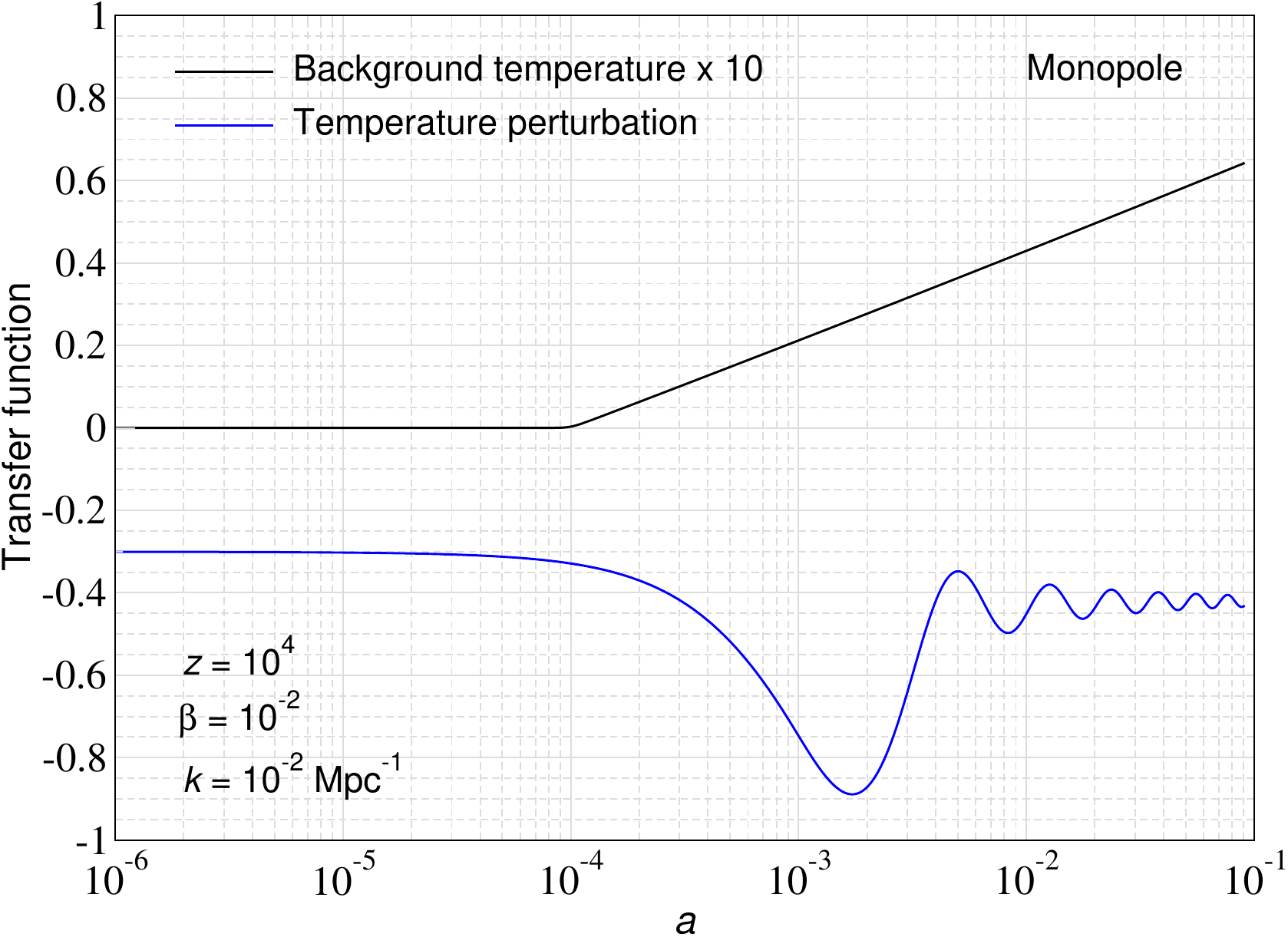}
\includegraphics[width=0.495\columnwidth]{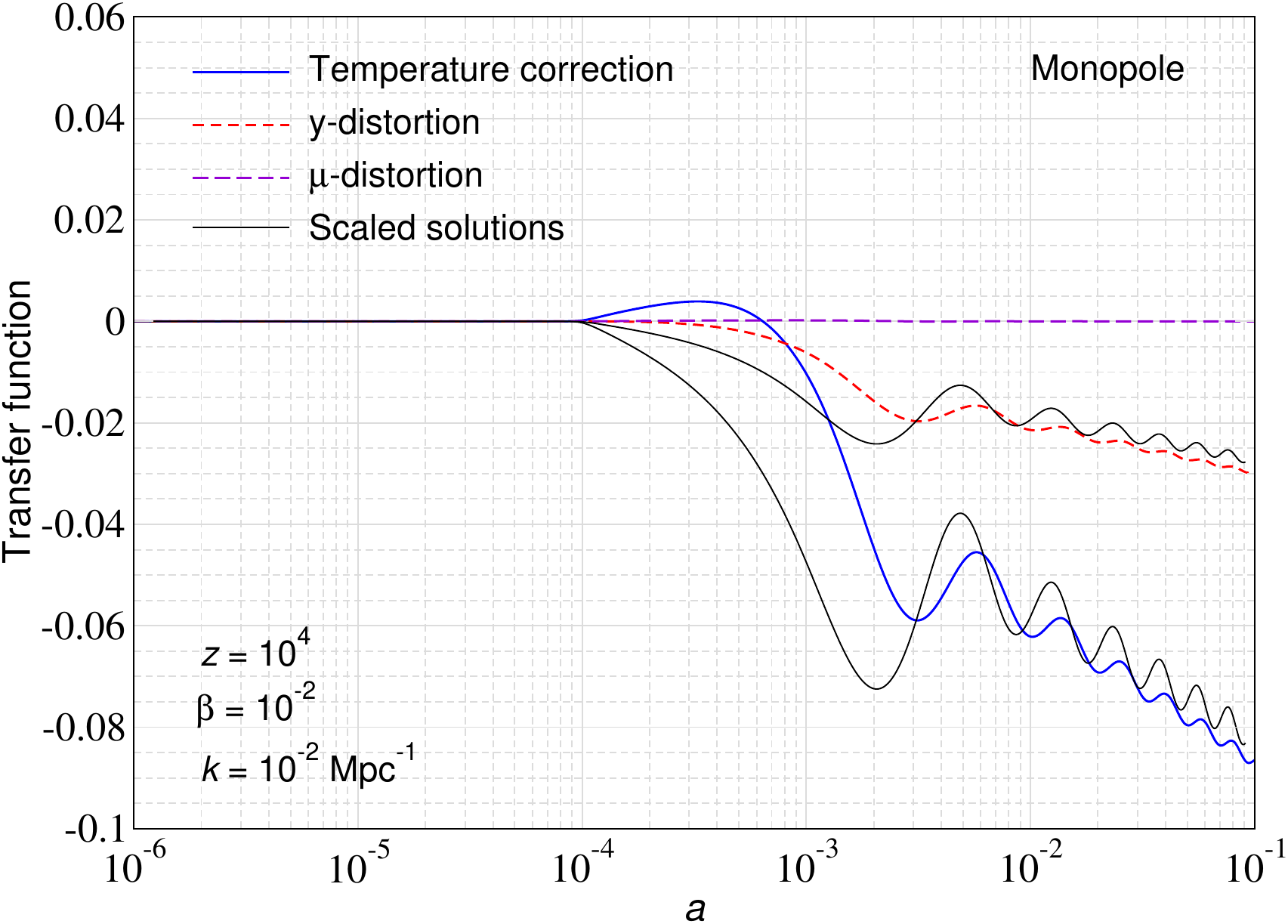}
\\
\caption{Transfer functions for the monopole perturbations assuming a temperature source term with parameters as annotated. The left column shows that evolution of the background temperature (no background distortions are sourced by construction) and temperature perturbations. The right column shows the induced temperature correction and distortion $y$ and $\mu$-parameter.
For comparison we show the expected equilibrium solutions $\delta\Theta^{(1)}_0 \approx 3 \bar{\Theta} \Theta^{(1)}_0$ and $y^{(1)}_{0,0} \approx \bar{\Theta} \Theta^{(1)}_0$, which match the solutions at late times very well for $k=10^{-1}\,\Mpc^{-1}$, while departures are visible for $k=10^{-2}\,\Mpc^{-1}$ due to the lower Thomson scattering rate when the mode is inside the horizon.
We note the varying change of scale on the $y$-axis.}
\label{fig:T-shift-I}
\vspace{-4mm}
\end{figure}

\subsubsection{Transfer functions and power spectra}
In Fig.~\ref{fig:T-shift-I} we show some example transfer functions for the monopole and varying values of the wavenumber $k$. We chose $\beta=10^{-2}$ and $z_{\rm s}=10^4$ as illustrative example. The change in the background temperature does not induce any background distortion and even at the perturbed level, small-scale modes equilibrate quickly to come close to the expected equilibrium solution. However, departures are visible as a phase-shift and overall offset in the amplitude, which becomes more pronounced for large-scale modes that enter the horizon after the rise in the background temperature has started. This stems from the fact that the rate of equilibration in the chosen case primarily depends on the Thomson scattering rate, which is longer for large-scale perturbation. The fluctuations of large-scale modes therefore lag behind and show a clearer record of when the change in the temperature redshift relation (TRR) has started.

\begin{figure}
    \centering
\includegraphics[width=0.495\columnwidth]{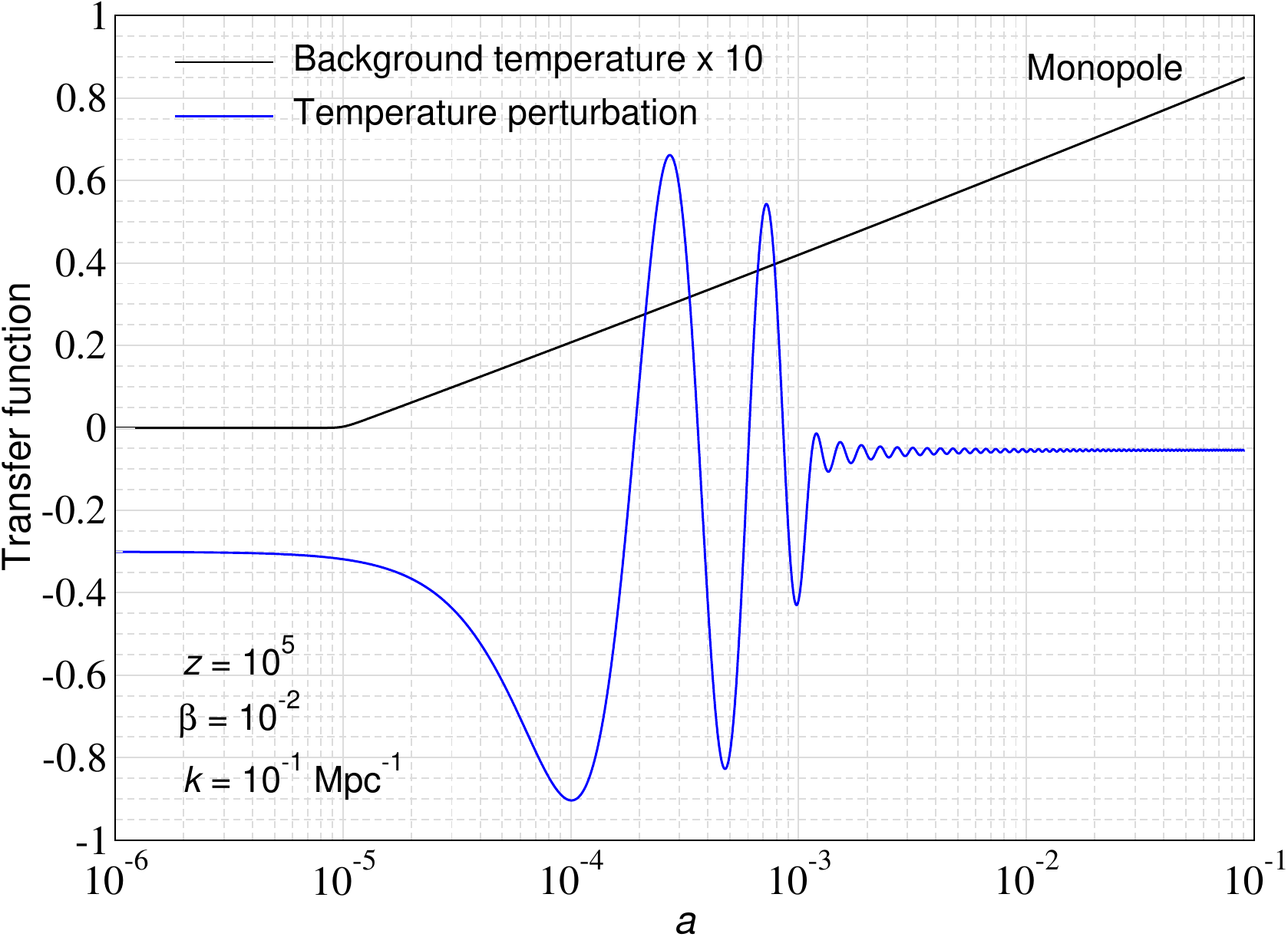}
\includegraphics[width=0.495\columnwidth]{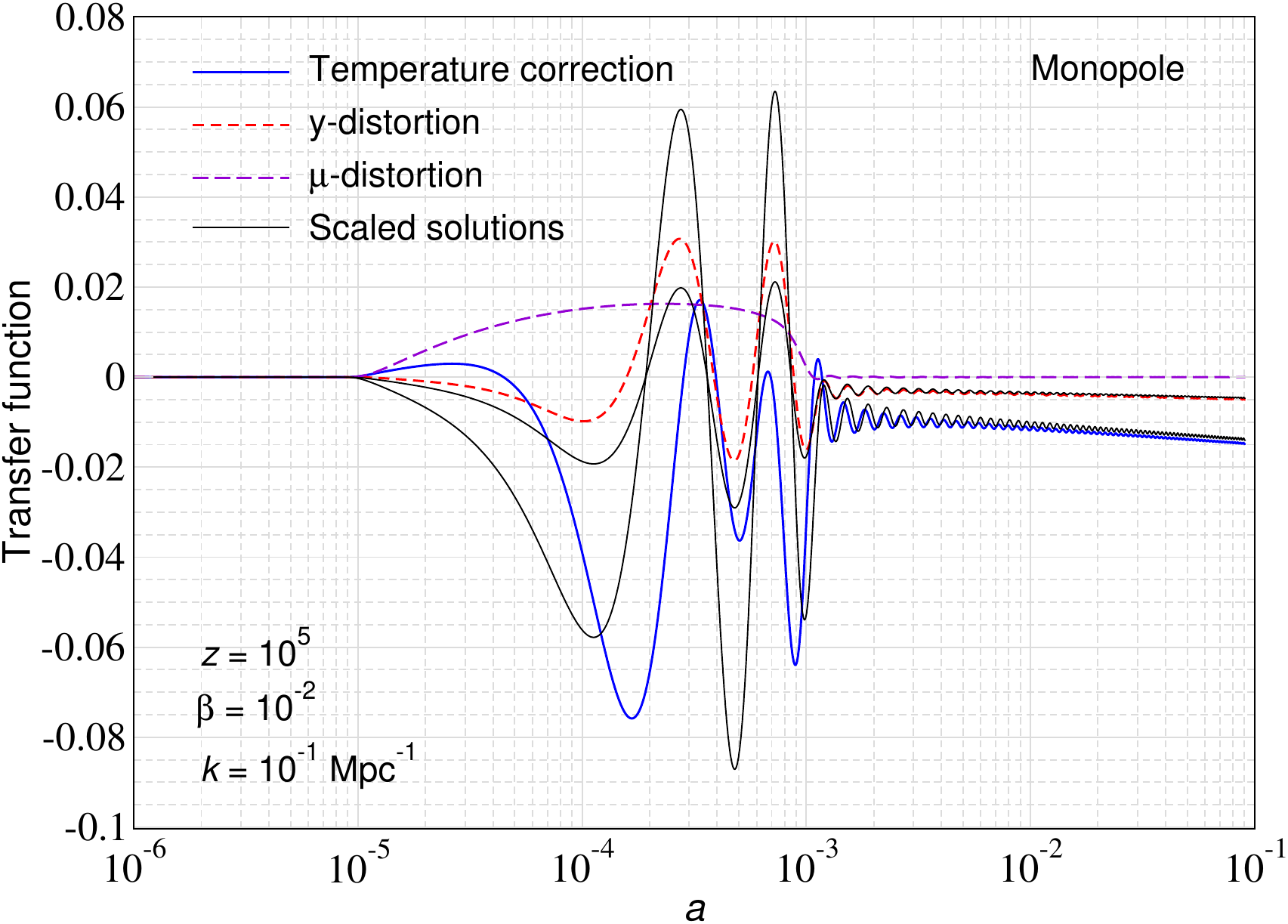}
\\
\includegraphics[width=0.495\columnwidth]{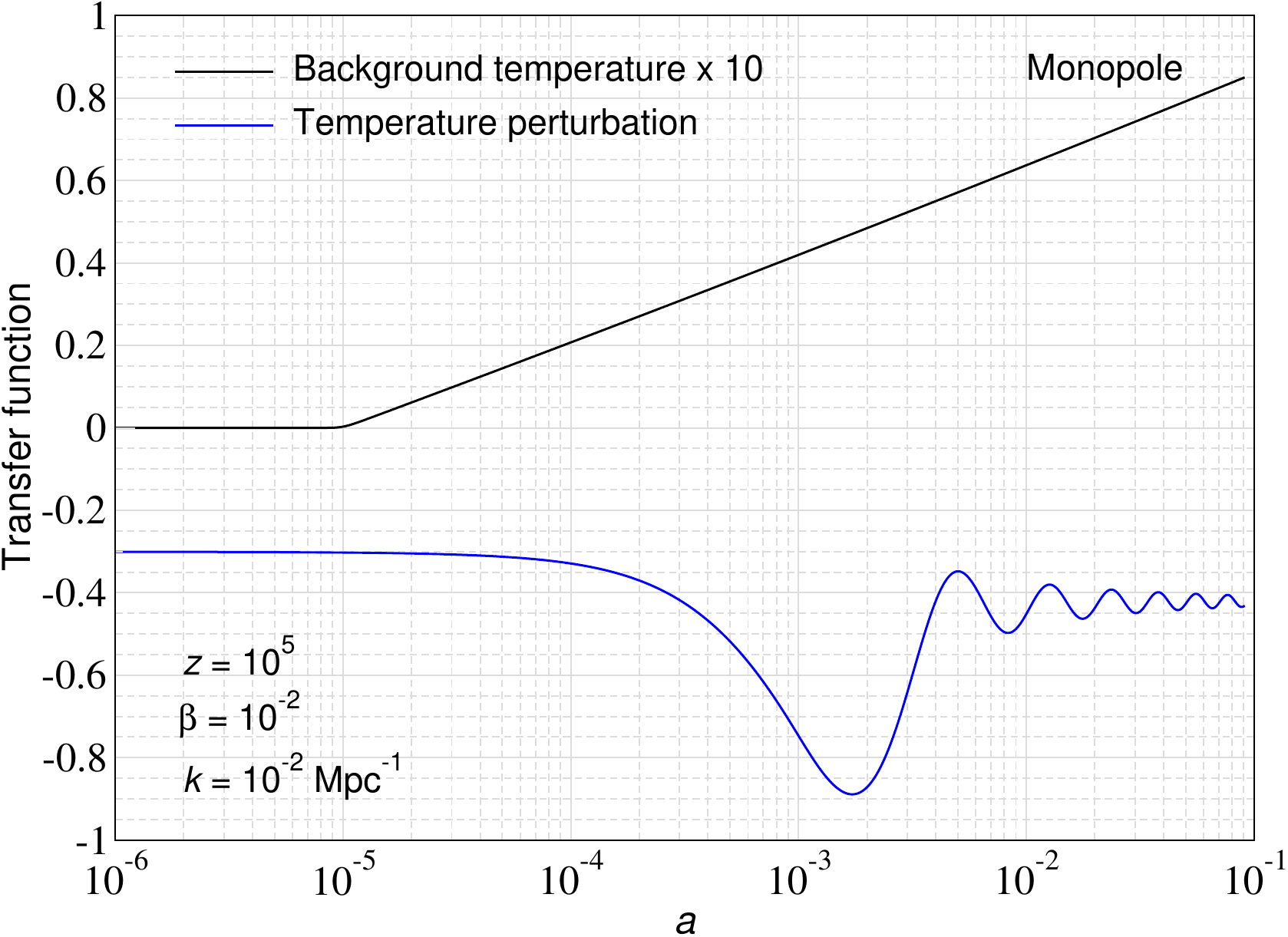}
\includegraphics[width=0.495\columnwidth]{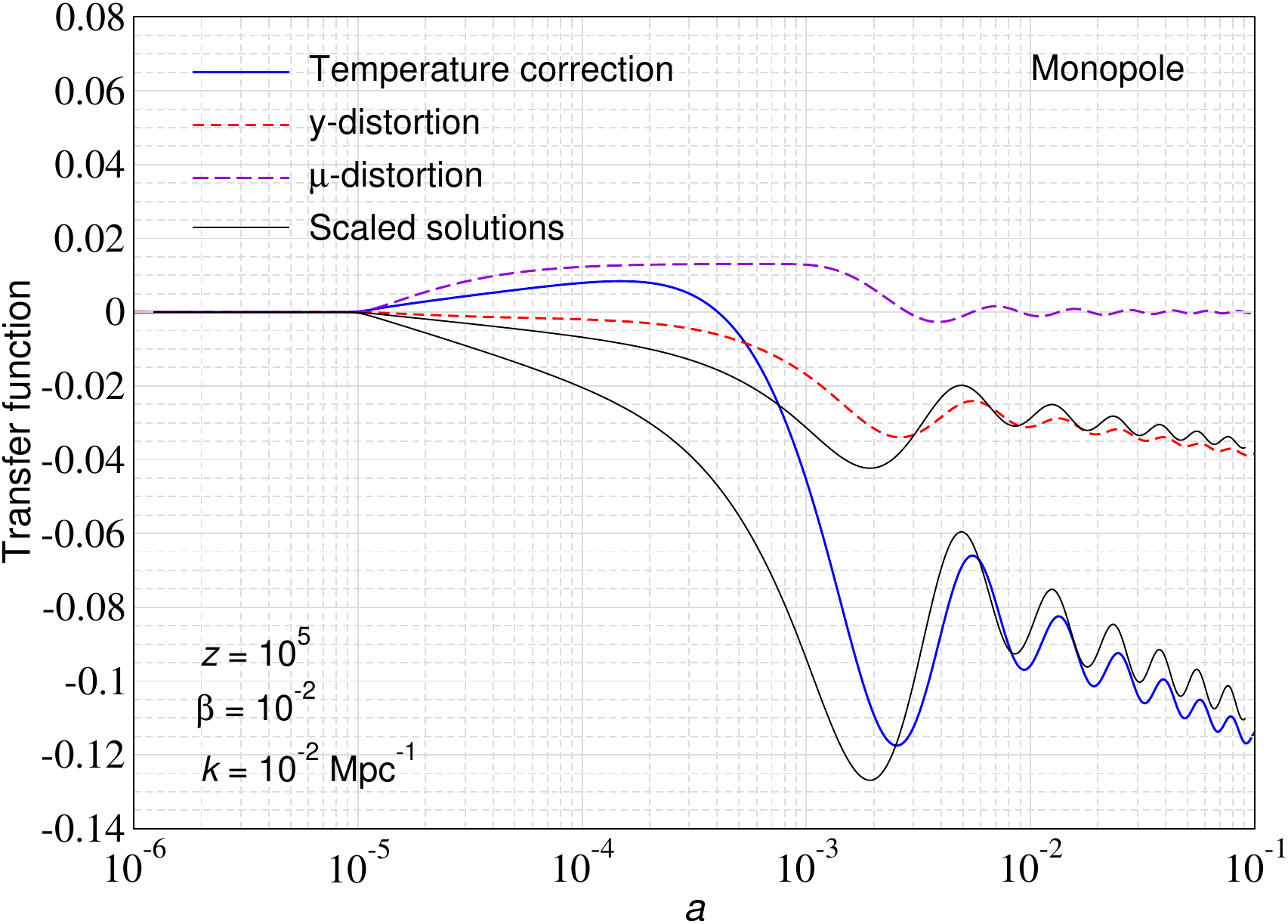}
\\
\caption{Transfer functions for the monopole perturbations assuming a temperature source term with parameters as annotated. The left column shows that evolution of the background temperature (no background distortions are sourced by construction) and temperature perturbations. The right column shows the induced temperature correction and distortion $y$ and $\mu$-parameter.
For comparison we show the expected equilibrium solutions $\delta\Theta^{(1)}_0 \approx 3 \bar{\Theta} \Theta^{(1)}_0$ and $y^{(1)}_{0,0} \approx \bar{\Theta} \Theta^{(1)}_0$, which match the solutions at late times very well for $k=10^{-1}\,\Mpc^{-1}$, while departures are visible for $k=10^{-2}\,\Mpc^{-1}$ due to the lower Thomson scattering rate when the mode is inside the horizon.
We note the varying change of scale on the $y$-axis.}
\label{fig:T-shift-II}
\vspace{-4mm}
\end{figure}
In Fig.~\ref{fig:T-shift-II} we repeat the calculation of Fig.~\ref{fig:T-shift-I} but for $z_{\rm s}=10^5$. Overall the picture is very similar, but for the small-scale modes we now also notice a $\mu$-distortion mode. The double Compton and Bremsstrahlung processes are not efficient, but Comptonization converts some of the energy from the heat exchange with the electrons into a $\mu$-distortion shape, possibly providing a way to determine the time when the temperature redshift relation started departing from the standard scaling.

\begin{figure}
    \centering
\includegraphics[width=0.8\columnwidth]{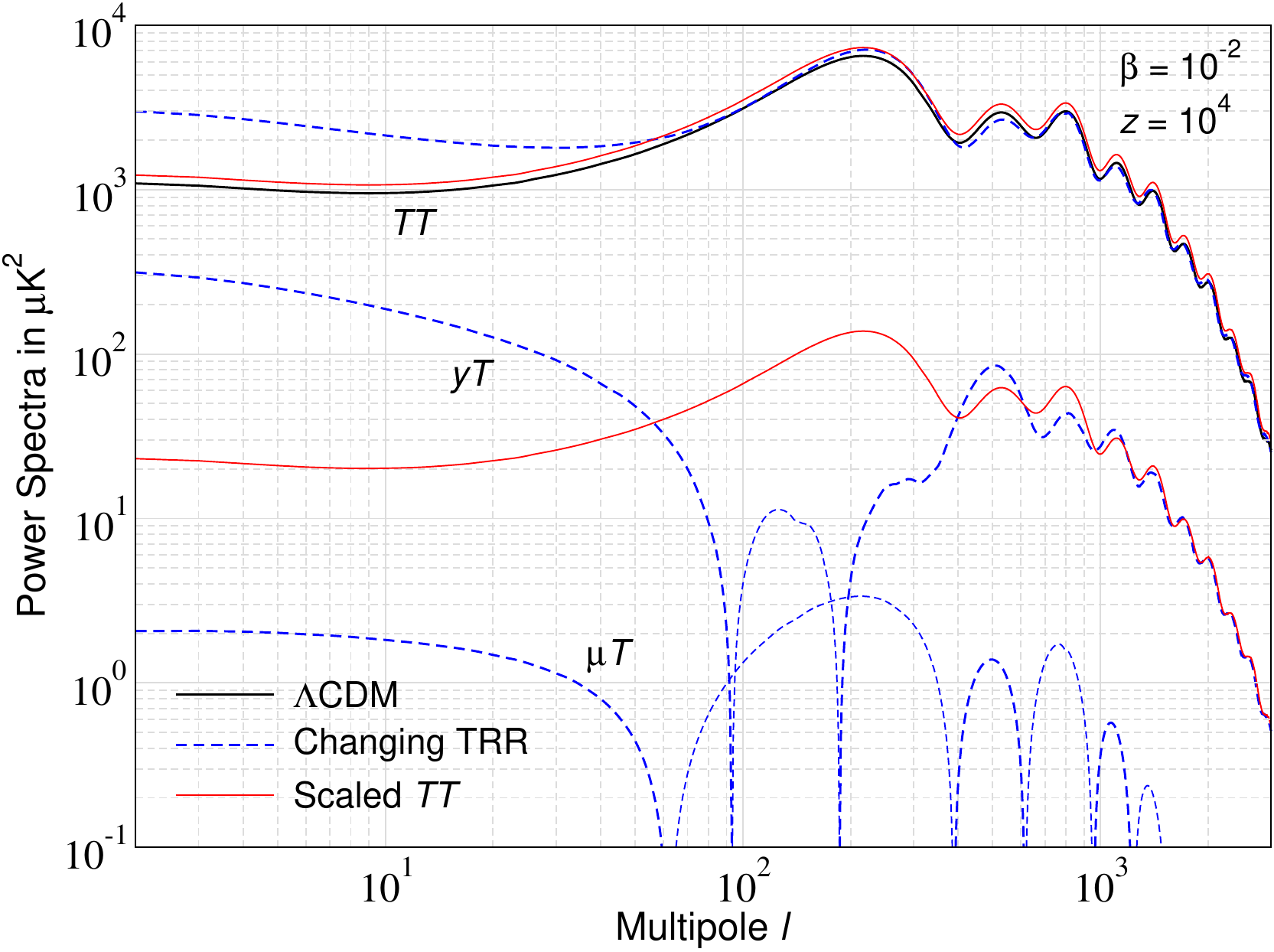}
\\
\caption{CMB power spectra for the settings of Fig.~\ref{fig:T-shift-I} [observational basis with up to $Y_5$.]. The solid black line is the $\Lambda$CDM case (which has no distortion contributions), while the blue-dashed lines are computed for $\beta=10^{-2}$ and $z_{\rm s}=10^4$, showing the total $TT$, $yT$ and $\mu T$ power spectra. Around the maximum of the Thomson visibility function at $z\simeq 1080$ one has $\bar{\Theta}\approx 0.02$. The scaled models (solid red lines) use the $\Lambda$CDM case to obtain $C_\ell^{TT-{\rm scaled}}=(1+3 \Thetabar)^2 \,C_\ell^{TT}$ and $C_\ell^{yY-{\rm scaled}}=(1+3 \Thetabar)\times \Thetabar \,C_\ell^{TT}$. For the {$\mu T$ and $y T$ signals}, negative branches are shown with a thinner line.}
\label{fig:TT-power-shift-I}
\end{figure}
Matters are even more apparent when looking at the power spectra with modified TRR. Assuming the same model as for Fig.~\ref{fig:T-shift-I}, we compare the $TT$, $yT$ and $\mu T$ power spectra with the $\Lambda$CDM $TT$ power spectrum in Fig.~\ref{fig:TT-power-shift-I}. If the power spectra were a reflection of full equilibrium blackbody radiation, one would expect the scalings $C_\ell^{TT-{\rm scaled}}=(1+3 \bar\Theta_{\rm eff})^2 \,C_\ell^{TT}$ and $C_\ell^{yT-{\rm scaled}}=(1+3 \bar\Theta_{\rm eff})\bar\Theta_{\rm eff} \,C_\ell^{TT}$. At small scales ($\ell\gtrsim 10^3$) we find this indeed to be approximately true for the $yT$ power spectrum, with $\bar\Theta_{\rm eff}=0.02$ corresponding roughly to the value of $\bar\Theta$ at last scattering. However, for $TT$ an additional off-set is present, which is likely due to the combination with changes in the potentials. However, we leave a detailed investigation of this point to future work. 

At larger angular scales clear departures from the scaled power spectra exist in both $TT$ and $yT$, showing that the radiation field has not fully mixed, and the CMB spectrum is found in different phases of its spectral evolution. The CMB power spectra and $yT$ cross-power spectra thus seem to carry information about late-time changes of the temperature redshift relation, with {\it real} distortion anisotropies being created, as also indicated by the presence of a small $\mu T$ contribution {that would otherwise be absent}. We note that for the considered case these signals are purely due to Thomson scattering and free streaming effects in a time-varying blackbody field and our choice of spectral decomposition focusing on $G$, $Y$ and $M$ in the signal processing. 

{We confirmed that neglecting the thermalization terms in the calculation does not modify the results in the considered scenario since the sources are only active at late times when these have already become extremely slow. We also note that we confirmed that the new terms in the line-of-sight approach due to $\partial_\eta \vek{b}^{(0)}_0$ (see Appendix~\ref{eq:add_LOS_term}) are indeed noticeable at large angular scales in the considered scenario.}

\begin{figure}
    \centering
\includegraphics[width=0.8\columnwidth]{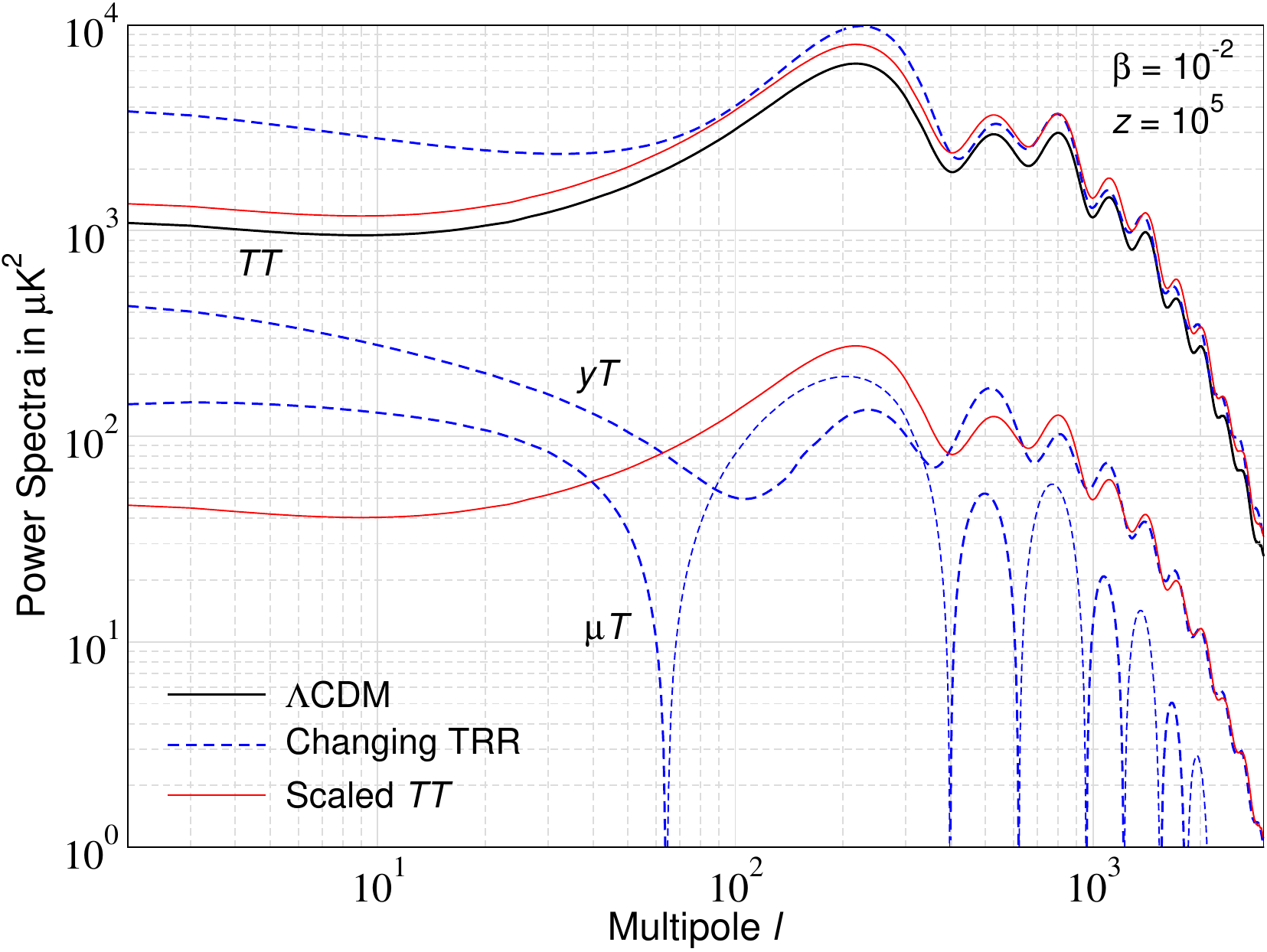}
\\
\caption{CMB power spectra for the settings of Fig.~\ref{fig:T-shift-II} [observational basis with $Y_k$ up to $Y_5$.]. The solid black line is the $\Lambda$CDM case (which has no distortion contributions), while the blue-dashed lines are computed for $\beta=10^{-2}$ and $z_{\rm s}=10^5$, showing the total $TT$, $yT$ and $\mu T$ power spectra. Around the maximum of the Thomson visibility function one has $\bar{\Theta}\approx 0.041$, but we used $\bar{\Theta}\approx 0.038$ in the scaling, which matches the result better. The scaled models (solid red lines) use the $\Lambda$CDM case to obtain $C_\ell^{TT-{\rm scaled}}=(1+3 \Thetabar)^2 \,C_\ell^{TT}$ and $C_\ell^{yY-{\rm scaled}}=(1+3 \Thetabar)\times \Thetabar \,C_\ell^{TT}$. For the $\mu T$ signal, negative branches are shown with a thinner line.}
\label{fig:TT-power-shift-II}
\end{figure}
In Fig.~\ref{fig:TT-power-shift-II}, we repeat the exercise for $z_{\rm s}=10^5$, for which we expect additional thermalization effects to become visible. Again we find the scaled predictions to be consistent with the $yT$ power spectrum at small scales, while in all cases departures from thermal power spectra are present at large angular scales. A more significant $\mu T$ contribution is now present, in this case being also created by partial thermalization effects, providing time-dependent information about the problem. 

Overall these examples imply that the CMB distortion anisotropies may be used to constrain scenarios with changing TRR, as also pointed out in \citep{Chluba2014TRR, chluba_spectro-spatial_2023-II}. However, simple predictions do not seem to be possible and detailed radiative transport effects have to be included. For specific scenarios, additional effects relating to changes to the average Hubble expansion rate will also have to be considered. We shall leave a more detailed investigation of this problem to another paper.

\begin{figure}
    \centering
\includegraphics[width=0.8\columnwidth]{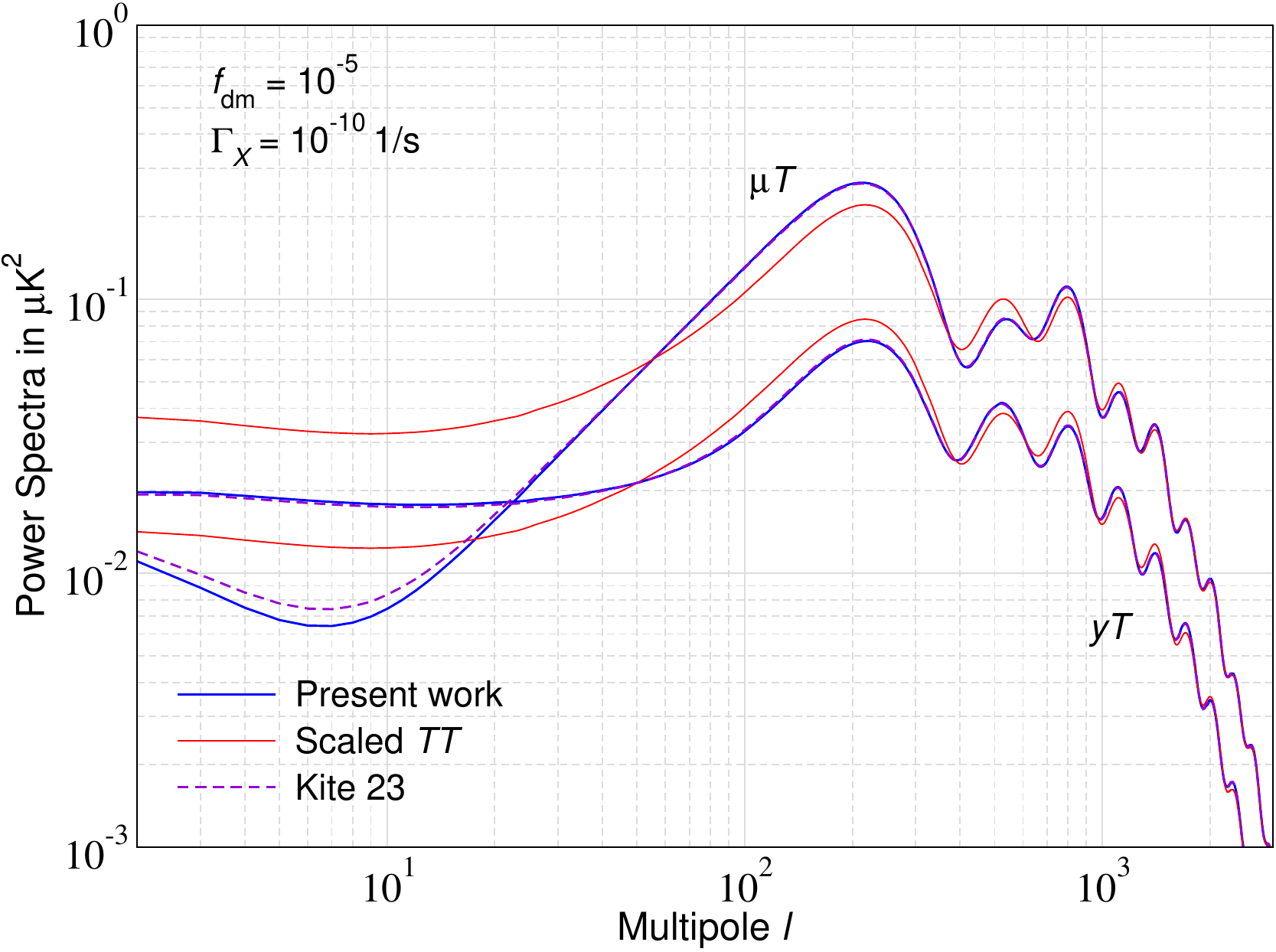}
\\
\caption{CMB power spectra for a decaying particle scenario with $f_{\rm dm}=10^{-5}$ and $\Gamma_X=10^{-10}\,{\rm s^{-1}}$ [observational basis with $Y_k$ up to $Y_5$.]. The solid blue lines show the results obtained with our improved frequency hierarchy setup, while the violet lines use the implementation of \citep{kite_spectro-spatial_2023-III}. For comparison we also show the scaled $TT$ power spectra using a factor $f=\pot{3.4}{-5}$ for the $\mu T$ case and $f=\pot{1.3}{-5}$ for $yT$. We note that the $\mu T$ signal is negative for the considered example, and thus was multiplied by $-1$.}
\label{fig:power_spectra_decay}
\end{figure}
\subsection{Energy release from decaying particles}
\label{sec:decay}
As a final example we consider a simple decaying particle scenario using the heating rates for
a constant lifetime, $t_X=1/\Gamma_X$, and mass of the particle, $m_X$ \citep[see][]{kite_spectro-spatial_2023-III}
\bsub
\begin{align}
\label{eq:decay_heating}
\frac{\id \mathcal{Q}^{(0)}}{\id t}
&\approx \frac{m_X c^2\,\Gamma_X\,N_X}{\rho_\gamma}=\frac{\rho_{X,0}\,\Gamma_X\,\expf{-\Gamma_X t}}{\rho_{\gamma,0} (1+z)}
\approx \pot{4.85}{3}\,
f_{\rm dm}\,
\left[\frac{\Omega_{\rm cdm}h^2}{0.12}\right]\,\frac{\Gamma_X\,\expf{-\Gamma_X t}}{1+z}
\\
\frac{\id \mathcal{Q}^{(1)}}{\id t}
&\approx (\delta_{\rm cdm}^{(1)}+\Psi^{(1)})\,\frac{\id \mathcal{Q}^{(0)}}{\id t},
\end{align}
\esub
where we introduced $f_{\rm dm}=\rho_{X,0}/\rho_{\rm cdm,0}$ to allow the variation of the dark matter fraction that is made up by the particle. As shown in \citep{kite_spectro-spatial_2023-III}, the detailed spectro-spatial signal created in these scenarios should in principle give an opportunity to directly constrain the lifetime of the particle.

In Fig.~\ref{fig:power_spectra_decay} we illustrate the CMB power spectra for a decaying particle scenario with $f_{\rm dm}=10^{-5}$ and $\Gamma_X=10^{-10}\,{\rm s^{-1}}$. 
The signals are consistent with the upper limits from \COBEF on the average distortions and hence are suppressed by a factor $\simeq 10^{-5}$ relative to the $TT$ signal.
In particular, we compare with the predictions obtained with the original setup\footnote{We did find a small bug in the original implementation with a factor of 4 missing in one of the heating terms, but it did not change the overall picture significantly.} of \citep{kite_spectro-spatial_2023-III}, which does not account for the new effects considered here. However, overall the agreement is extremely good, showing that the approximation were indeed mostly justified.

\section{Conclusions}
\label{sec:conclusions}
We have studied improvements to the frequency hierarchy treatment of CMB spectral distortions developed previously in \citep{chluba_spectro-spatial_2023-II, kite_spectro-spatial_2023-III}. The initial formulation dropped some of the stimulated scattering effects and also omitted kinematic corrections to the thermalization terms, which we now included here. In addition, we reformulated the problem to separately capture changes to the standard perturbation variables and also explained how to account for sources of photons.

The new effects are extremely important for reaching exact equilibrium distributions in the presence of changes in the average CMB temperature. Related scenarios can only be consistently treated with the new setup developed here, while not increasing the computational time much. We illustrate the solutions for a few examples in Sect.~\ref{sec:illustrations} with a focus on the new effects. We also demonstrate that the new effects do not alter the predictions for the distortion anisotropies {from decaying particles} significantly, even if now the formulation is internally much more sound. {The new framework has now been applied} to problems with photon to dark photon and axion conversions, providing the opportunity to predict anisotropic distortion signals from the pre-recombination {era~\citep{Evangelista2026DP}}.

We note that the improved frequency hierarchy formulation given here is still not exact. Firstly, the basis does not capture some aspects of the distortion evolution at low-frequencies ($\nu\lesssim 10\,\GHz$) as already noted in \citep{chluba_spectro-spatial_2023-I}. Recent work \citep{Evangelista2025} improved the treatment by more carefully matching the $\mu$-distortion template numerically and also considering modifications to the evolution of the critical frequency; however, a more general extension of the frequency basis may be required. One option could be related to the decomposition discussed in \citep{Barenboim2025}, but here one will have to keep the dimensionality of the problem constrained, since for common problems not all parts of the function space are accessed.
In addition, we have not included the effects of thermalization terms caused by anisotropic Compton scattering \citep[e.g.,][]{Chluba2012, chluba_spectro-spatial_2023-II}. These could further alter the spectral evolution, potentially in unexpected ways, which we shall explore in the future also by solving {\it full} single mode transfer problems.
Nevertheless, the overall framework developed here should provide a good starting point for future investigations of the potential that anisotropic CMB distortions hold.

\section*{Acknowledgments}
{We would like to thank the referee for their thorough reading of the manuscript and very constructive comments that helped improve the text.
This work was supported by the UKSA grant: LiteBIRD UK ST/Y005945/1. 
SE is thankful to Dean's Doctoral Scholarship awarded by the University of Manchester.} TD is part of the project The Magnetic Universe with file number OCENW.XL.23.147 of the research programme XL which is (partly) financed by the Dutch Research Council (NWO) under the grant [grant ID \url{https://doi.org/10.61686/ZDGZL85263}].

{\small
\bibliographystyle{plain}
\bibliography{Lit-all}
}

\appendix

\newpage

\vspace{-2mm}
\section{Operators of the thermalization problem}
\vspace{-2mm}

\label{app:operator_props}
We can somewhat reduce the complexity of the calculations by studying the properties of the operators $\boostO=-x\partial_x$ and 
\bsub
\bealf{
\DiffO&=\frac{1}{x^2}\partial_x x^4 \partial_x = 4\boost + x^2\,\partial_x^2
= \boostO(\boostO-3)
\\
\DiffstarO
&=\frac{1}{x^2}\partial_x x^4  = 4x + x^2\,\partial_x
= x (4-\boostO) 
\\
\KompO&=\DiffO+\DiffstarO A= \DiffO + \DiffstarO \frac{w_y+4}{x}
=(\boostO-3)\left[(\boostO-4)-w_y\right]
=(\boostO-3)\left[\boostO-w\right]
}
\esub
involved in the scattering problems. Here we used the functions $w=x(1+2\nbb)=x (\expf{x}+1)/(\expf{x}-1)$, $\nbb=1/(\expf{x}-1)$, $w_y=Y/G=w -4 $, $G=\boostO \nbb$ and $Y=\DiffO \nbb=w_y G = G (w -4)$. Let us start by looking for the operator eigenfunctions. For $\boostO$ we have
\bealf{
\boostO f_\lambda(x)&=\lambda f_\lambda(x)\qquad \rightarrow \qquad f_\lambda(x)= x^{-\lambda}.
}
This directly implies that $(\alpha + \boostO) x^\alpha = 0$, $\DiffO x^{-3}=0$ and $\DiffstarO x^{-4}=0$. We also have
\bealf{
\DiffO f_\lambda&= \lambda(\lambda-3) f_\lambda 
\qquad \leftrightarrow \qquad 
\DiffO g_\lambda = \lambda g_\lambda 
\qquad \rightarrow \qquad 
g^{\pm}_\lambda(x)= x^{-\frac{3}{2}\pm\frac{\sqrt{9+4\lambda}}{2}}.
}
Similarly, we find
\bealf{
\DiffstarO h_\lambda &= \lambda h_\lambda 
\qquad \rightarrow \qquad 
h_\lambda(x)= \frac{\exp{-\lambda/x}}{x^{4}}.
}
We also mention the useful identities
\bealf{
\KompO \frac{G}{x} &= 0 \qquad \KompO G = - Y \qquad  
\KompO M = - \frac{Y}{\beta_M} = -\eta_M Y 
\\
\boostO \frac{G}{x}&=w\,\frac{G}{x}
\qquad 
\boostO w_y = \boostO w = - (w-2 x G)
\\
\boostO G &=3G+Y 
\qquad 
\boostO M = 3M + \eta_M Y - \frac{G+Y}{x} = 3M + \eta_M Y - (w_y+1) \,\frac{G}{x}
\\
\DiffO G &=4Y_1 
\qquad 
\DiffO M = 4 \eta_M Y_1 - \frac{4G+2Y+4Y_1}{x}
\qquad 
\DiffO Y_n = 4 Y_{n+1}-Y_n
\\
\DiffstarO G &=x(G-Y) 
\qquad 
\DiffstarO M = x\left[ M -\eta_M Y + \frac{G+Y}{x}\right]
\qquad 
\DiffstarO Y_n = 4x\left[ Y_n - Y_{n+1}\right]
}
with $\beta_M=3G_2 / [2 G_1]=3\zeta(3)/\zeta(2)=2.1923$, where $G_k=\int x^k \nbb \id x$. 

\newpage

\section{Simplified derivation for the perturbed emission term}
\label{sec:simplified_pert_em}
In \cite{chluba_spectro-spatial_2023-II} a detailed derivation of the perturbed emission and absorption term is given. This treatment can be simplified in the following manner. For $x=h\nu/\kB\Tz$, the net photon emission and absorption term takes the explicit form \citep{Hu1993, Chluba2011therm, Chluba2014}
\begin{align}
\label{eq:gen_emissionterm}
\frac{\partial n_0}{\partial \tau}\Bigg|_{\rm em}&=\frac{\Lambda(x, \The,\Thg)\,\expf{-x\,\Thz/\The}}{x^3}\left[1-n_0\left(\expf{x\,\Thz/\The}-1\right)\right]
=-\frac{\Lambda(x, \The,\Thg)\,(1-\expf{-x\,\Thz/\The})}{x^3}\left[n_0-n_{\rm e}\right]
\end{align}
in the local inertial frame with $n_{\rm e}=\nbb(x \Thz/\The)=1/[\exp(x \Thz/\The)-1]$ being a blackbody at the temperature of the electrons. We introduced the photon emissivity, $\Lambda(x, \The, \Thg)$, which for double Compton scales as $\Lambda(x, \The, \Thg)\propto \Thg^2$ being driven by the high-frequency blackbody photons \citep{Lightman1981, Danese1982, Ravenni2020DC}. For convenience, we shall use the shorthand notation $\Lambda_z\equiv \Lambda(x, \Thz, \Thz)$ below. 

By writing $\The = \Thz (1+\Theta_{\rm e})$, $\Thg = \Thz (1+\Theta_0)$ and $n_0=\nbb(x) + \Delta n$, we can readily evaluate the emission term at the background level (i.e., the uniform contribution)
\bealf{
\label{eq:em_zeroth}
\frac{\partial n^{(0)}_0}{\partial \tau}\Bigg|_{\rm em}
&\approx -\frac{\Lambda_z}{x^2}\,\frac{\nbb}{G}\left[\Delta n^{(0)}_0-
\Theta^{(0)}_{\rm e}\,G\right],
}
where we used $(1-\expf{-x})/x=\nbb/G$ and expanded $n_{\rm e}\approx \nbb(x)+ G(x)\,\Theta_{\rm e}$. The relevant spatially-varying terms can then be seen by thinking about the emission coefficient as
\begin{align}
\label{eq:emissionterm_rewrite}
\frac{\Lambda(x, \The,\Thg)\,(1-\expf{-x\,\Thz/\The})}{x^3}
&=\frac{\Lambda_z}{x^2}\,\frac{\nbb}{G} \times
\frac{\Lambda(x, \The,\Thg)\,(1-\expf{-x\,\Thz/\The})}{\Lambda_z (1-\expf{-x})}.
\end{align}
We can then expand the second ratio into first order of $\Theta_{\rm e}$ and $\Theta_0$ finding
\begin{align}
\label{eq:expansion_emission}
\frac{\Lambda(x, \The,\Thg)\,(1-\expf{-x\,\Thz/\The})}{\Lambda_z (1-\expf{-x})}
&\approx 1
+ \frac{\id \ln  \Lambda}{\id \ln \Thg}\Bigg|_{\Thg=\The=\Thz}\,\Theta_0+
\left[\frac{\id \ln  \Lambda}{\id \ln \The}\Bigg|_{\Thg=\The=\Thz}-x\, \nbb\right]\Theta_{\rm e}.
\end{align}
For the average evolution, we neglected the terms $\Theta^{(0)}_{\rm e}$ and $\Theta^{(0)}_0$ at higher order. 
We can now include the spatial variations in both terms of Eq.~\eqref{eq:gen_emissionterm}, finding the simple expression:
\begin{align}
\label{eq:em_first}
\frac{\partial n^{(1)}_0}{\partial \tau}\Bigg|_{\rm em}
&\approx 
-\frac{\Lambda_z}{x^2}\,\frac{\nbb}{G}\Bigg\{\left[\Delta n^{(1)}_0-
\Theta^{(1)}_{\rm e}\,G\right] 
+
\left[\delta^{(1)}_{\rm b}+\Psi^{(1)}
\right]\left[\Delta n^{(0)}_0-
\Theta^{(0)}_{\rm e}\,G\right]
\nonumber\\
&\qquad\qquad +
  \left[
\Theta^{(1)}_0 \frac{\partial \ln \Lambda}{\partial\ln\Thg}\Bigg|_{\Thz}+\left(\frac{\partial \ln \Lambda}{\partial\ln\The}\Bigg|_{\Thz}
-x\, \nbb \right) \Theta^{(1)}_{\rm e}
\right]\left[\Delta n^{(0)}_0-
\Theta^{(0)}_{\rm e}\,G\right]\Bigg\},
\end{align}
where we did not include any higher order uniform terms (i.e., $\propto [\Theta^{(0)}_{\rm e}]^2$), as these will be a correction to the evolution of the average distortion. Using $\Theta_{\rm e}^{(1)}\approx \Theta_0^{(1)}$ (well justified because of Compton scattering) and taking the low-frequency limit (i.e., $\nbb/G\rightarrow 1$ and $x\,\nbb\rightarrow 1$) we then obtain
\begin{align}
\label{eq:em_first_low}
\frac{\partial n^{(1)}_0}{\partial \tau}\Bigg|^{\rm low}_{\rm em}
&\approx 
-\frac{\Lambda_z}{x^2}\Bigg\{\Delta n^{(1)}_{0,\rm d} 
+
\left(\delta^{(1)}_{\rm b}+\Psi^{(1)}
+
  \Theta^{(1)}_0  \left[
\frac{\partial \ln \Lambda}{\partial\ln\Thg}\Bigg|_{\Thz}+\frac{\partial \ln \Lambda}{\partial\ln\The}\Bigg|_{\Thz}
-1 
\right]\right)\Delta n^{(0)}_{0,\rm d}
\Bigg\},
\end{align}
where $\Delta n^{(i)}_{0,\rm d}=\Delta n^{(i)}_0-
\Theta^{(i)}_{\rm e}\,G$ is the distortion part of the CMB spectrum. This is the result given in section C.3.2 of \citep{chluba_spectro-spatial_2023-II} and is also used here in Sect.~\ref{sec:photon_prod}. 

\section{Additional Line-of-sight terms}
\label{eq:add_LOS_term}
In the line-of-sight approach one has to compute the Legendre expansion of the source term, which explicitly reads \citep[compare][]{chluba_spectro-spatial_2023-II}
\begin{align}
\mathcal{\vek{S}}_\ell(\eta, \eta_f, k)&=\frac{\i^\ell}{2}\int P_\ell(\chi)\,\expf{-\i k \chi \Delta \eta}\,\frac{\vek{S}_\text{LOS}(\eta, \chi, k)}{\tau'}\id \chi. 
\end{align}
One of the terms that appears in the sources is $\propto -\i k \chi \Psi^{(1)}\,\vek{b}_0^{(0)}$. This term is simplified using
\begin{align}
&\frac{\i^\ell}{2}\int P_\ell(\chi)\,(-\i \kB\chi)\,\expf{-\i k \chi \Delta \eta}\id \chi
=
\partial_{\Delta\eta} j_\ell(\kB\Delta\eta)
=k j_\ell^{(1,0)}(\kB\Delta\eta)=-\partial_\eta j_\ell(\kB\Delta\eta)
\end{align}
to obtain the related source term contribution:
\begin{align}
\frac{\i^\ell}{2}\int P_\ell(\chi)\,\expf{-\i k \chi \Delta \eta}\,\frac{-\i k \chi \Psi^{(1)}\,\vek{b}_0^{(0)}}{\tau'}\id \chi=\frac{k \Psi^{(1)}\,\vek{b}_0^{(0)}}{\tau'} j_\ell^{(1,0)}(\kB\Delta\eta)\equiv - \frac{\Psi^{(1)}\,\vek{b}_0^{(0)}}{\tau'} \partial_\eta j_\ell(\kB\Delta\eta).
\end{align}
It is common to relate this back to $j_\ell(\kB\Delta\eta)$ by partial integration over time. This works as follows: 
\begin{align}
&\int_0^{\eta_f} \id \eta \,g(\eta)\, \frac{\kB\Psi^{(1)}\,\vek{b}_0^{(0)}}{\tau'} j_\ell^{(1,0)}(\kB\Delta\eta)=-\int_0^{\eta_f} \id \eta \,\frac{g(\eta)}{\tau'}\, \Psi^{(1)}\,\vek{b}_0^{(0)} \partial_{\eta} j_\ell(\kB\Delta\eta)
\nonumber\\
&\qquad =-\int_0^{\eta_f} \id \eta \,\frac{g(\eta)}{\tau'}\, \left\{\partial_{\eta}\left[
\Psi^{(1)}\,\vek{b}_0^{(0)} j_\ell(\kB\Delta\eta)\right]-j_\ell(\kB\Delta\eta)\,\partial_{\eta}\left[
\Psi^{(1)}\,\vek{b}_0^{(0)}\right] \right\}
\\ \nonumber
&\qquad =\int_0^{\eta_f} \id \eta \,\frac{g(\eta)}{\tau'}\left[\vek{b}_0^{(0)} \partial_{\eta}
\Psi^{(1)}+\Psi^{(1)}\partial_{\eta}\vek{b}_0^{(0)}\,\right]\,j_\ell(\kB\Delta\eta)
-\int_0^{\eta_f} \id \eta \,\expf{-\tau_{\rm b}}\, \partial_{\eta}\left[
\Psi^{(1)}\,\vek{b}_0^{(0)} j_\ell(\kB\Delta\eta)\right]. 
\end{align}
The last integral can be further simplified using
\begin{align}
&\int_0^{\eta_f} \id \eta \,\expf{-\tau_{\rm b}}\, \partial_{\eta}\left[
\Psi^{(1)}\,\vek{b}_0^{(0)} j_\ell(\kB\Delta\eta)\right]
=
\left[\expf{-\tau_{\rm b}}\, 
\Psi^{(1)}\,\vek{b}_0^{(0)} j_\ell(\kB\Delta\eta)\right]_0^{\eta_f}
-\int_0^{\eta_f} \id \eta \,\left[\partial_{\eta}\expf{-\tau_{\rm b}}\right]\,\Psi^{(1)}\,\vek{b}_0^{(0)} j_\ell(\kB\Delta\eta)
\nonumber\\
&\quad =
\delta_{\ell 0}\,\Psi^{(1)}(\eta_f)\,\vek{b}_0^{(0)}(\eta_f) -
\int_0^{\eta_f} \id \eta \,g(\eta)\,\Psi^{(1)}\,\vek{b}_0^{(0)} j_\ell(\kB\Delta\eta),
\end{align}
where we used $g(\eta)=\partial_{\eta}\expf{-\tau_{\rm b}}=\tau' \expf{-\tau_{\rm b}}$.
It is commonplace to drop the boundary term, $\Psi^{(1)}(\eta_f)\,\vek{b}_0^{(0)}(\eta_f)$, related to the potential fluctuations at the location of the observer as for ensembles of observers this term would vanish on average.
Putting things together this then gives the terms
\begin{align}
\int_0^{\eta_f} \id \eta \,g(\eta)\, \frac{\kB\Psi^{(1)}\vek{b}_0^{(0)}}{\tau'} j_\ell^{(1,0)}(\kB\Delta\eta)
=\int_0^{\eta_f} \id \eta \,g(\eta)\left[
\Psi^{(1)}\left(\vek{b}_0^{(0)}+\frac{\partial_{\eta}\vek{b}_0^{(0)}}{\tau'}\right)+\frac{\vek{b}_0^{(0)}}{\tau'}\,\partial_{\eta}
\Psi^{(1)}\right]\,j_\ell(\kB\Delta\eta),
\end{align}
with a new term $\propto \partial_{\eta}\vek{b}_0^{(0)}$ that was previously omitted because only early energy release scenarios were being considered. Since from Eq.~\eqref{eq:evol_transport_yn_final_bg} we have
\begin{align}
\label{app:new_term}
\frac{\partial_{\eta}\vek{b}_0^{(0)}}{\tau'}=\frac{M_{\rm B} \partial_{\eta}\vek{y}_0^{(0)}}{\tau'}=\Thz M_{\rm B} M_{\rm T} \vek{y}_0^{(0)} +M_{\rm B} \left[\frac{{\vek{Q}'}^{(0)}}{4\tau'}+\frac{\vek{S}'^{(0)}_0}{\tau'}\right]
\end{align}
this term is then added to the thermalization and photon source contributions in Eq.~\eqref{eq:formal_sol_Leg_fin}.

\end{document}